\documentclass[journal, onecolumn]{IEEEtran}
\usepackage{authblk}
\usepackage{amsthm,amsmath,amssymb,color,fullpage,hyperref,cleveref,cite}
\usepackage[algo2e,ruled,vlined]{algorithm2e}
\usepackage{natbib}
\usepackage{multicol}
\usepackage{mathrsfs}
\usepackage{graphicx}
\graphicspath{ {./images/} }
\usepackage{bbm}
\usepackage{algorithmic}
\usepackage{algorithm2e}
\usepackage{subcaption}
\RestyleAlgo{ruled}

\newtheorem{lem}{Lemma}
\newtheorem{thm}{Theorem}
\newtheorem*{thm*}{Theorem}
\newtheorem{prop}{Proposition}

\newtheorem{rem}{Remark}

\def\erdos{Erd\H os}
\def\renyi{R\'enyi }
\def\E{{\mathsf E}}
\def\P{{\mathsf P}}
\def\cg{{\mathcal{G}}}
\def\vuone{{\mathcal{V}^\mathrm{u}_1}}
\def\vutwo{{\mathcal{V}^\mathrm{u}_2}}
\def\euone{{\mathcal{E}^\mathrm{u}_1}}
\def\eutwo{{\mathcal{E}^\mathrm{u}_2}}
\def\eaone{{\mathcal{E}^\mathrm{a}_1}}
\def\eatwo{{\mathcal{E}^\mathrm{a}_2}}
\def\eu{{\mathcal{E}^\mathrm{u}}}
\def\ea{{\mathcal{E}^\mathrm{a}}}
\def\vu{{\mathcal{V}^\mathrm{u}}}
\def\va{{\mathcal{V}^\mathrm{a}}}
\def\var{{\mathrm{Var}}}
\def\cov{{\mathrm{Cov}}}
\def\rmu{{\mathrm{u}}}
\def\rma{{\mathrm{a}}}
\def\ra{{\rho_\mathrm{a}}}
\def\ru{{\rho_\mathrm{u}}}
\def\qa{{q_\mathrm{a}}}
\def\qu{{q_\mathrm{u}}}
\def\aa{{A^\mathrm{a}}}
\def\au{{A^\mathrm{u}}}
\def\ba{{B^\mathrm{a}}}
\def\bu{{B^\mathrm{u}}}

\def\cA{{\mathcal{A}}}
\def\cB{{\mathcal{B}}}

\def\L{\Lambda}

\newcommand{\AttrRich}{\textsc{AttrRich}}
\newcommand{\AttrSparse}{\textsc{AttrSparse}}
\begin{document}
\title{Efficient Algorithms for Attributed Graph Alignment with Vanishing Edge Correlation}
\date{}
\author{Ziao Wang,\thanks{Ziao Wang is with the Department of Electrical and Computer Engineering, University of British Columbia, Vancouver, BC V6T1Z4, Canada (email: ziaow@ece.ubc.ca).}  Weina Wang, and Lele Wang
\thanks{Weina Wang is with the Computer Science Department, Carnegie Mellon University, Pittsburgh, PA 15213, USA, (email: weinaw@cs.cmu.edu).}
\thanks{Lele Wang is with the Department of Electrical and Computer Engineering, University of British Columbia, Vancouver, BC V6T1Z4, Canada (email: lelewang@ece.ubc.ca).}
\thanks{Accepted for presentation at the
Conference on Learning Theory (COLT) 2024.}}

% \author[1]{Ziao Wang}
% \author[2]{Weina Wang}
% \author[1]{Lele Wang}
% \affil[1]{Department of Electrical and Computer Engineering, University of British Columbia}
% \affil[2]{Department of Computer Science, Carnegie Mellon University}

\maketitle

\begin{abstract}
    Graph alignment refers to the task of finding the vertex correspondence between two correlated graphs of $n$ vertices. Extensive study has been done on polynomial-time algorithms for the graph alignment problem under the \erdos--\renyi graph pair model, where the two graphs are \erdos--\renyi graphs with edge probability $\qu$, correlated under certain vertex correspondence. To achieve exact recovery of the correspondence, all existing algorithms at least require the edge correlation coefficient $\ru$ between the two graphs to be \emph{non-vanishing} as $n\rightarrow\infty$. Moreover, it is conjectured that no polynomial-time algorithm can achieve exact recovery under vanishing edge correlation $\ru<1/\mathrm{polylog}(n)$. In this paper, we show that with a vanishing amount of additional \emph{attribute information}, exact recovery is polynomial-time feasible under \emph{vanishing} edge correlation $\ru \ge n^{-\Theta(1)}$. We identify a \emph{local} tree structure, which incorporates one layer of user information and one layer of attribute information, and apply the subgraph counting technique to such structures. A polynomial-time algorithm is proposed that recovers the vertex correspondence for most of the vertices, and then refines the output to achieve exact recovery. The consideration of attribute information is motivated by real-world applications like LinkedIn and Twitter, where user attributes like birthplace and education background can aid alignment.

\end{abstract}
% \begin{IEEEkeywords}
% Graph theory, statistics, inference algorithms.
% \end{IEEEkeywords}
\section{Introduction}
\begin{sloppypar}
\emph{\textbf{Graph alignment for the \erdos--\renyi graph pair.}}
The graph alignment problem, also referred to as the graph matching or noisy graph isomorphism problem, is the problem of finding the correspondence between the vertices of two correlated graphs. This problem has garnered significant attention due to its widespread use in various real-world applications~\citep{singh2007pairwise,Cho-Lee-progressive2012,Hag-Ng-robust2005,Kor-Lat-Reconciliation2014,Nar-Shm-deanonymizing2009}.
% In the context of social network deanonymization~\citep{Kor-Lat-Reconciliation2014,Nar-Shm-deanonymizing2009}, an agent is tasked with inferring user identities in an anonymized social network, given two graphs depicting friendship connections in separate networks: one anonymized (e.g., Twitter, Netflix), the other with publicly available user identities (e.g., LinkedIn). The graph alignment problem is central to understanding the circumstances under which the user identities in the anonymized network can be correctly inferred.
The graph alignment problem has been studied under various graph generation models. Among those models, the most well-studied one is the \emph{\erdos--\renyi graph pair model} $\mathcal{G}(n,\qu,\ru)$ by~\citet{Ped-Gro-privacy2011}. In this model, let $\Pi^*$ denote an underlying permutation on $[n]\triangleq \{1,\ldots,n\}$ drawn uniformly at random. We generate two graphs $G_1$ and $G_2'$ with vertex set $[n]$ as follows: For integers $(i,j)$ with $1\le i<j\le n$, the pair of indicators $(\mathbbm{1}_{\{(i,j)\in G_1\}},\mathbbm{1}_{\{(\Pi^*(i),\Pi^*(j))\in G_2'\}})$ are i.i.d pairs of Bernoulli random variables with mean $\qu$ and correlation coefficient $\ru$. Given a pair of graphs $(G_1,G_2')\sim \mathcal{G}(n,\qu,\ru)$, the goal is to exactly recover the underlying permutation $\Pi^*$.
\end{sloppypar}

\begin{figure}[htbp]
% \hspace{-30pt}
\centering
% \hfill
\begin{subfigure}{0.45\textwidth}
    \def\svgwidth{1.1\columnwidth}
    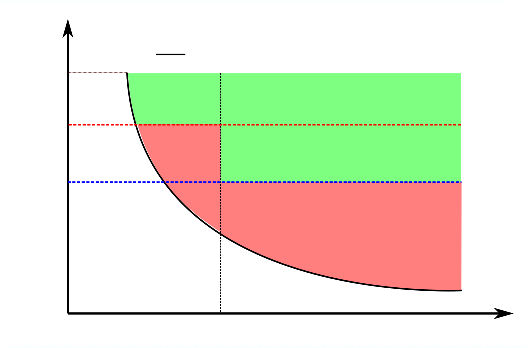
    \caption{}
    \label{fig:no_attri}
\end{subfigure}
\hfill
\begin{subfigure}{0.45\textwidth}
    \def\svgwidth{1.1\columnwidth}
    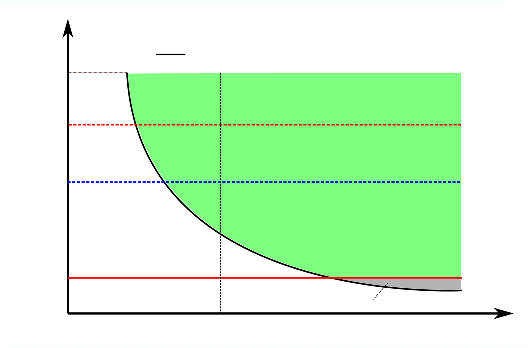
    \caption{}
    \label{fig:with_attri}
\end{subfigure}
\caption{Comparison between polynomial-time feasible regions with (a) no attribute information and (b) vanishing attribute information. Figure~\ref{fig:no_attri} is when $m\qa\ra=0$. The green region represents the best-known polynomial-time feasible region. The portion above the red dotted line is attainable using the algorithm by \cite{mao2023}, while the section to the right of the black dotted line is achievable with the algorithm introduced by \cite{ding2023polynomialtime}. It is further conjectured by~\cite{sophie2023matching} and \cite{mao2022testing} that no polynomial-time algorithm achieves exact recovery in the red region. Figure~\ref{fig:with_attri} is an example of vanishing attribute information $m=\sqrt{n}$, $\qa=\frac{n^{-7/16}}{\sqrt{\log n}}$ and $\ra=n^{-1/16}$. The green region is feasible by the proposed algorithm in this work. Note that for clarity the $\ru$ axis in the plots are not scaled linearly.}
\label{fig:attri-effect}
\end{figure}

Two main lines of study have emerged in the graph alignment problem. In the first line, researchers assume unlimited computational resources and study the range of graph parameters $\qu$ and $\ru$ such that recovering the permutation $\Pi^*$ is feasible or infeasible~\citep{Ped-Gro-privacy2011,Cul-Kiy-improved2016,Cullina-Kiyavash2017,settling-TIT}. This is commonly known as the \emph{information-theoretic limit}. The second line of study focuses on the region of graph statistics where exact alignment is feasible with polynomial-time algorithms~\citep{Dai-Cullina-Kiyavash-Grossglauser2019,fan-20a,ding2021efficient,mao2023exact,mao21a-pmlr,mao2023,ding2023polynomialtime}. This region is commonly known as the polynomial-time feasible region. 
Compared to the information-theoretic feasible region, the best-known polynomial-time feasible region requires further assumptions regarding the edge correlation coefficient $\ru$. To the best of our knowledge, the best-known polynomial-time feasible region is achieved together by \citet{mao2023} and \citet{ding2023polynomialtime}. The proposed polynomial-time algorithm by \citet{mao2023} achieves exact recovery, provided that $\ru$ exceeds a threshold of $\sqrt{\alpha}$, where $\alpha\approx 0.338$ corresponds to Otter's tree counting constant \citep{otter1948}. In the most recent development, \citet{ding2023polynomialtime} specifically focus on the dense regime, characterized by $n\qu= n^{\Theta(1)}$, and propose a novel polynomial-time algorithm that further relaxes the correlation requirement to $\ru>\beta$, where $\beta$ represents an arbitrarily small constant.
% Compared to the information-theoretic feasible region, the best-known polynomial-time feasible region, achieved by an algorithm proposed in~\citep{mao2023}, requires the additional assumption that $\ru>\sqrt{\alpha}$, where $\alpha\approx 0.338$ is Otter's tree counting constant~\citep{otter1948}.
% the existing polynomial-time algorithms require more stringent conditions on the correlation coefficient $\ru$ to exactly recover the permutation $\Pi^*$. To the best of our knowledge, the state-of-the-art feasible region of exact recovery by polynomial-time algorithms is provided in recent work~\citep{mao2023}. In this work, a polynomial-time algorithm is proposed and shown to achieve exact recovery in the information-theoretic \lele{feasible} region with the additional assumption that $\ru>\sqrt{\alpha}$, where $\alpha\approx 0.338$ is Otter's tree counting constant~\citep{otter1948}. 
We delineate the polynomial-time feasible region together with the information-theoretic limit in Figure~\ref{fig:no_attri}.
% It is conjectured in~\citep{sophie2023matching} that no polynomial-time algorithm can achieve exact recovery under a weak correlation $\ru<\sqrt{\alpha}$.
The question of whether there exists any polynomial-time algorithm that achieves exact recovery in the information-theoretic feasible region with $\ru< \sqrt{\alpha}$ in the sparse regime $n\qu=n^{o(1)}$ or with $\ru=o(1)$ in the dense regime $n\qu=n^{\Theta(1)}$ remains open, and the non-existence of such polynomial-time algorithms is conjectured by~\citet{sophie2023matching} and \citet{mao2022testing}. 

% \begin{figure}[!t]
% \floatconts
%   {fig:attri-effect}
%   {\caption{Comparison between polynomial-time feasible regions with (a) no attribute information and (b) vanishing attribute information. Figure~\ref{fig:no_attri} is when $m\qa\ra=0$. The green region represents the best-known polynomial-time feasible region. The portion above the red dotted line is attainable using the algorithm by \cite{mao2023}, while the section to the right of the black dotted line is achievable with the algorithm introduced by \cite{ding2023polynomialtime}. It is further conjectured by~\cite{sophie2023matching} and \cite{mao2022testing} that no polynomial-time algorithm achieves exact recovery in the red region. Figure~\ref{fig:with_attri} is an example of vanishing attribute information $m=\sqrt{n}$, $\qa=\frac{n^{-7/16}}{\sqrt{\log n}}$ and $\ra=n^{-1/16}$. The green region is feasible by the proposed algorithm in this work. Note that for clarity the $\ru$ axis in the plots are not scaled linearly.}}
%   {\vspace{-1.5em}
%     \subfigure[Without attribute information]{\label{fig:no_attri}%
%       \def\svgwidth{0.5\textwidth}
%    \hspace{-10pt} \input{./images/new_region_COLT.eps_tex}}%
%     % \qquad
%     \subfigure[With attribute information]{\label{fig:with_attri}%
%       % \def\svgwidth{1.1\columnwidth}
%       \def\svgwidth{0.5\textwidth}
%     \input{./images/new_region_2_COLT.eps_tex}}
%   }
% \end{figure}

\emph{\textbf{Attributed graph alignment.}}
As mentioned above, it is computationally challenging to achieve exact recovery in the low-correlation regime using only the vertex connection information. This motivates us to exploit additional side information about vertices, termed \emph{attributes}, which are often widely available in real-world applications, to facilitate graph alignment.
% However, we notice that in many real-world applications, in addition to the edges between vertices, there is often additional side information available related to the vertices themselves. 
Examples of publicly available attribute information include (1) LinkedIn user profiles containing age, birthplace, and affiliation and (2) Netflix users' movie-watching histories and ratings.
% For example, LinkedIn provides user profiles on their website that includes information such as age, birthplace, and affiliation. Similarly, anonymized networks such as Netflix make users' movie-watching histories and ratings publicly available. This additional information is commonly referred to as attributes, and it is often publicly available. 
The utilization of attribute information has been considered in various of graph inference tasks such as community detection~\citep{Deshpande2018,Saad2020,Lu2023}, link prediction~\citep{Feng2023,Nie2024} and so on.
In the context of graph alignment, a natural question is how does the existence of such attribute information facilitate the matching process?

To theoretically study the attributed graph alignment problem, the \emph{attributed \erdos--\renyi graph pair model} $\mathcal{G}(n,\qu,\ru; m,\qa,\ra)$ is proposed by~\citet{zhang2024TIT}. In the proposed model, two attributed graphs $G_1$ and $G_2'$ both with $n+m$ vertices are generated. The vertex set of both graphs $G_1$ and $G_2'$ can be partitioned into two subsets. The first subset includes vertices $[n]\triangleq\{1,\ldots,n\}$ which are called the users. 
The second subset includes vertices $\{a_1,\ldots,a_m\}$, which are called the attributes. 
% In both graphs, edges are defined between user and user, and between user and attribute.
Let $\Pi^*$ denote a permutation on $[n]$ drawn uniformly at random, which represents the relabelling of user vertices. In the two graphs, the user-user edges are randomly generated in the same manner as the edges in the \erdos--\renyi graph pair model $\mathcal{G}(n,\qu,\ru)$. 
In this model, we also generate edges between the attributes and users to represent the attachment of those attributes to users. For a user vertex $i$ and an attribute vertex $a_j$, the pair of user-attribute edge indicators $(\mathbbm{1}_{\{(i,a_j)\in G_1\}},\mathbbm{1}_{\{(\Pi^*(i),a_j)\in G_2'\}})$ are i.i.d pairs of $\mathrm{Bern}(q_a)$ random variables with correlation coefficient $\ra$\footnote{We comment that an alternative view of the attribute information defined here is considering each user possesses a length-$m$ binary latent vector, where entry $i$ is the indicator of whether the user is with attribute $i$.}. Similarly to the \erdos--\renyi graph pair model, our goal is to recover the underlying permutation $\Pi^*$.

\emph{\textbf{Constant user-user edge correlation in prior work.}}
The study by \cite{zhang2024TIT} provides an extensive investigation of the information-theoretic limit for graph alignment. The authors almost tightly characterize both the feasible and infeasible regions, with some gaps remaining in certain regimes. Computationally efficient algorithms for attributed graph alignment have been studied by~\cite{wang2023feasible}, where two polynomial-time algorithms are proposed and shown to achieve exact recovery with any $\ru=\Theta(1)$.\footnote{We comment that efficient algorithms for attributed graph alignment are also studied in the line of work~\cite{Zhang-Tong2016,Zhang-Tong-2019,Zhou-Li-Wu-Cao-Ying-Tong2021}, where the focus is the empirical performance rather than the theoretically feasible regions.}
Both algorithms proposed by~\cite{wang2023feasible} take a two-step approach. In the first step, user-attribute connection information is used to align a part of the users which are called the anchors, and in the second step, user-user connections between the anchors and the remaining users are used to complete the alignment.
Compared to the state-of-the-art polynomial-time feasible region under the \erdos--\renyi graph pair model \emph{without attributes}, the benefit of attribute information is reflected by lowering the requirement on $\ru$ from $\ru>\sqrt{\alpha}$ to $\ru=\Theta(1)$. Despite the progress, the performance of algorithms by~\cite{wang2023feasible} is largely constrained by the two-step procedure. Although both user-user and user-attribute edge information are available, they are \emph{separately} used in the two steps for alignment. Moreover, this two-step procedure  requires the exact accuracy of both steps, further restricting the feasible region.

% \emph{\textbf{Our results.}}
\emph{\textbf{Our results.}}
In this paper, we advance the state-of-the-art polynomial-time feasible region
% of efficient algorithms for attributed graph alignment 
and propose a polynomial-time algorithm that achieves exact recovery with edge correlation $\ru\ge n^{-\Theta(1)}$ by leveraging a small amount of attribute information.
% (See Theorem~\ref{thm:exact} for the formal statement). \lele{Theorem~\ref{thm:exact} summarizes the exact requirement... To illustrate...  we show in Figure...}
% \lele{[Could we summarize the improvement in general before going into the detailed comparison below?]}
Theorem~\ref{thm:exact} presents the detailed conditions for the proposed algorithm to achieve exact recovery. To demonstrate a direct comparison to a few existing results on polynomial-time algorithms for graph alignment, we consider an example with attribute-related graph parameters fixed to $m=\sqrt{n}$, $\qa=\frac{n^{-7/16}}{\sqrt{\log n}}$ and $\ra=n^{-1/16}$, and delineate the feasible region of the proposed algorithm in terms of $\ru$ and $n\qu$ in Figure~\ref{fig:with_attri}.
In this example, the amount of attribute side information in the graph generation model is small in the sense that for two correctly matched users $i$ and $\Pi^*(i)$, the expected number of common attribute neighbors is given by $m\qu\ru(1+o(1))=\frac{1+o(1)}{\sqrt{\log n}}$, which is vanishing as $n\rightarrow\infty$. 
% It is shown in the figure that such a small amount of attribute information would not change the information theoretic threshold of exact recovery in terms of $\ru$ and $n\qu$. 
In the following, we separately compare the feasible region in Figure~\ref{fig:with_attri} to the best-known polynomial-time feasible region for graph alignment without attributes by~\cite{mao2023} and the best-known polynomial-time feasible region for attributed graph alignment by~\cite{wang2023feasible}. In addition, we specialize the attributed graph alignment problem to the seeded graph alignment problem, which is another variation of graph alignment, and compare the feasible region of the proposed algorithm to that of the best-known polynomial-time algorithms proposed by~\cite{Mossel-seeded}.
\begin{itemize}[leftmargin=1.5em]
    \item Compared to the feasible region by~\cite{mao2023} shown in Figure~\ref{fig:no_attri}, we observe that the feasible region of the proposed algorithm shown in Figure~\ref{fig:with_attri} allows a vanishing correlation coefficient $\ru=\Omega(n^{-3/16+\delta})$, where $\delta$ is an arbitrarily small constant, instead of requiring $\ru$ to be greater than certain constant. In other words, a small amount of attribute information significantly facilitates the computation in the graph alignment problem by lowering the requirement on $\ru$ in the feasible region of polynomial-time algorithms.

    \item Compared to the feasible region by~\cite{wang2023feasible}, we observe that none of the feasible region shown in Figure~\ref{fig:with_attri} can be achieved by the algorithms by~\cite{wang2023feasible} for two reasons. Firstly, the feasible region presented by~\cite{wang2023feasible} does not allow such a small amount of attribute information $m\qa\ra=\frac{1}{\sqrt{\log n}}$. Secondly, the algorithms by~\cite{wang2023feasible} require both $\ra$ and $\ru$ to be constants, while in this example we set $\ra=n^{-1/16}$ and allow any $\ru=\Omega(n^{-3/16+\delta})$. However, to make the comparison fair, we point out there also exists certain feasible region achieved by the algorithms by~\cite{wang2023feasible}, but not by the proposed algorithm in this work. The feasible region in this work requires a necessary condition $m=n^{\Theta(1)}$, while that in~\cite{wang2023feasible} allows certain feasible region with $m=n^{o(1)}$.
    \item 
    % \lele{[Maybe also compare to the seeded graph alignment work?]}
    % Consider the attributed \erdos--\renyi graph pair model $\cg(n,\qu,\ru;m,\qa,\ra)$ with $\qu=\qa$ and $\ru=\ra$. Then the $m$ attributes can be viewed as $m$ a seeds, and the graphs $(G_1,G_2')$ can be view as a pair of graphs generated from the \emph{seeded \erdos--\renyi graph pair model} $\cg(m+n,p;s,\alpha)$~\cite{Mossel-seeded} with $p= \frac{\qu}{\qu+\ru(1-\qu)}$, $s= \qu+\ru(1-\qu)$ and $\alpha= \frac{m}{m+n}$. In the following, we compare the feasible region of the proposed algorithms to the polynomial-time feasible region presented in~\cite{Mossel-seeded}. For simplicity, we focus on the case with two mild conditions $m=o(n)$ and $\qu=o(1)$ throughout the comparison. Firstly, the feasible region of the proposed algorithm allows a wider range of $s$. While $s=\Theta(1)$ appears as a necessary condition in the feasible region by all the algorithms in~\cite{Mossel-seeded}, the value of $s$ in the feasible region of the proposed algorithm ranges from $s=\Theta(n^{-\Theta(1)})$ to $s=\Theta(1)$. Secondly, in the case of $s=\Theta(1)$ and $np=n^{c}$ for some constant $c$ with $\frac13<c<1$ and $c\neq \frac12$, the number of seeds $m$ required by the proposed algorithm is strictly smaller than that required by the algorithms in~\cite{Mossel-seeded}. To make the comparison fair, we comment that if $np\le n^{1/3}$ or $np=n^{1/2}$, there exists some algorithm in~\cite{Mossel-seeded} that requires less seeds than the proposed algorithm to achieve exact recovery.
    Consider the attributed \erdos--\renyi graph pair model $\mathcal{G}(n,\qu,\ru;m,\qa,\ra)$ with $\qu=\qa$ and $\ru=\ra$. In this model, the $m$ attributes can be interpreted as $m$ seeds in the seeded graph alignment problem. Compared to the best-known existing algorithms for seeded graph alignment proposed by~\cite{Mossel-seeded}, the proposed algorithm in this work can tolerate a weaker edge correlation $\ru$ between the two graphs. While the edge correlation $\ru$ being a constant appears as a necessary condition in the feasible region of all the algorithms by~\cite{Mossel-seeded}, the proposed algorithm's feasible region covers values of $\ru$ ranging from $\Theta(n^{-\Theta(1)})$ to $\Theta(1)$. Moreover, if we restrict the consideration to the case of $\ru=\Theta(1)$, 
    the proposed algorithm requires strictly fewer seeds than the algorithms by~\citet{Mossel-seeded} if $n\qu=n^c$ for $1/3<c<1$ and $c\neq 1/2$. To provide a fair comparison, it is worth mentioning that when $c$ is a constant not satisfying these two conditions or $c=o(1)$, the proposed algorithm requires strictly more seeds. We provide the detailed justification for the above comparison in Appendix~\ref{appd:seeded-comparison}.

\end{itemize}

\begin{figure}[htbp]
% \hspace{-30pt}
\centering
\begin{subfigure}{0.35\textwidth}
    \def\svgwidth{1\columnwidth}
    %% Creator: Inkscape 1.2.2 (b0a84865, 2022-12-01), www.inkscape.org
%% Accompanies image file 'attri_subgraph.eps' (pdf, eps, ps)
\begingroup%
  \makeatletter%
  \providecommand\color[2][]{%
    \errmessage{(Inkscape) Color is used for the text in Inkscape, but the package 'color.sty' is not loaded}%
    \renewcommand\color[2][]{}%
  }%
  \providecommand\transparent[1]{%
    \errmessage{(Inkscape) Transparency is used (non-zero) for the text in Inkscape, but the package 'transparent.sty' is not loaded}%
    \renewcommand\transparent[1]{}%
  }%
  \providecommand\rotatebox[2]{#2}%
  \newcommand*\fsize{\dimexpr\f@size pt\relax}%
  \newcommand*\lineheight[1]{\fontsize{\fsize}{#1\fsize}\selectfont}%
  \ifx\svgwidth\undefined%
    \setlength{\unitlength}{105.69398998bp}%
    \ifx\svgscale\undefined%
      \relax%
    \else%
      \setlength{\unitlength}{\unitlength * \real{\svgscale}}%
    \fi%
  \else%
    \setlength{\unitlength}{\svgwidth}%
  \fi%
  \global\let\svgwidth\undefined%
  \global\let\svgscale\undefined%
  \makeatother%
  \begin{picture}(1,0.75480928)%
    \lineheight{1}%
    \setlength\tabcolsep{0pt}%
    \put(0,0){\includegraphics[width=\unitlength]{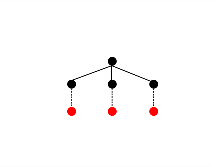}}%
    \put(0.48346369,0.52835045){\color[rgb]{0,0,0}\makebox(0,0)[lt]{\lineheight{1.25}\smash{\begin{tabular}[t]{l}$i$\end{tabular}}}}%
  \end{picture}%
\endgroup%

    \caption{Local attributed tree}
    \label{fig:attri_subgraph}
\end{subfigure}
\begin{subfigure}{0.35\textwidth}
    \def\svgwidth{1\columnwidth}
    %% Creator: Inkscape 1.2.2 (b0a84865, 2022-12-01), www.inkscape.org
%% Accompanies image file 'chandelier.eps' (pdf, eps, ps)
\begingroup%
  \makeatletter%
  \providecommand\color[2][]{%
    \errmessage{(Inkscape) Color is used for the text in Inkscape, but the package 'color.sty' is not loaded}%
    \renewcommand\color[2][]{}%
  }%
  \providecommand\transparent[1]{%
    \errmessage{(Inkscape) Transparency is used (non-zero) for the text in Inkscape, but the package 'transparent.sty' is not loaded}%
    \renewcommand\transparent[1]{}%
  }%
  \providecommand\rotatebox[2]{#2}%
  \newcommand*\fsize{\dimexpr\f@size pt\relax}%
  \newcommand*\lineheight[1]{\fontsize{\fsize}{#1\fsize}\selectfont}%
  \ifx\svgwidth\undefined%
    \setlength{\unitlength}{105.69524028bp}%
    \ifx\svgscale\undefined%
      \relax%
    \else%
      \setlength{\unitlength}{\unitlength * \real{\svgscale}}%
    \fi%
  \else%
    \setlength{\unitlength}{\svgwidth}%
  \fi%
  \global\let\svgwidth\undefined%
  \global\let\svgscale\undefined%
  \makeatother%
  \begin{picture}(1,0.75480035)%
    \lineheight{1}%
    \setlength\tabcolsep{0pt}%
    \put(0,0){\includegraphics[width=\unitlength]{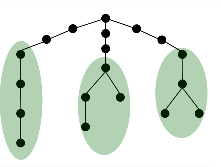}}%
    \put(0.45426261,0.71843263){\color[rgb]{0,0,0}\makebox(0,0)[lt]{\lineheight{1.25}\smash{\begin{tabular}[t]{l}$i$\end{tabular}}}}%
  \end{picture}%
\endgroup%

    \caption{Chandelier}
\label{fig:chandelier}
\end{subfigure}
% \hfill

\caption{Comparison of subgraphs counted in~\cite{mao2023} and in this work. Figure~\ref{fig:attri_subgraph} provides an example of the attributed subgraph we propose to count. In this example, we have $k=3$ branches, and each red node represents an attribute. Figure~\ref{fig:chandelier} provides an example of a chandelier. The structures enclosed by the ovals are the bulbs. }
\label{fig:subgraphs}
\end{figure}

\emph{\textbf{Why can we achieve diminishing correlation coefficient with attributes?}}
The proposed attributed graph alignment algorithm is based on the technique of \emph{subgraph counting}, which is widely applied to various network analysis problems including cluster identification~\citep{mossel2014}, hypothesis testing~\citep{mao2022testing} and graph alignment under the \erdos--\renyi graph pair model~\citep{barak2019,mao2023}. 
In this work, we identify a family of local tree structures involving both users and attributes. The family of trees are indexed by the $\binom{m}{k}$ size-$k$ attribute subsets, and each of them comprises $k$ independent length-$2$ paths connecting the root node $i$ to the $k$ distinct attributes within its index subset, passing through some user vertices. An example of an attributed tree from this family is depicted in Figure~\ref{fig:attri_subgraph}.
By counting the occurrence of these simple tree structures rooted at each user $i$, we are able to construct a feature vector of length $\binom{m}{k}$ for each $i$. Utilizing the similarity of these feature vectors, we can successfully align almost all users under a vanishing edge correlation $\ru=n^{-\Theta(1)}$.

% \begin{figure}[!tbp]
% \floatconts
%   {fig:subgraphs}
%   {\caption{Comparison of subgraphs counted in~\cite{mao2023} and in this work. 
% Figure~\ref{fig:attri_subgraph} provides an example of the attributed subgraph we propose to count. In this example, we have $k=3$ branches, and each red node represents an attribute.
% Figure~\ref{fig:chandelier} provides an example of a chandelier. The structures enclosed by the ovals are the bulbs. }}
%   {
%     \subfigure[Local attributed tree]{\label{fig:attri_subgraph}%
%       \def\svgwidth{0.3\textwidth}
%    \hspace{-10pt} \input{./images/attri_subgraph.eps_tex}}%
%     % \qquad
%     \subfigure[Chandelier]{\label{fig:chandelier}%
%       % \def\svgwidth{1.1\columnwidth}
%       \def\svgwidth{0.3\textwidth}
%     \input{./images/chandelier.eps_tex}}
%   }
% \end{figure}

% Counting these simple tree structures allows us to construct a feature vector of length $\binom{m}{k}$ for each user $i$, and successfully align almost all the users under a vanishing edge correlation $\ru=n^{-\Theta(1)}$ based on the similarity of their feature vectors.

In the context of graph alignment without attributes, a polynomial-time algorithm based on subgraph counting is proposed by~\cite{mao2023}. This algorithm relies on counting a family of subgraph structures called the \emph{chandeliers}. Chandeliers are delicately designed trees rooted at $i$, and each branch of the tree consists of a path starting at $i$ followed by a rooted tree structure called a \emph{bulb}. See Figure~\ref{fig:chandelier} for an example of a chandelier. By counting the chandeliers, the algorithm proposed by~\cite{mao2023} can correctly align almost all the vertices under a constant correlation $\ru> \sqrt{\alpha}$.

% To compare the chandeliers and local attributed tree structures counted in this work, we observe that while the local attributed trees connect the root $i$ to a set of \emph{labelled} attributes, the chandeliers connects the root $i$ to a set of \emph{unlabelled} bulbs. Because the bulbs are unlabelled trees, whether a bulb observed in $G_1$ matches to a same bulb observed in $G_2'$ is uncertain. Therefore, bulbs can be viewed as a form of \emph{noisy} signal for detecting the vertex correspondence, while the attributes in the proposed tree structures can viewed \emph{noiseless} signal. This is the main reason why the proposed algorithm is more robust against edge perturbation between the two graphs and allows a much wider range of edge correlation $\ru$. 
When comparing chandeliers to the local attributed tree structures counted in this work, we observe a key distinction: while the local attributed trees connect the root $i$ to a set of \emph{labelled} attributes, the chandeliers connect the root $i$ to a set of \emph{unlabelled} bulbs. As the bulbs are unlabelled trees, it remains uncertain whether a bulb observed in $G_1$ matches a same bulb observed in $G_2'$. Consequently, bulbs can be considered as a form of \emph{noisy} signal for detecting vertex correspondence, whereas the attributes in the proposed tree structures provide a \emph{noiseless} signal. This crucial distinction renders the proposed algorithm more robust against edge perturbations between the two graphs, enabling a wider range of edge correlation $\ru$ in the feasible region.
Another distinction between the attributed trees and the chandeliers lies in the length of the paths. In the attributed trees, the length of each path is fixed to $2$, while in the chandeliers, the path length between the root and a bulb goes to infinity as $n$ grows. 
In subgraph structures, the vertices on the paths are unlabelled, and thus, a longer wire length would introduce higher noise levels in the subgraph counting algorithm. In chandeliers, the wire length needs to be large enough to counterbalance the noise induced by the overlap in bulb structures. But in our case, as the attributes are individual vertices, we can set the path length to the minimum value of $2$, thereby reducing noise in the alignment process.

% \begin{figure}[htbp]
% % \hspace{-30pt}
% \centering
% \begin{subfigure}{0.35\textwidth}
%     \def\svgwidth{1\columnwidth}
%     \input{./images/attri_subgraph.eps_tex}
%     \caption{Proposed local attributed tree}
%     \label{fig:attri_subgraph}
% \end{subfigure}
% \begin{subfigure}{0.35\textwidth}
%     \def\svgwidth{1\columnwidth}
%     \input{./images/chandelier.eps_tex}
%     \caption{Chandelier}
% \label{fig:chandelier}
% \end{subfigure}
% % \hfill
% \caption{Comparison of subgraphs counted in~\cite{mao2023} and in this work. 
% Figure~\ref{fig:attri_subgraph} provides an example of the attributed subgraph we propose to count. In this example, we have $k=3$ branches, and each red node represents an attribute.
% Figure~\ref{fig:chandelier} provides an example of a chandelier. The structures enclosed by the ovals are the bulbs. }
% \label{fig:subgraphs}
% \end{figure}

% \emph{\textbf{More details on attributed graph alignment.}}
\emph{\textbf{More details on attributed graph alignment.}}
Recall that $G_1$ and $G_2'$ are the two attributed graphs generated from $\cg(n,\qu,\ru; m,\qa,\ra)$. For graph $G_1$, its user-user connection and user-attribute connection are captured by two adjacency matrices $\au$ and $\aa$ respectively. The matrix $\au$ is an $n\times n$ symmetric matrix with all-zero diagonal entries. For a pair of distinct users $(i,j)\in[n]\times[n]$, the entry $A^\mathrm{u}_{ij}$ denotes the indicator of whether the edge $(i,j)$ exists in $G_1$. The matrix $\aa$ is an $n\times m$ matrix. For a pair of user and attribute $(i,a_j)\in [n]\times \{a_1,\ldots,a_m\}$, entry $A^\mathrm{a}_{ij}$ denotes the indicator of whether the edge $(i,a_j)$ exists in $G_1$. For graph $G_2'$, we similarly use two adjacency matrices $\bu$ and $\ba$ to capture its user-user connection and user-attribute connection respectively. 
Let $\sigma_\rmu^2\triangleq\qu(1-\qu)$ denote the variance of a non-diagonal entry in $\au$ or $\bu$, and $\sigma_\rma^2\triangleq\qa(1-\qa)$ denote the variance of an entry in $\aa$ or $\ba$. For a given attributed graph $G$, let $\vu(G)$ denote the set of user vertices in $G$, $\va(G)$ denote the set of attribute vertices in $G$, and $\mathcal{V}(G)\triangleq\vu(G)\cup \va(G)$. For the edges in $G$, let $\eu(G)$ denote the set of all user-user edges in $G$, $\ea(G)$ denote the set of all user-attribute edges in $G$, and $\mathcal{E}(G)\triangleq\ea(G)\cup\eu(G)$.
Let $G^\rmu$ denote the induce subgraph of $G$ on its set of users $\vu(G)$. Let $G^\rma$ be the graph such that $\mathcal{V}(G^\rma)=\mathcal{V}(G)$ and $\mathcal{E}(G^\rma)=\ea(G)$. In other words, $G^\rmu$ and $G^\rma$ denote the user-user part and the user-attribute part of $G$ respectively.
For graphs $G_1$ and $G_2'$, we define short-hand notation $\vuone\triangleq\vu(G_1)$,  $\vutwo\triangleq\vu(G_2')$, 
$\mathcal{V}^\mathrm{a}(G_1)=\mathcal{V}^\mathrm{a}(G_2')\triangleq\va$, $\mathcal{V}_1\triangleq \mathcal{V}(G_1)$, $\mathcal{V}_2\triangleq \mathcal{V}(G_2')$, $\euone\triangleq\eu(G_1)$, $\eutwo\triangleq\eu(G_2')$, $\eaone\triangleq\ea(G_1)$, $\eatwo\triangleq\ea(G_2')$, $\mathcal{E}_1\triangleq\mathcal{E}(G_1)$, and $\mathcal{E}_2\triangleq\mathcal{E}(G_2')$.

\section{Main Results and Proposed Algorithms}
    In this section, we present characterizations of polynomial-time feasible regions and our proposed polynomial-time algorithms that achieve these feasible regions.
In Section~\ref{subsec:results-regions}, we provide polynomial-time feasible regions for almost exact recovery and exact recovery, in Theorem~\ref{thm:almost-exact} and Theorem~\ref{thm:exact}, respectively.
In Section~\ref{subsec:alg-almost}, we describe the proposed algorithm for almost exact recovery, which is based on subgraph counting.
In Section~\ref{subsec:alg-exact}, we present two refinement algorithms, which build upon the output of the almost exact recovery algorithm to achieve exact recovery.
These two algorithms work in two different regimes defined based on how much attribute information is available.

% In this section, we propose three polynomial-time algorithms. The first algorithm is based on subgraph counting. We present its feasible region for almost exact recovery in Theorem~\ref{thm:almost-exact}. The second and third algorithms are designed for refining the almost exact recovery result to achieve exact recovery. They work in \lele{the attribute information sparse regime $m\qa\ra = o(\log n)$ and the attribute information rich regime $m\qa\ra = \Omega(\log n)$} 
% % user-attribute connection and dense user-attribute information 
% respectively. We present the union of the feasible region for exact recovery achieved by the proposed algorithms in Theorem~\ref{thm:exact}.

\subsection{Polynomial-time feasible regions}\label{subsec:results-regions}
\begin{thm}[Almost exact recovery]
    \label{thm:almost-exact}
    Consider an attributed \erdos--\renyi graph pair model \\$\cg(n,\qu,\ru;m,\qa,\ra)$. Suppose
\begin{multicols}{3}
\noindent
    \begin{equation}
        n\qu\ru=\omega(1), 
        \label{eq:almost-cond-1}
    \end{equation}
    \begin{equation}
       m\rho_\rma^2\rho_\rmu^2=\omega(n^{2/k}),
    \label{eq:almost-cond-4}
    \end{equation}
    \begin{equation}
        n\rho_\rmu^2=\omega(n^{2/k}),\label{eq:almost-cond-5}
    \end{equation}
\end{multicols}   
\vspace{-1em}
\begin{equation}
    m\qa\ra=\omega\left(\frac{n^{2/k}}{n\rho_\rmu^2}\right)
    \;\;\text{and}\;\;m\qa\ra=\omega\left(\frac{1}{n\qu\ru}\right)\label{eq:almost-cond-2}
\end{equation}
    % \begin{align}
    %     n\qu\ru&=\omega(1),\label{eq:almost-cond-1}\\
    %     m\qa\ra&=\omega\left(\frac{1}{n\qu\ru}\right),\label{eq:almost-cond-2}\\
    %     m\qa\ra&=\omega\left(\frac{n^{2/k}}{n\rho_\rmu^2}\right)\label{eq:almost-cond-3},\\
    %     m\rho_\rma^2\rho_\rmu^2&=\omega(n^{2/k}),\label{eq:almost-cond-4}\\
    %     n\rho_\rmu^2&=\omega(n^{2/k})\label{eq:almost-cond-5}
    % \end{align}
    for some positive integer $k\ge 3$. Then with high probability, Algorithm~\ref{alg:subgraph-count} with parameters $k$ and 
    \begin{equation}
    \label{eq:choice-tau}
        \tau=c\binom{m}{k}(\ru\sigma_\rmu^2)^k(\ra\sigma_\rma^2)^k\binom{n-1}{k}k!
    \end{equation}
    for some constant $0<c<1$ outputs an index set $ I$ and a vertex correspondence estimate $\hat{\Pi}:I\rightarrow [n]$ satisfying that
    \begin{enumerate}
        \item $\hat{\Pi}(i)=\Pi^*(i), \forall i\in I$;
        \item $|I|\ge n-o(n)$.
    \end{enumerate}
\end{thm}
We outline the key steps for establishing Theorem~\ref{thm:almost-exact} in Section~\ref{sec:main-steps}, while deferring the detailed proof to Appendices~\ref{appd:first-moment} and~\ref{appd:second-moment}.
\begin{sloppypar}
\begin{thm}[Exact recovery]
    \label{thm:exact}
    Consider an attributed \erdos--\renyi graph pair $\cg(n,\qu,\ru;m,\qa,\ra)$. Suppose conditions~\eqref{eq:almost-cond-1}--\eqref{eq:almost-cond-2} in Theorem~\ref{thm:almost-exact} are satisfied. Additionally, assume there exists some arbitrarily small positive constant $\epsilon$ such that
    \begin{equation}
    \label{eq:exact-cond-1}
        n\qu(\qu+\ru(1-\qu))+m\qa(\qa+\ra(1-\qa))\ge (1+\epsilon)\log n,
    \end{equation}
    \begin{equation}
    \label{eq:exact-cond-2}
    \frac{\ru(1-\qu)}{\qu}\ge \epsilon\;\;\text{and}\;\;\frac{\ra(1-\qa)}{\qa}\ge \epsilon.
    \end{equation}
    % and
    % \begin{equation}
    % \label{eq:exact-cond-3}
    % \frac{\ra(1-\qa)}{\qa}\ge \epsilon.
    % \end{equation}
    Then, there exists some polynomial-time algorithm that outputs a permutation $\hat\Pi$, which exactly recovers $\Pi^*$ with high probability, i.e., $\P(\hat\Pi=\Pi^*)=1-o(1)$.
\end{thm}
The proof of Theorem~\ref{thm:exact} is presented in Appendix~\ref{appd:exact}.
\end{sloppypar}

\begin{rem}[Interpretation of conditions for exact recovery]
    In this remark, we provide an interpretation for conditions Theorems~\ref{thm:almost-exact} and~\ref{thm:exact}. Condition \eqref{eq:almost-cond-1} is known as the information theoretic limit of almost exact recovery under the \erdos--\renyi graph pair model \citep{Cullina2019}. As previously explained in our introduction, the quantity $m\qa\ra$ represents the expected number of common attribute neighbors for two correctly matched users, which characterizes the amount of attribute information in the model. Equation \eqref{eq:almost-cond-2} provides lower bounds on the amount of attribute information. Equations \eqref{eq:almost-cond-4} and \eqref{eq:almost-cond-5} are lower bounds for edge correlations $\ru$ and $\ra$. These are in line with the constant correlation assumption (for example $\ru=\Theta(1)$ and $\ra=\Theta(1)$ by \cite{wang2023feasible}).  In our later discussion in Appendix~\ref{appd:tightness}, we illustrate the necessity of conditions~\eqref{eq:almost-cond-1}-\eqref{eq:almost-cond-2} for the proposed algorithm to achieve exact recovery.
    Conditions~\eqref{eq:exact-cond-1} and~\eqref{eq:exact-cond-2} can be better understood under an equivalent subsampling formulation of the attributed \erdos--\renyi graph pair model $\mathcal{G}(n,p,s_\mathrm{u}; m,q,s_\mathrm{a})$ by~\cite{wang2023feasible}, where $p$ and $q$ are the edge probabilities in the base graph for user-user connection and user-attribute connection, and $s_\mathrm{u}$ and $s_\mathrm{a}$ are the subsampling probabilities for user-user edges and user-attribute edges. Under this model, condition~\eqref{eq:exact-cond-1} is equivalent to $nps_\mathrm{u}^2 + mqs_\mathrm{a}^2 \ge (1+\epsilon) \log n$, which is the information-theoretic limit for exact recovery for a certain range of graph parameters~\citep{zhang2024TIT}. Condition~\eqref{eq:exact-cond-2} represents the mild condition that the base graph edge probabilities $p$ and $q$ are bounded away from~$1$.
    % For conditions~\eqref{eq:exact-cond-2} and~\eqref{eq:exact-cond-3}, if we consider a different subsampling formulation of the attributed \erdos--\renyi graph pair model $\mathcal{G}(n,p,s_\mathrm{u}; m,q,s_\mathrm{a})$ in~\cite{wang2023feasible}, they represent the mild conditions that base graph edge probabilities $p$ and $q$ are bounded away from $1$. 
    % The rest of conditions~\eqref{eq:almost-cond-1}--\eqref{eq:almost-cond-5} in Theorem~\ref{thm:almost-exact} are lower bounds on the correlation coefficients $\ru$ and~$\ra$.
\end{rem}

\subsection{Proposed subgraph counting algorithm for almost exact recovery}\label{subsec:alg-almost}
In this section, we introduce the proposed subgraph counting algorithm. For graph $G_1$, let $\tilde{A}^\rmu\triangleq\au-\qu$ denote the $n\times n$ normalized matrix with $\tilde A^\rmu_{ij}=A^\rmu_{ij}-\qu$, and $\tilde{A}^\rma\triangleq\aa-\qa$ denote the $n\times m$ normalized matrix with $\tilde A^\rma_{ij}=A^\rma_{ij}-\qa$.
% normalized adjacency matrices. 
For graph $G_2'$, we similarly define the two normalized adjacency matrices $\tilde{B}^\rmu\triangleq B^\rmu-\qu$ and $\tilde{B}^\rma\triangleq B^\rma-\qa$. As a result, all the off-diagonal entries of $\tilde{A}^\rmu$ and $\tilde{B}^\rmu$ and all entries of $\tilde{A}^\rma$ and $\tilde{B}^\rma$ have zero means.

\begin{figure}[!tbp]
\centering
    \def\svgwidth{0.35\columnwidth}
    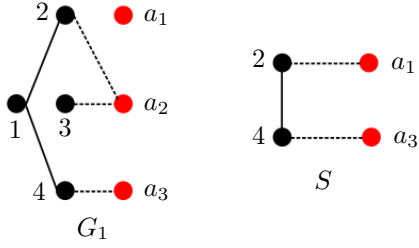
    \caption{An example of computing $\omega_1(S)$. In graph $S$, edge $(4,a_3)$ appears in $G_1$, while $(2,4)$ and $(2,a_1)$ do not. Therefore, these three edges have weights $1-\qa, -\qu$ and $-\qa$ respectively, and the weight of $S$ in $G_1$ is $\omega_1(S)=(1-\qa)\qu\qa$.}
    \label{fig:sub_count}
\end{figure}

Suppose we have an attributed graph $S$ with $\vu(S)\subseteq [n]$ and $\va(S)\subseteq \va$. We define the weight of $S$ in graph $G_1$ as 
\begin{equation}
\label{eq:omega1}
    \omega_1(S)\triangleq\prod_{e\in \eu(S)}\tilde{A}^\rmu_e\prod_{f\in \ea(S)}\tilde{A}^\rma_f.
\end{equation}
% \ziao{Add remark clarifying the term subgraph.}
Similarly, we define the weight of $S$ in graph $G_2'$ as
\begin{equation}
\label{eq:omega2}
    \omega_2(S)\triangleq\prod_{e\in \eu(S)}\tilde{B}^\rmu_e\prod_{f\in \ea(S)}\tilde{B}^\rma_f.
\end{equation}
Note in the definitions above, graph $S$ is not necessarily a subgraph of $G_1$ or $G_2'$. Instead, $S$ is a subgraph of the complete attributed graph defined on user set $[n]$ and attribute set $\va$. We provide an example of computing the weight $\omega_1(S)$ for some $S\not\subseteq G_1$ in Figure~\ref{fig:sub_count}.

With the weight of a subgraph defined, we introduce the family of subgraphs we propose to count. Fix a user vertex $i\in [n]$ and an attribute subset $\mathcal{A}\subseteq \va$ of size $k$, we define $\cg_{i,\mathcal{A}}$ as the collection of all attributed graphs of the following structure: graphs in $\cg_{i,\mathcal{A}}$ are trees rooted at $i$. Each tree has $k$ branches, and each branch is a path of length $2$. The $k$ vertices one hop away from the root $i$ are user vertices from $[n]\setminus\{i\}$, and the $k$ vertices two hops away from the root $i$ are the $k$ attributes in $\mathcal{A}$. In other words, $\cg_{i,\mathcal{A}}$ includes all the ways of forming $k$ length-$2$ independent paths from $i$ to the $k$ attributes in $\mathcal{A}$. We notice that for any $i\in [n]$ and $\mathcal{A}\subseteq \va$ of size $k$, 
\begin{equation}
\label{eq:size-of-subgraphs}
|\cg_{i,\mathcal{A}}|=\binom{n-1}{k}k!.
\end{equation}
This is because in the graph structure we mentioned above, we only have the flexibility to choose the $k$ user vertices that are one hop away from $i$ out of $n-1$ users in $[n]\setminus\{i\}$, and the order of choosing those user vertices matters. Given a fixed user vertex $i\in [n]$ and an attribute subset $\mathcal{A}\subseteq \va$ of size $k$, we define the weighted count of the collection $\cg_{i,\mathcal{A}}$ in $G_1$ as 
\begin{equation}
\label{eq:w1}
    W_{i,\mathcal{A}}(G_1)\triangleq \sum_{S\in \cg_{i,\mathcal{A}}}\omega_1(S).
\end{equation}
Similarly we define the subgraph count in $G_2'$ as
\begin{equation}
\label{eq:w2}
    W_{i,\mathcal{A}}(G_2')\triangleq \sum_{S\in \cg_{i,\mathcal{A}}}\omega_2(S).
\end{equation}

In the proposed graph counting algorithm, for each user vertex $i\in\vuone$, we compute $W_{i,\mathcal{A}}(G_1)$ for each attribute subset $\mathcal{A}$ of size $k$, and use them to form a feature vector of length $\binom{m}{k}$. For each user vertex $j\in\vutwo$, a feature vector is constructed similarly by computing $W_{j,\mathcal{A}}(G_2')$ for each $\mathcal{A}\subseteq \va$ of size $k$.
Given the feature vectors, for a pair of user vertices $(i,j)\in \vuone\times \vutwo$, we define their similarity score $\Phi_{ij}$ as the inner product of their feature vectors:
\begin{equation}    \label{eq:similarity}\Phi_{ij}\triangleq\sum_{\mathcal{A}\in\va:|\mathcal{A}|=k}W_{i,\mathcal{A}}(G_1)W_{j,\mathcal{A}}(G_2').
\end{equation}
In the proposed algorithm, a threshold is set on the similarity scores to determine if user $i$ should be mapped to $j$.
% We then take the inner product of the feature vectors of each pair $(i,j)\in \vuone\times \vutwo$. These inner products serve as the similarity scores for us to match those user vertices. 
We provide a detailed description of the proposed algorithm in Algorithm~\ref{alg:subgraph-count}.

\begin{algorithm2e}[t]
\caption{Attributed graph alignment based on subgraph counting} \label{alg:subgraph-count}
\DontPrintSemicolon
\SetKwInOut{Input}{Input}
\SetKwInOut{Output}{Output}
\Input{Graphs $G_1$ and $G_2'$, graph parameters $\qu,\ru,\qa,\ra$, threshold $\tau$, parameter $k$}
\Output{User index set $I$ and mapping $\hat{\pi}:I\rightarrow [n]$}
\nl $I \gets \emptyset$.\\
\nl For each $i\in \vuone$ and each $\mathcal{A} \subseteq\va$ of size $k$, compute
$W_{i,\mathcal{A}}(G_1)=\sum_{S\in\cg_{i,\mathcal{A}}}\omega_1(S).$\\
\nl For each $j\in \vutwo$ and each $\mathcal{A} \subseteq\va$ of size $k$, compute
$W_{j,\mathcal{A}}(G_2')=\sum_{S\in\cg_{j,\mathcal{A}}}\omega_2(S).$\\
\For{$(i,j)\in \vuone\times\vutwo$}{
\nl    Compute the similarity score between $i$ and $j$: $\Phi_{ij}=\sum_{\mathcal{A}\subseteq\va:|\mathcal{A}|=k}W_{i,\mathcal{A}}(G_1)W_{j,\mathcal{A}}(G_2')$.\\
\If{$\Phi_{ij}\ge \tau$}{
\nl $\hat{\pi}(i)\gets j$.\\
\nl $I\gets I\cup\{i\}$.\\
}
}
\nl \Return{$I$ and $\hat{\pi}$.}
\end{algorithm2e}

\begin{rem}[Complexity of Algorithm~\ref{alg:subgraph-count}]
\label{rem:complexity}
    The time complexity of Algorithm~\ref{alg:subgraph-count} is $O(n^{k+1}m^k)$. To see this, notice that computing $W_{i,\mathcal{A}}(G_1)$ for a vertex $i\in \mathcal{V}_1^\rmu$ requires $O(n^k)$ complexity because $|\cg_{i,\mathcal{A}}|=\binom{n-1}{k}k!$. Moreover, there are $\binom{m}{k}$ attribute subsets of size $k$. Therefore, computing the feature vectors for all the users in $G_1$ requires $O(n^{k+1}m^k)$ time complexity. Similarly, the same time complexity is required for computing the feature vectors for all users in $G_2'$. For computing the similarity scores, the time complexity is $O(n^2m^k)$, because we need to compute the score for $n^2$ pairs of users and each feature vector has length $\binom{m}{k}$. The total time complexity of the algorithm is $O(n^{k+1}m^k)$ because $k\ge 3$. Under our choice of $k=\Theta(1)$ in Theorem~\ref{thm:almost-exact}, Algorithm~\ref{alg:subgraph-count} runs in polynomial-time.
\end{rem}

\begin{rem}[Choice of parameter $k$]
    Regarding the choice of $k$, there is a trade-off between the time complexity and the size of the feasible region. As elaborated in Remark \ref{rem:complexity}, the time complexity of the proposed algorithm scales as $O(n^{k+1}m^k)$, which increases as 
    $k$ assumes larger values. Conversely, as indicated by the conditions~\eqref{eq:almost-cond-4}-\eqref{eq:almost-cond-2}, a larger value of $k$ expands the feasible parameter range. Therefore, to minimize the time complexity of the algorithm, we can optimally choose $k$ to be the smallest integer such that~\eqref{eq:almost-cond-4}-\eqref{eq:almost-cond-2} are satisfied.
\end{rem}

\begin{rem}[Intuition behind the proposed statistics]
    The intuition behind the proposed statistic becomes clearer if we consider the \emph{unnormalized} adjacency matrices ${A}^\mathrm{u}, {A}^\mathrm{a}, {B}^\mathrm{u}, {B}^\mathrm{a}$ when computing the subgraph weights in equations~\eqref{eq:omega1} and~\eqref{eq:omega2}. In this case, the weights $\omega_1(S)$ and $\omega_2(S)$ simply indicate whether $S$ is a subgraph of $G_1$ and $G_2'$ respectively. For instance, with $k=2$ and attribute set $\mathcal{A}=\{a_1,a_2\}$, the feature vector entry $W_{i,\mathcal{A}}(G_1)$ represents the number of desired trees with root $i$ and leaves $a_1$ and $a_2$ in $G_1$. When $(i,j)$ is the correct pair, $W_{i,\mathcal{A}}(G_1)$ and $W_{j,\mathcal{A}}(G_2')$ are positively correlated, whereas they are nearly independent for incorrect pairs. This leads to a higher mean of the product $W_{i,\mathcal{A}}(G_1)W_{j,\mathcal{A}}(G_2')$ for correct pairs. Aggregating the positive correlation over all attribute subsets $\mathcal{A}$ via taking the inner product of feature vectors yields a robust similarity metric resilient to edge perturbation. 
\end{rem}

\subsection{Refinement algorithms for exact recovery}\label{subsec:alg-exact}
In this section, we introduce two algorithms for refining the permutation $\hat\pi$ found by Algorithm~\ref{alg:subgraph-count} to recover the correspondence of all users. The two algorithms work in the attribute information sparse regime $m\qa\ra = o(\log n)$ and the attribute information rich regime $m\qa\ra = \Omega(\log n)$ respectively. We later refer to these two regimes as the \AttrSparse{} regime and the \AttrRich{} regime.

Before we introduce the two algorithms, we first set up essential notation for the neighborhood of user vertices. Given $I'\subseteq [n]$ and injection $\pi:I'\rightarrow [n] $, for a pair of users $i\in \vuone$ and $j\in \vutwo$, we define $N^\rmu_{\pi}(i,j)$ as the number of common user neighbors of $i$ in $G_1$ and $j$ in $G_2'$ under the mapping $\pi$. In other words, $N^\rmu_{\pi}(i,j)$ represents the number of vertices $u\in I'$ such that $i$ is connected to $u$ in $G_1$ and $j$ is connected to $\pi(u)$ in $G_2'$. Moreover, for a pair of user $i\in \vuone$ and $j\in \vutwo$, we define $N^\rma(i,j)$ as the number of common attribute neighbors of $i$ and $j$, i.e., the number of attributes $a\in \va$ such that $a$ is connected to $i$ in $G_1$ and $a$ is connected to $j$ in $G_2'$. We are now ready to introduce the two refinement algorithms.

\emph{\textbf{Refinement algorithm in \AttrSparse{} regime.}}
In this regime, since the amount of attribute information in the two graphs is limited, we use only the user-user edges to align the unmatched users. More explicitly, a pair of users are matched if they have enough matched common user neighbors. The detailed refinement algorithm is stated as Algorithm~\ref{alg:refine-sparse}.
    \begin{algorithm2e}[th]
\caption{Refinement algorithm in \AttrSparse{} regime} \label{alg:refine-sparse}
\DontPrintSemicolon
\SetKwInOut{Input}{Input}
\SetKwInOut{Output}{Output}
\Input{Graphs $G_1$ and $G_2'$, graph parameters $\qu$, threshold parameter $\gamma_1$, outputs $I$ and $\hat{\pi}$ from Algorithm~\ref{alg:subgraph-count}}
\Output{Refined mapping $\tilde{\pi}$}
\nl $J\gets I$, $\tilde{\pi}\gets \hat{\pi}$.\\
\While{$\exists i\notin J, j\notin \tilde{\pi}(J):N^\rmu_{\hat{\pi}}(i,j)\ge \gamma_1 (n-2)q_\rmu^2$}{
\nl $\tilde{\pi}(i)\gets j$, $J\gets J\cup \{i\}$.
}
\nl \Return{$\tilde{\pi}$.}
\end{algorithm2e}
\begin{rem}[Complexity of Algorithm~\ref{alg:refine-sparse}]
    The time complexity of Algorithm~\ref{alg:refine-sparse} is $O(n^3)$. 
    % In the regime $n\qu=\Omega(\log n)$ under which we prove the achievability of Algorithm~\ref{alg:refine-sparse}, the time complexity is $O(n^3q_\rmu^2)$ with high probability. 
    To see the time complexity of Algorithm~\ref{alg:refine-sparse}, we need to analyze the complexity of updating $N^\rmu_{\hat{\pi}}(u,v)$ for all pairs of users $(u,v)$ throughout the process. Notice that in each execution of the while loop in Algorithm~\ref{alg:refine-sparse}, we just need to update $N^\rmu_{\hat{\pi}}(u,v)$ for those user pairs $(u,v)$ such that $u$ is connected to $i$ in $G_1$ and $v$ is connected to $\tilde\pi(i)$ in $G_2$, where $i$ is the vertex newly added to $I$. This implies that in each while loop, the time complexity of updates is bounded by $d_{G_1}(i)d_{G_2'}(\tilde\pi(i))$, where $d_{G_1}(i)$ and $d_{G_2'}(\tilde\pi(i))$ represent the degree of $i$ in $G_1$ and the degree of $\tilde\pi(i)$ in $G_2'$ respectively. Because the number of iterations is $O(n)$ and the order of the user degrees are $O(n)$, the total time complexity is $O(n^3)$. 
    % Moreover, when $n\qu=\Omega(\log n)$, the highest degree in both graphs are $O(n)$ with high probability. Therefore, the time complexity is $O(n^3q_\rmu^2)$ with high probability.
\end{rem}
We comment that Algorithm~\ref{alg:refine-sparse} is first proposed by~\cite{mao2023}. In this work, we adapt the original analysis to allow the vanishing correlation regime $\ru=o(1)$.

\emph{\textbf{Refinement algorithm in \AttrRich{} regime.}}
In this regime, because we have a considerable amount of attribute information, we collaboratively consider user-user edges and  user-attribute edges to align the unmatched users. More specifically, each pair of user vertices are matched if they have enough matched common user neighbors \emph{or} enough common attribute neighbors. The detailed refinement algorithm is delineated as Algorithm~\ref{alg:refine-rich}.
\begin{algorithm2e}[ht]
\caption{Refinement algorithm in \AttrRich{} regime} \label{alg:refine-rich}
\DontPrintSemicolon
\SetKwInOut{Input}{Input}
\SetKwInOut{Output}{Output}
\Input{Graphs $G_1$ and $G_2'$, graph parameters $\qu,\qa$, threshold parameters $\gamma_2,\gamma_3$, outputs $I$ and $\hat{\pi}$ from Algorithm~\ref{alg:subgraph-count}}
\Output{Refined mapping $\tilde{\pi}$}
\nl $J\gets I$, $\tilde{\pi}\gets \hat{\pi}$.\\
\While{$\exists i\notin J, j\notin \tilde{\pi}(J):N^\rmu_{\hat{\pi}}(i,j)\ge \gamma_2 (n-2)q_\rmu^2\text{ or }  N^\rma(i,j)\ge \gamma_3 mq_\rma^2$}{
\nl $\tilde{\pi}(i)\gets j$, $J\gets J\cup \{i\}$.
}
\nl \Return{$\tilde{\pi}$.}
\end{algorithm2e}
\begin{rem}[Time complexity of Algorithm~\ref{alg:refine-rich}]
    The time complexity of Algorithm~\ref{alg:refine-rich} is $O(n^3+n^2m)$. Compared to Algorithm~\ref{alg:refine-sparse}, Algorithm~\ref{alg:refine-rich} additionally needs to compute $N^\rma(u,v)$ for each pair of users $u,v$, which requires $O(n^2m)$ time complexity. Therefore, the total time complexity is $O(n^3+n^2m)$.
\end{rem}

\section{Main Ideas in Achieving Almost Exact Recovery}\label{sec:main-steps}
In this section, we establish Theorem~\ref{thm:almost-exact} based on three intermediate propositions. We defer the proofs of the intermediate propositions to Appendices~\ref{appd:first-moment} and~\ref{appd:second-moment}. From this point, we assume without loss of generality that the underlying permutation $\Pi^*$ is the identity permutation.
\subsection{First moment of similarity score}
For the similarity score $\Phi_{ij}$ to be an appropriate measure of the similarity between two user vertices, we need the expected similarity score between a correct pair $\E[\Phi_{ii}]$ to be larger than that between a wrong pair $\E[\Phi_{ij}]$. The following proposition shows that $\E[\Phi_{ii}]>0$, while $\E[\Phi_{ij}]=0$.
\begin{prop}[Expectation of similarity score]
    \label{prop:expectation}
    For a pair of users $(i,j)\in\vuone\times\vutwo$, the expectation of the similarity score $\Phi_{ij}$ is given by
    \begin{equation}
        \E[\Phi_{ij}]=\binom{m}{k}(\ru\sigma_\rmu^2)^k(\ra\sigma_\rma^2)^k\binom{n-1}{k}k!\mathbbm{1}_{\{i=j\}}.
    \end{equation}
    Moreover, if $k=\Theta(1)$, then
    \begin{equation}
    \label{eq:prop-1-order}
        \E[\Phi_{ii}]=\Omega(m^k n^k(\ru\sigma_\rmu^2)^k(\ra\sigma_\rma^2)^k).
    \end{equation}
\end{prop}
\subsection{Second moment of similarity score}
In the proposed algorithm, we separate almost all the correct pairs of matching from the wrong pairs by setting a threshold on the similarity score. 
For this algorithm to achieve almost exact recovery, requiring only that the correct pairs have a higher expected similarity score is not enough. We further need that the second moment of the similarity score is small enough, so that the similarity scores concentrate well around their means. In the following two propositions, we provide upper bounds on the similarity score between correct pairs and between wrong pairs, respectively.
% \begin{proposition}
%     \label{prop:var-correct-pair}
%     Suppose
%     \begin{align}
%         n\qu\ru&=\omega(1),\label{eq:correct-pair-1}\\
%         n\rho_\rmu^2&=\omega(1),\label{eq:correct-pair-2}\\
%         m\qa\ra&=\omega\left(\frac{1}{n\qu\ru}\right),\label{eq:correct-pair-3}\\
%         m\qa\ra&=\omega\left(\frac{1}{n\rho_\rmu^2}\right),\label{eq:correct-pair-4}\\
%         m\rho_\rma^2\rho_\rmu^2&=\omega(1),\label{eq:correct-pair-5}\\
%         k&=\Theta(1)\label{eq:correct-pair-6}.
%     \end{align}
%     Then for any $i\in [n]$, we have
%     \begin{equation}
%         \frac{\var(\Phi_{ii})}{\E[\Phi_{ii}]^2}=o(1).
%     \end{equation}
% \end{proposition}
\begin{prop}
    \label{prop:var-correct-pair}
    Suppose graph parameters $n,m,\qu,\qa,\ru,\ra$ satisfy conditions $n\qu\ru=\omega(1),$ $n\rho_\rmu^2=\omega(1),$ $m\qa\ra=\omega\left(\frac{1}{n\qu\ru}\right),$ $m\qa\ra=\omega\left(\frac{1}{n\rho_\rmu^2}\right)$ and $m\rho_\rma^2\rho_\rmu^2=\omega(1)$. Assume parameter $k$ in Algorithm~\ref{alg:subgraph-count} is set to a constant. Then for any $i\in [n]$, we have $\frac{\var(\Phi_{ii})}{\E[\Phi_{ii}]^2}=o(1).$
    % \begin{align}
    %     \label{eq:correct-pair-1}\\
    %     n\rho_\rmu^2&=\omega(1),\label{eq:correct-pair-2}\\
    %     m\qa\ra&=\omega\left(\frac{1}{n\qu\ru}\right),\label{eq:correct-pair-3}\\
    %     m\qa\ra&=\omega\left(\frac{1}{n\rho_\rmu^2}\right),\label{eq:correct-pair-4}\\
    %     m\rho_\rma^2\rho_\rmu^2&=\omega(1),\label{eq:correct-pair-5}\\
    %     k&=\Theta(1)\label{eq:correct-pair-6}.
    % \end{align}
    % Then for any $i\in [n]$, we have
    % \begin{equation}
    %     \frac{\var(\Phi_{ii})}{\E[\Phi_{ii}]^2}=o(1).
    % \end{equation}
\end{prop}
\begin{prop}
    \label{prop:var-wrong-pair}
    Suppose graph parameters $n,m,\qu,\qa,\ru,\ra$ satisfy conditions $n\rho_\rmu^2=\omega(n^{2/k}),$ $n^2\qu=\Omega(1),$ $m\qa\ra=\omega\left(\frac{n^{2/k}}{n\rho_\rmu^2}\right)$ and $m\rho_\rma^2\rho_\rmu^2=\omega(n^{2/k}).$ Assume parameter $k$ in Algorithm~\ref{alg:subgraph-count} is set to a constant.
    Then for any $(i,j)\in [n]\times [n]$ with $i\neq j$, we have
        $\frac{\var(\Phi_{ij})}{\E[\Phi_{ii}]^2}=o(1/n^2).$
    % Suppose
    % \begin{align}
    %     n\rho_\rmu^2&=\omega(n^{2/k}),\label{eq:wrong-pair-1}\\
    %     n^2\qu&=\Omega(1),\label{eq:wrong-pair-add}\\
    %     m\qa\ra&=\omega\left(\frac{n^{2/k}}{n\rho_\rmu^2}\right),\label{eq:wrong-pair-2}\\
    %     m\rho_\rma^2\rho_\rmu^2&=\omega(n^{2/k}),\label{eq:wrong-pair-3}\\
    %     k&=\Theta(1).\label{eq:wrong-pair-4}
    % \end{align}
    % Then for any $(i,j)\in [n]\times [n]$ with $i\neq j$, we have
    % \begin{equation}
    %     \frac{\var(\Phi_{ij})}{\E[\Phi_{ii}]^2}=o(1/n^2).
    % \end{equation}
\end{prop}
We comment that the conditions presented in Propositions~\ref{prop:var-correct-pair} and~\ref{prop:var-wrong-pair} are not only sufficient, but also necessary for establishing the variance bounds for the similarity score. We discuss more details about the tightness of these conditions in Appendix~\ref{appd:tightness}.
% \begin{figure}[htbp]
% \floatconts
%   {fig:true-overlap}
%   {\caption{Three overlapping patterns of $(S_1,S_2,T_1,T_2)$ for a correct pair of users $(i,i)$: In these figures we assume $k=3$. The black nodes represent users and the red nodes represent attributes. The solid lines represent user-user edges and the dotted lines represent user-attribute edges. The subgraph names in the figures indicate which subgraphs do the vertices belong to.}}
%   {
%     \subfigure[]{\label{fig:true_overlap1}%
%       \def\svgwidth{0.2\textwidth}
%    \hspace{-30pt}  \input{./images/true_overlap1.eps_tex}}%
%     % \qquad
%     \hspace{5em}
%     \subfigure[]{\label{fig:true_overlap2}%
%       % \def\svgwidth{1.1\columnwidth}
%       \def\svgwidth{0.18\textwidth}
%      \input{./images/true_overlap2.eps_tex}}
%      \hspace{5em}
%     \subfigure[]{\label{fig:true_overlap3}%
%       % \def\svgwidth{1.1\columnwidth}
%       \def\svgwidth{0.22\textwidth}
%      \input{./images/true_overlap3.eps_tex}}
%   }
% \end{figure}
\subsection{Proof of Theorem~\ref{thm:almost-exact}}
Given Propositions~\ref{prop:expectation}-\ref{prop:var-wrong-pair}, we are ready to prove Theorem~\ref{thm:almost-exact}. It is easy to check that all conditions in Propositions~\ref{prop:var-correct-pair} and~\ref{prop:var-wrong-pair} are satisfied under conditions~\eqref{eq:almost-cond-1}-\eqref{eq:almost-cond-2}.
% To prove Theorem~\ref{thm:almost-exact}, we define two error events $\mathcal{E}_1\triangleq \{\exists (i,j)\in [n]\times[n]\colon i\neq j, \Phi_{ij}\ge \tau\}$ and $\mathcal{E}_2\triangleq \{X\ge\lambda n\}$,
% for some $\lambda=o(1)$, which will be specified later, where $X\triangleq |\{i\in [n]\colon\Phi_{ii}<\tau\}|$. 
To prove Theorem~\ref{thm:almost-exact}, we define error event $\mathcal{E}_1\triangleq \{\exists (i,j)\in [n]\times[n]\colon i\neq j, \Phi_{ij}\ge \tau\}$. Moreover, let $X\triangleq |\{i\in [n]\colon\Phi_{ii}<\tau\}|$. We define error event $\mathcal{E}_2\triangleq \{X\ge\lambda n\}$ for some $\lambda=o(1)$, which will be specified later.
We claim that it suffices to show $\P(\mathcal{E}_1\cup \mathcal{E}_2)=o(1)$. To see this, we observe that $\mathcal{E}_1^c$ represents the event that all wrong pairs have similarity score less than the threshold, which implies that $\hat\Pi(i)=\Pi^*(i),\forall i\in I$. On event $\mathcal{E}_1^c$, $\mathcal{E}_2^c$ implies the event that $|I|\ge n-o(n)$.

Firstly, we show that $\P(\mathcal{E}_1)=o(1)$. For a wrong pair $(i,j)\in [n]\times[n]$ with $i\neq j$, we have
\begin{align*}
\P(\Phi_{ij}\ge\tau)&\stackrel{(a)}{=}\P(\Phi_{ij}-\E[\Phi_{ij}]\ge c\E[\Phi_{ii}])\stackrel{(b)}{\le}\frac{\var(\Phi_{ij})}{c^2\E[\Phi_{ii}]^2}\stackrel{(c)}{=}o(1/n^2),
\end{align*}
where (a) follows because $\tau= c\E[\Phi_{ii}]$ from Proposition~\ref{prop:expectation}, (b) follows by Chebyshev's inequality and (c) follows by Proposition~\ref{prop:var-wrong-pair}.
Applying a union bound over the $n^2-n$ wrong pairs of users yields that $\P(\mathcal{E}_1)=o(1)$.

Secondly, let $\lambda=\sqrt{\P(\Phi_{ii}<\tau)}$. We show that $\P(\mathcal{E}_2)=o(1)$ in the following. 
% Recall that $I$ is the set of user vertices mapped by $\hat{\Pi}$. On event $\{\Phi_{ij}<\tau,\forall (i,j)\in[n]\times[n],i\neq j\}$, set $I$ reduces to the set of users $i\in[n]$ such that $\Phi_{ii}\ge \tau$. Let $X$ denote the number of user vertices with $\Phi_{ii}< \tau$. Since we have already shown that $\{\Phi_{ij}<\tau,\forall (i,j)\in[n]\times[n],i\neq j\}$ happens w.h.p, it suffices to show that $\P(X=\epsilon n)=o(1)$ for any positive constant $\epsilon$. 
For the similarity score $\Phi_{ii}$ for a correct pair, we have
\begin{equation}
    \P(\Phi_{ii}<\tau)\stackrel{(d)}{=}\P(\Phi_{ii}-\E[\Phi_{ii}]\le (c-1)\E[\Phi_{ii}])\stackrel{(e)}{\le}\frac{\var(\Phi_{ii})}{(c-1)^2\E[\Phi_{ii}]^2}\stackrel{(f)}{=}o(1)\label{eq:correct-pair-low},
\end{equation}
where (d) follows by Proposition~\ref{prop:expectation}, (e) follows by Chebyshev's inequality and (f) follows by Proposition~\ref{prop:var-correct-pair}. From~\eqref{eq:correct-pair-low}, we have $\lambda=o(1)$. By the linearity of expectation, we know that $\E[X]=n\P(\Phi_{ii}<\tau)$. By Markov's inequality, it follows that
\begin{equation*}
    \P(X\ge \lambda n)\le \frac{\E[X]}{\lambda n}=\sqrt{\P(\Phi_{ii}<\tau)}=o(1).
\end{equation*}
Finally, applying a union bound over $\mathcal{E}_1$ and $\mathcal{E}_2$ completes the proof.

\section*{Acknowledgement}
This work was supported in part by the Natural Sciences and Engineering Research Council of Canada (NSERC) Discovery under Grant RGPIN-2019-05448, in part by the NSERC Collaborative Research and Development under Grant CRDPJ 54367619, and in part by the U.S. National Science Foundation (NSF) under Grant CNS-2007733 and Grant ECCS-2145713. 

\bibliographystyle{apalike}
\bibliography{bibliography.bib}
\newpage
\appendices

\section{First Moment Computation}\label{appd:first-moment}
In this Section, we prove Proposition~\ref{prop:expectation} by computing the expectation of similarity score $\Phi_{ij}$ for correct pairs and wrong pairs.
\renewcommand*{\proofname}{Proof of Proposition~\ref{prop:expectation}.}
\begin{proof}
    By the definition of $\Phi_{ij}$ given in~\eqref{eq:similarity}, we have
    \begin{align}
        \E[\Phi_{ij}]&=\E\left[\sum_{\mathcal{A}\subseteq\vu:|\mathcal{A}|=k}W_{i,\cA}(G_1)W_{j,\cA}(G_2')\right]\nonumber\\
        &\stackrel{(a)}{=}\E\left[\sum_{\mathcal{A}\subseteq\vu:|\mathcal{A}|=k}\sum_{S\in\cg_{i,\cA}}\omega_1(S)\sum_{T\in\cg_{j,\cA}}\omega_2(T)\right]\nonumber\\
        &\stackrel{(b)}{=}\sum_{\mathcal{A}\subseteq\vu:|\mathcal{A}|=k}\sum_{S\in\cg_{i,\cA}}\sum_{T\in\cg_{j,\cA}}\E[\omega_1(S)\omega_2(T)],\label{eq:expectation-expand}
    \end{align}
    where (a) follows because by definitions~\eqref{eq:w1} and~\eqref{eq:w2}, and (b) follows by the linearity of expectation. Now we focus on one term in the summation $\E[\omega_1(S)\omega_2(T)]$. We claim that a necessary condition for $\E[\omega_1(S)\omega_2(T)]$ to be non-zero is that $S=T$. Towards contradiction, suppose $S\neq T$, i.e., there exists an edge $e^*\in E(S)$, but $e^*\notin E(T)$, or  $e^*\in E(T)$, but $e^*\notin E(S)$. Without loss of generality, we only consider the case of $e^*\in E(S)$, but $e^*\notin E(T)$.  We firstly assume that $e^*\in\eu(S)$. By definitions~\eqref{eq:omega1} and~\eqref{eq:omega2}, we have
    \begin{align}
        \E[\omega_1(S)\omega_2(T)]&=\E\left[\prod_{e\in \eu(S)}\tilde{A}^\rmu_e\prod_{f\in \ea(S)}\tilde{A}^\rma_f\prod_{g\in \eu(T)}\tilde{B}^\rmu_g\prod_{h\in \ea(T)}\tilde{B}^\rma_h\right]\nonumber\\
        &\stackrel{(c)}{=}\E[\tilde{A}^\rmu_{e^*}]\E\left[\prod_{e\in \eu(S)\setminus\{e^*\}}\tilde{A}^\rmu_e\prod_{f\in \ea(S)}\tilde{A}^\rma_f\prod_{g\in \eu(T)}\tilde{B}^\rmu_g\prod_{h\in \ea(T)}\tilde{B}^\rma_h\right]\nonumber\\
        &\stackrel{(d)}{=}0,\label{eq:expectation-zero}
    \end{align}
    where (c) follows because $e^*$ only appears in $S$, so $\tilde{A}^\rmu_{e^*}$ in independent of the rest of the terms in the product, and (d) follows because each single term in adjacency matrix $\tilde{A}^\mathrm{u}$ has zero mean. Similarly, if $e^*\in\ea(S)$, we still have $\E[\omega_1(S)\omega_2(T)]=0$. Therefore, the claim holds. The claim further implies that $\E[\Phi_{ij}]=0$ for any pair $i\neq j$. To see this, we notice that $\cg_{i,\cA}\cap\cg_{j,\cA}=\emptyset$ if $i\neq j$, and it follows that each term in the summation~\eqref{eq:expectation-expand} equals to zero. 

    Now, we can rewrite the summation~\eqref{eq:expectation-expand} as 
    \begin{equation}
        \label{eq:expectation-sum}
        \sum_{\mathcal{A}\subseteq\vu:|\mathcal{A}|=k}\sum_{S\in\cg_{i,\cA}}\sum_{T\in\cg_{j,\cA}}\E[\omega_1(S)\omega_2(T)]=\mathbbm{1}_{\{i=j\}}\sum_{\mathcal{A}\subseteq\vu:|\mathcal{A}|=k}\sum_{S\in\cg_{i,\cA}}\E[\omega_1(S)\omega_2(S)].
    \end{equation}
    By symmetry, we notice that for any $|\mathcal{A}|=k$ and any $S\in\cg_{i,\cA}$, the term $\omega_1(S)\omega_2(S)$ has exactly the same distribution. Therefore, we fix a set $\cA$ and a graph $S\in\cg_{i,\cA}$, and consider the expectation $\E[\omega_1(S)\omega_2(S)]$:
    \begin{align}
    \label{eq:expectation-one-term}
        \E[\omega_1(S)\omega_2(S)]&=\E\left[\prod_{e\in \eu(S)}\tilde{A}^\rmu_e\tilde{B}^\rmu_e\prod_{f\in \ea(S)}\tilde{A}^\rma_f\tilde{B}^\rma_f\right]\nonumber\\
        &\stackrel{(e)}{=}\prod_{e\in \eu(S)}\E[\tilde{A}^\rmu_e\tilde{B}^\rmu_e]\prod_{f\in \ea(S)}\E[\tilde{A}^\rma_f\tilde{B}^\rma_f]\nonumber\\
        &\stackrel{(f)}{=}(\sigma_\rmu^2\ru)^k(\sigma_\rma^2\ra)^k,
    \end{align}
    where (e) follows by the independence of the terms in the product and (f) follows because 
    for each $e\in \eu(S)$,
    \begin{equation*}
        \E[\tilde{A}^\rmu_e\tilde{B}^\rmu_e]= \cov(\tilde{A}^\rmu_e,\tilde{B}^\rmu_e) +\E[\tilde{A}^\rmu_e]\E[\tilde{B}^\rmu_e] =\sigma_\rmu^2\ru
    \end{equation*}
    and for each $f\in \ea(S)$,
    \begin{equation*}
        \E[\tilde{A}^\rma_f\tilde{B}^\rma_f] =\cov(\tilde{A}^\rma_f,\tilde{B}^\rma_f)+\E[\tilde{A}^\rma_f]\E[\tilde{B}^\rma_f]=\sigma_\rma^2\ra.
    \end{equation*}  
    Finally, plugging~\eqref{eq:expectation-one-term} into~\eqref{eq:expectation-sum} yields
    \begin{align}
        \E[\Phi_{ij}]&=\mathbbm{1}_{\{i=j\}}\sum_{\mathcal{A}\subseteq\vu:|\mathcal{A}|=k}\sum_{S\in\cg_{i,\cA}}(\sigma_\rmu^2\ru)^k(\sigma_\rma^2\ra)^k\nonumber\\
        &\stackrel{(g)}{=}\binom{m}{k}\binom{n-1}{k}k!(\sigma_\rmu^2\ru)^k(\sigma_\rma^2\ra)^k\mathbbm{1}_{\{i=j\}},
    \end{align}
    where (g) follows because there are $\binom{m}{k}$ attribute subsets of size $k$ and for each $|\cA|=k$, $|\cg_{i,\cA}|=\binom{n-1}{k}k!$ as given in~\eqref{eq:size-of-subgraphs}.

    To see~\eqref{eq:prop-1-order}, we simply need to notice that $\binom{m}{k}=\Omega(m^k)$ and $\binom{n-1}{k}k!=\Omega(n^k)$ when $k=\Theta(1)$.
\end{proof}

\section{Second Moment Computation}\label{appd:second-moment}
In this section, we prove Propositions~\ref{prop:var-correct-pair} and~\ref{prop:var-wrong-pair}, which provide upper bounds on the variances of correct user pairs and wrong user pairs respectively. Towards that goal, we first perform some preprocessing on the the variance expression of the similarity score. By definitions~\eqref{eq:w1}-~\eqref{eq:similarity}, we have
\begin{align}
    \var(\Phi_{ij})&=\var\left(\sum_{\cA\subseteq\va:|\cA|=k} W_{i,\cA}(G_1)W_{j,\cA}(G_2')\right)\nonumber\\
    &=\var\left(\sum_{\cA\subseteq\va:|\cA|=k} \sum_{S_1\in\cg_{i,\cA}} \sum_{S_2\in\cg_{i,\cA}} \omega_1(S_1)\omega_2(S_2)\right)\nonumber\\
    &=\sum_{\cA,\cB\subseteq\va:|\cA|=|\cB|=k} \sum_{S_1\in\cg_{i,\cA}} \sum_{S_2\in\cg_{i,\cA}}\sum_{T_1\in\cg_{i,\cB}} \sum_{T_2\in\cg_{i,\cB}}\cov(\omega_1(S_1)\omega_2(S_2),\omega_1(T_1)\omega_2(T_2)).\label{eq:var-expansion}
\end{align}
Now, we focus on one single term $\cov(\omega_1(S_1)\omega_2(S_2),\omega_1(T_1)\omega_2(T_2))$ in the summation.
Define the indicator $f_1(S_1,S_2,T_1,T_2)\triangleq\mathbbm{1}_{\{S_1\neq S_2 \text{ or }T_1\neq T_2 \text{ or } V(S_1)\cap V(T_1)=\{i\}\}}$. We claim that 
\begin{equation}
\label{eq:claim-cov}
\cov(\omega_1(S_1)\omega_2(S_2),\omega_1(T_1)\omega_2(T_2))=\cov(\omega_1(S_1)\omega_2(S_2),\omega_1(T_1)\omega_2(T_2))f_1(S_1,S_2,T_1,T_2).
\end{equation}
To see the claim, we observe that if $f_1(S_1,S_2,T_1,T_2)=0$, then we have $S_1=S_2$, $S_2=T_2$ and $V(S_1)\cap V(T_1)=\{i\}$, which together imply that $(E(S_1)\cup E(S_2))\cap (E(T_1)\cup E(T_2))=\emptyset$. So $\omega_1(S_1)\omega_2(S_2)$ and $\omega_1(T_1)\omega_2(T_2)$ are two independent random variables with zero covariance. For simplicity, we write $f_1$ instead of $f_1(S_1,S_2,T_1,T_2)$ when the choice of $S_1,S_2,T_1,T_2$ is clear from context.
% Define event $\cC(S_1,S_2,T_1,T_2)\triangleq \{S_1=S_2\}\cap\{T_1=T_2\}\cap\{V(S_1)\cap V(T_1)=\{i\}\}$, i.e., $\cC(S_1,S_2,T_1,T_2)$ denotes the event that $S_1$ and $S_2$ are the same, $T_1$ and $T_2$ are the same, and $S_1$, $T_1$ intersect only at the root vertex $i$. For simplicity, we write $\cC$ instead of $\cC(S_1,S_2,T_1,T_2)$ when the choice of $S_1,S_2,T_1,T_2$ is clear from context. We claim that 
% \begin{equation}
% \label{eq:claim-cov}
% \cov(\omega_1(S_1)\omega_2(S_2),\omega_1(T_1)\omega_2(T_2))=\cov(\omega_1(S_1)\omega_2(S_2),\omega_1(T_1)\omega_2(T_2))\mathbbm{1}_{\{\cC^c\}}.
% \end{equation}
% To see this, we notice that if event $\cC$ happens, then $E(S_1)\cap E(T_1)=\emptyset$ as they intersect only at one vertex. As a result, $(E(S_1)\cup E(S_2))\cap(E(T_1)\cup E(T_2))=\emptyset$, which further implies that $\omega_1(S_1)\omega_2(S_2)$ and $\omega_1(T_1)\omega_2(T_2)$ are two independent variables with zero covariance. 

Furthermore, we notice that by the definition of covariance, we have
\begin{align*}
    &\cov(\omega_1(S_1)\omega_2(S_2),\omega_1(T_1)\omega_2(T_2))\\&=\E[\omega_1(S_1)\omega_2(S_2)\omega_1(T_1)\omega_2(T_2)]-\E[\omega_1(S_1)\omega_2(S_2)]\E[\omega_1(T_1)\omega_2(T_2)]\\
    &\stackrel{(a)}{\le}\E[\omega_1(S_1)\omega_2(S_2)\omega_1(T_1)\omega_2(T_2)],
\end{align*}
where (a) follows by equations~\eqref{eq:expectation-zero} and~\eqref{eq:expectation-one-term}. 
\begin{sloppypar}
Now, we focus on the cross moment $\E[\omega_1(S_1)\omega_2(S_2)\omega_1(T_1)\omega_2(T_2)]$. We define indicator $f_2(S_1,S_2,T_1,T_2)\triangleq \mathbbm{1}_{\{(E(S_1)\triangle E(T_1))\subseteq(E(S_2)\cup E(T_2))\text{ and }(E(S_2)\triangle E(T_2))\subseteq(E(S_1)\cup E(T_1))\}}$, where $\triangle$ denotes the symmetric difference operation. In other words, $f_2(S_1,S_2,T_1,T_2)$ is the indicator that every edge in the union graph of $S_1\cup S_2\cup T_1\cup T_2$ appears in at least two graphs out of $S_1,S_2,T_1,T_2$. We claim that 
\begin{equation}
\label{eq:claim-cross-moment}
\E[\omega_1(S_1)\omega_2(S_2)\omega_1(T_1)\omega_2(T_2)]=\E[\omega_1(S_1)\omega_2(S_2)\omega_1(T_1)\omega_2(T_2)]f_2(S_1,S_2,T_1,T_2).
\end{equation}
The reason for the claim is similar to the reason for~\eqref{eq:expectation-zero} in the proof of Proposition~\ref{prop:expectation}. Suppose there exists an edge $e^*$ that only appears in one of the four graphs $S_1,S_2,T_1,T_2$, then the entry corresponding to $e^*$ in the adjacency matrix is independent of all other terms in the product $\omega_1(S_1)\omega_2(S_2)\omega_1(T_1)\omega_2(T_2)$. Therefore, the expectation of the entry, which is zero, can be pulled out of the expectation of the product, and result in a zero expectation of the product. For simplicity, we write $f_2$ instead of $f_2(S_1,S_2,T_1,T_2)$ when the choice of $S_1,S_2,T_1,T_2$ is clear from context.
\end{sloppypar}
% We define event $\cD(S_1,S_2,T_1,T_2)\triangleq\{(E(S_1)\triangle E(T_1))\subset(E(S_2)\cup E(T_2))\}\cap \{(E(S_2)\triangle E(T_2))\subset(E(S_1)\cup E(T_1))\}$, where $\triangle$ denotes the symmetric difference operation. In other words, $\cD(S_1,S_2,T_1,T_2)$ is the event that every edge in the union graph of $S_1\cup S_2\cup T_1\cup T_2$ appears in at least two graphs out of $S_1,S_2,T_1,T_2$. For simplicity, we write $\cD$ instead of $\cD(S_1,S_2,T_1,T_2)$ when the choice of $S_1,S_2,T_1,T_2$ is clear from context. We claim that 
% \begin{equation}
% \label{eq:claim-cross-moment}
% \E[\omega_1(S_1)\omega_2(S_2)\omega_1(T_1)\omega_2(T_2)]=\E[\omega_1(S_1)\omega_2(S_2)\omega_1(T_1)\omega_2(T_2)]\mathbbm{1}_{\{\cD\}}.
% \end{equation}
% The reason for the claim is similar to the reason for~\eqref{eq:expectation-zero} in the proof of Proposition~\ref{prop:expectation}. Suppose there exists an edge $e^*$ that only appears in one of the four graphs $S_1,S_2,T_1,T_2$, then entry corresponding to $e^*$ in the adjacency matrix is independent of all other terms in the product $\omega_1(S_1)\omega_2(S_2)\omega_1(T_1)\omega_2(T_2)$. Therefore, the expectation of the entry, which is zero, can be pulled out of the expectation of the product, and result in a zero expectation of the product. 
By claims~\eqref{eq:claim-cov} and~\eqref{eq:claim-cross-moment}, we have
\begin{equation}
    \cov(\omega_1(S_1)\omega_2(S_2),\omega_1(T_1)\omega_2(T_2))\le \E[\omega_1(S_1)\omega_2(S_2)\omega_1(T_1)\omega_2(T_2)]f_1 f_2.
    \label{eq:cov-bound}
\end{equation}

Plugging~\eqref{eq:cov-bound} into~\eqref{eq:var-expansion} gives
\begin{align}
    &\var(\Phi_{ij})\nonumber\\
    &\le \sum_{\cA,\cB\subseteq\va:|\cA|=|\cB|=k} \sum_{S_1\in\cg_{i,\cA}} \sum_{S_2\in\cg_{i,\cA}}\sum_{T_1\in\cg_{i,\cB}} \sum_{T_2\in\cg_{i,\cB}}\E[\omega_1(S_1)\omega_2(S_2)\omega_1(T_1)\omega_2(T_2)]f_1 f_2.
    \label{eq:var-expansion-bound}
\end{align}
To simplify~\eqref{eq:var-expansion-bound}, we define $\L_{ij}$ as the set of all 4-tuples $(S_1,S_2,T_1,T_2)$ such that 
\begin{enumerate}
    \item $S_1\in\cg_{i,\cA}$, $S_2\in\cg_{j,\cA}$, $T_1\in\cg_{i,\cB}$, $T_2\in\cg_{j,\cB}$, for some $\cA,\cB\subseteq \va$ with $|\cA|=|\cB|=k$;
    \item $f_1(S_1,S_2,T_1,T_2)=1$;
    \item $f_2(S_1,S_2,T_1,T_2)=1$.
\end{enumerate}
As a result of this definition, we have 
\begin{equation}
    \var(\Phi_{ij})\le\sum_{(S_1,S_2,T_1,T_2)\in \L_{ij}}\E[\omega_1(S_1)\omega_2(S_2)\omega_1(T_1)\omega_2(T_2)].
    \label{eq:var-sum}
\end{equation}
To bound the variance, we enumerate all those 4-tuples $(S_1,S_2,T_1,T_2)\in \L_{ij}$ and bound their corresponding cross moment $\E[\omega_1(S_1)\omega_2(S_2)\omega_1(T_1)\omega_2(T_2)]$. For a fixed 4-tuple $(S_1,S_2,T_1,T_2)$, let $H$ denote the union graph $S_1\cup S_2\cup T_1\cup T_2$. Consider the user-user edges and user-attribute edges separately. Firstly, we partition the set of user-user edges in $H$ into six disjoint subsets:
\begin{equation}
    \eu(H)=K_{11}^\rmu\cup K_{21}^\rmu\cup K_{12}^\rmu\cup K_{22}^\rmu\cup K_{02}^\rmu\cup K_{20}^\rmu, 
\end{equation}
where
\begin{align*}
    K_{11}^\rmu&\triangleq (\eu(S_1)\triangle\eu(T_1))\cap (\eu(S_2)\triangle\eu(T_2)),\\
    K_{21}^\rmu&\triangleq (\eu(S_1)\cap\eu(T_1))\cap (\eu(S_2)\triangle\eu(T_2)),\\
    K_{12}^\rmu&\triangleq (\eu(S_1)\triangle\eu(T_1))\cap (\eu(S_2)\cap\eu(T_2)),\\
    K_{22}^\rmu&\triangleq (\eu(S_1)\cap\eu(T_1))\cap (\eu(S_2)\cap\eu(T_2)),\\
    K_{02}^\rmu&\triangleq (\eu(S_1)\cup\eu(T_1))^c\cap (\eu(S_2)\cap\eu(T_2)),\\
    K_{20}^\rmu&\triangleq (\eu(S_1)\cap\eu(T_1))\cap (\eu(S_2)\cup\eu(T_2))^c.
\end{align*}
We notice that the sets $K_{11}^\rmu, K_{21}^\rmu, K_{12}^\rmu, K_{22}^\rmu, K_{02}^\rmu, K_{20}^\rmu$ are indeed disjoint by their definition, and they include all the user-user edges in $H$ because each edge in $H$ appears in at least two out of $S_1,S_2,T_1,T_2$. Moreover, note that in the product $\omega_1(S_1)\omega_2(S_2)\omega_1(T_1)\omega_2(T_2)$, we are considering the weights of $S_1$ and $T_1$ in $G_1$ and the weights of $S_2$ and $T_2$ in $G_2'$. As a result, for an edge $e\in K_{m_1,m_2}^\rmu$, its contribution to the product $\omega_1(S_1)\omega_2(S_2)\omega_1(T_1)\omega_2(T_2)$ is given by $(\tilde{A}^\rmu_e)^{m_1}(\tilde{B}^\rmu_e)^{m_2}$. 

For the user-attribute edges in $H$, similarly, we partition it into six disjoint sets:
\begin{equation}
    \ea(H)=K_{11}^\rma\cup K_{21}^\rma\cup K_{12}^\rma\cup K_{22}^\rma\cup K_{02}^\rma\cup K_{20}^\rma, 
\end{equation}
where
\begin{align*}
    K_{11}^\rma&\triangleq (\ea(S_1)\triangle\ea(T_1))\cap (\ea(S_2)\triangle\ea(T_2)),\\
    K_{21}^\rma&\triangleq (\ea(S_1)\cap\ea(T_1))\cap (\ea(S_2)\triangle\ea(T_2)),\\
    K_{12}^\rma&\triangleq (\ea(S_1)\triangle\ea(T_1))\cap (\ea(S_2)\cap\ea(T_2)),\\
    K_{22}^\rma&\triangleq (\ea(S_1)\cap\ea(T_1))\cap (\ea(S_2)\cap\ea(T_2)),\\
    K_{02}^\rma&\triangleq (\ea(S_1)\cup\ea(T_1))^c\cap (\ea(S_2)\cap\ea(T_2)),\\
    K_{20}^\rma&\triangleq (\ea(S_1)\cap\ea(T_1))\cap (\ea(S_2)\cup\ea(T_2))^c.
\end{align*}
For the similar reason as the user-user edges, for an edge $e\in K_{m_1,m_2}^\rma $ its contribution to the product $\omega_1(S_1)\omega_2(S_2)\omega_1(T_1)\omega_2(T_2)$ is given by $(\tilde{A}^\rma_e)^{m_1}(\tilde{B}^\rma_e)^{m_2}$. With the partition of the user-user edges and user-attribute edges, we have
\begin{align}
    &\E[\omega_1(S_1)\omega_2(S_2)\omega_1(T_1)\omega_2(T_2)]\nonumber\\
    &=\E\left[\left(\prod_{m_1,m_2}\prod_{e\in K^\rmu_{m_1,m_2}}(\tilde{A}^\rmu_e)^{m_1}(\tilde{B}^\rmu_e)^{m_2}\right) \left(\prod_{m_1,m_2}\prod_{e\in K^\rma_{m_1,m_2}}(\tilde{A}^\rma_e)^{m_1}(\tilde{B}^\rma_e)^{m_2}\right)\right]\nonumber\\
    &=\left(\prod_{m_1,m_2}\prod_{e\in K^\rmu_{m_1,m_2}}\E[(\tilde{A}^\rmu_e)^{m_1}(\tilde{B}^\rmu_e)^{m_2}]\right) \left(\prod_{m_1,m_2}\prod_{e\in K^\rma_{m_1,m_2}}\E[(\tilde{A}^\rma_e)^{m_1}(\tilde{B}^\rma_e)^{m_2}]\right)\nonumber\\
    &=\left(\prod_{m_1,m_2}\left(\E[(\tilde{A}^\rmu_e)^{m_1}(\tilde{B}^\rmu_e)^{m_2}]\right)^{|K^\rmu_{m_1,m_2}|}\right)\left(\prod_{m_1,m_2}\left(\E[(\tilde{A}^\rma_e)^{m_1}(\tilde{B}^\rma_e)^{m_2}]\right)^{|K^\rma_{m_1,m_2}|}\right)\label{eq:cross-moment-1}
\end{align}
Now, we need to consider the cross moments $\E[(\tilde{A}^\rmu_e,)^{m_1}(\tilde{B}^\rmu_e)^{m_2}]$ and $\E[(\tilde{A}^\rma_e)^{m_1}(\tilde{B}^\rma_e)^{m_2}]$. Recall that $(\tilde{A}^\rmu_e,\tilde{B}^\rmu_e)$ is a pair of Bernoulli random variables with correlation coefficient $\ru$ and marginal probability $\qu$, and $(\tilde{A}^\rma_e,\tilde{B}^\rma_e)$ is a pair of Bernoulli random variables with correlation coefficient $\ra$ and marginal probability $\qa$. 
We bound the cross moments in terms of the graph parameters $\qu,\ru,\qa,\ra$ in the following lemma.
\begin{lem}[Cross moments upper bound]
\label{lem:cross-moment}
    Suppose $\qu\le \frac12$ and $\qa\le \frac12$, then there exists a constant $c$ such that
\begin{equation}
\label{eq:user-cross-moment}
    \sigma_\rmu^{-m_1-m_2}\E[(\tilde{A}^\rmu_e)^{m_1}(\tilde{B}^\rmu_e)^{m_2}]\le \begin{cases}
  \ru  & \text{ if } (m_1,m_2)=(1,1),\\
  1 & \text{ if } (m_1,m_2)=(2,0) \text{ or } (0,2),\\
  \frac{1}{\sqrt{\qu}} & \text{ if } (m_1,m_2)=(2,1) \text{ or } (1,2),\\
  c\max\{1,\frac{\ru}{\qu}\} & \text{ if } (m_1,m_2)=(2,2),
\end{cases}
\end{equation}    
and
\begin{equation}
\label{eq:attribute-cross-moment}
    \sigma_\rma^{-m_1-m_2}\E[(\tilde{A}^\rma_e)^{m_1}(\tilde{B}^\rma_e)^{m_2}]\le \begin{cases}
  \ra  & \text{ if } (m_1,m_2)=(1,1),\\
  1 & \text{ if } (m_1,m_2)=(2,0) \text{ or } (0,2),\\
  \frac{1}{\sqrt{\qa}} & \text{ if } (m_1,m_2)=(2,1) \text{ or } (1,2),\\
  c\max\{1,\frac{\ra}{\qa}\} & \text{ if } (m_1,m_2)=(2,2).
\end{cases}
\end{equation}
\end{lem}
\begin{rem}
    % The first three inequalities of~\eqref{eq:user-cross-moment} and~\eqref{eq:attribute-cross-moment} are first proven in~\cite{mao2022testing}. In Lemma~\ref{lem:cross-moment}, we tighten the bound for the case of $(m_1,m_2)=(2,2)$, so that the bounds are tight up to constant for any $\ru=O(1)$ and $\ra=O(1)$.
    The first three inequalities of \eqref{eq:user-cross-moment} and \eqref{eq:attribute-cross-moment} were originally established by \cite{mao2022testing}. In Lemma \ref{lem:cross-moment}, we tighten the bounds specifically for the case of $(m_1, m_2) = (2, 2)$, ensuring that the bounds are tight up to a constant factor for any $\ru = O(1)$ and $\ra = O(1)$.
\end{rem}
\renewcommand*{\proofname}{Proof of Lemma~\ref{lem:cross-moment}.}
\begin{proof}
    We firstly show~\eqref{eq:user-cross-moment} by separately consider different pairs of $(m_1,m_2)$. 
    \begin{itemize}
        \item Suppose $(m_1,m_2)=(0,2)\text{ or }(2,0)$, then we have
        \begin{equation*}
           \sigma_\rmu^{-2}\E[(\tilde{A}^\rmu_e)^2]=\sigma_\rmu^{-2}\E[(\tilde{B}^\rmu_e)^2]=1.
        \end{equation*}
        \item Suppose $(m_1,m_2)=(1,1)$, then we have
        \begin{equation*}
            \sigma_\rmu^{-2}\E[\tilde{A}^\rmu_e\tilde{B}^\rmu_e]=\sigma_\rmu^{-2}\cov(\tilde{A}^\rmu_e,\tilde{B}^\rmu_e)=\ru.
        \end{equation*}
        \item Suppose $(m_1,m_2)=(1,2)\text{ or }(2,1)$, then we have
        \begin{equation*}
            \sigma_\rmu^{-3}\E[(\tilde{A}^\rmu_e)^2\tilde{B}^\rmu_e]=\sigma_\rmu^{-3}\E[\tilde{A}^\rmu_e(\tilde{B}^\rmu_e)^2]=\frac{\ru(1-2\qu)}{\sqrt{\qu(1-\qu)}}.
        \end{equation*}
        Because $\qu\le \frac12$, we have $\frac{1-2\qu}{\sqrt{\qu(1-\qu)}}\le \frac{1}{\sqrt{\qu}}$. It follows that
        $$\frac{\ru(1-2\qu)}{\sqrt{\qu(1-\qu)}}\le \frac{\ru}{\sqrt{\qu}}\le \frac{1}{\sqrt{\qu}}.$$
        \item Suppose $(m_1,m_2)=(2,2)  $, then we have
        \begin{equation*}
            \sigma_\rmu^{-4}\E[(\tilde{A}^\rmu_e)^2(\tilde{B}^\rmu_e)^2]=\frac{q_\rmu^2(1-\qu)^2+\ru\qu(1-\qu)(1-2\qu)^2}{q_\rmu^2(1-\qu)^2}=1+\frac{\ru(1-2\qu)^2}{\qu(1-\qu)}.
        \end{equation*}
        Because $\qu\le \frac12$, we have $\frac{(1-2\qu)^2}{1-\qu}\le 1$, and it follows that
        $$\sigma_\rmu^{-4}\E[(\tilde{A}^\rmu_e)^2(\tilde{B}^\rmu_e)^2]\le 1+\frac{\ru}{\qu}.$$ To complete the proof of~\eqref{eq:user-cross-moment}, we just need to realize that $1+\frac{\ru}{\qu}=\Theta(\frac{\ru}{\qu})$ when $\frac{\ru}{\qu}\ge 1$, and $1+\frac{\ru}{\qu}=\Theta(1)$ when $\frac{\ru}{\qu}<1$.
    \end{itemize}
    The proof of~\eqref{eq:attribute-cross-moment} follows similarly as the proof of~\eqref{eq:user-cross-moment}.
\end{proof}

% Therefore, for each pair of $m_1$ and $m_2$, the cross moment can be calculated and bounded in terms of the graph parameters $\qu,\ru,\qa,\ra$. By Lemma 14 in~\cite{mao2023}, we have
% \begin{equation}
% \label{eq:user-cross-moment}
%     \sigma_\rmu^{-m_1-m_2}\E[(\tilde{A}^\rmu_e)^{m_1}(\tilde{C}^\rmu_e)^{m_2}]\le \begin{cases}
%   \ru  & \text{ if } (m_1,m_2)=(1,1),\\
%   1 & \text{ if } (m_1,m_2)=(2,0) \text{ or } (0,2),\\
%   \frac{1}{\sqrt{\qu}} & \text{ if } (m_1,m_2)=(2,1) \text{ or } (1,2),\\
%   \frac{1}{\qu} & \text{ if } (m_1,m_2)=(2,2),
% \end{cases}
% \end{equation}
% and
% \begin{equation}
% \label{eq:attribute-cross-moment}
%     \sigma_\rma^{-m_1-m_2}\E[(\tilde{A}^\rma_e)^{m_1}(\tilde{C}^\rma_e)^{m_2}]\le \begin{cases}
%   \ra  & \text{ if } (m_1,m_2)=(1,1),\\
%   1 & \text{ if } (m_1,m_2)=(2,0) \text{ or } (0,2),\\
%   \frac{1}{\sqrt{\qa}} & \text{ if } (m_1,m_2)=(2,1) \text{ or } (1,2),\\
%   \frac{1}{\qa} & \text{ if } (m_1,m_2)=(2,2).
% \end{cases}
% \end{equation}
Moreover, we note that because edges in $K^\rmu_{m_1,m_2}$ or $K^\rma_{m_1,m_2}$ appear in $m_1+m_2$ out of the four graphs $S_1,S_2,T_1,T_2$, we have
\begin{equation}
    \label{eq:user-sum}
    4k=2(K^\rmu_{11}+K^\rmu_{02}+K^\rmu_{20})+3(K^\rmu_{12}+K^\rmu_{21})+4(K^\rmu_{22}),
\end{equation}
and
\begin{equation}
    \label{eq:attri-sum}
    4k=2(K^\rma_{11}+K^\rma_{02}+K^\rma_{20})+3(K^\rma_{12}+K^\rma_{21})+4(K^\rma_{22}).
\end{equation}
By equations~\eqref{eq:user-cross-moment}-\eqref{eq:attri-sum}, we can rewrite~\eqref{eq:cross-moment-1} as 
\begin{align}
    &\E[\omega_1(S_1)\omega_2(S_2)\omega_1(T_1)\omega_2(T_2)]\nonumber\\
    &\le\sigma_\rmu^{4k}\sigma_\rma^{4k}\ru^{|K^\rmu_{11}|}\ra^{|K^\rma_{11}|}\qu^{-\frac12|K^\rmu_{21}|-\frac12|K^\rmu_{12}|}\qa^{-\frac12|K^\rma_{21}|-\frac12|K^\rma_{12}|}\nonumber\\
    &\;\;\;\cdot c^{|K^\rmu_{22}|+|K^\rma_{22}|}\max\left\{1,\tfrac{\ru^{|K^\rmu_{22}|}}{\qu^{|K^\rmu_{22}|}}\right\}\max\left\{1,\tfrac{\ra^{|K^\rma_{22}|}}{\qa^{|K^\rma_{22}|}}\right\}.\label{eq:cross-moment-2}
\end{align}
For simplicity of notation, for a 4-tuple $(S_1,S_2,T_1,T_2)\in \L_{ij}$, we define 
\begin{align}
    \Theta^\rmu(S_1,S_2,T_1,T_2)\triangleq 
    \ru^{|K^\rmu_{11}|}\qu^{-\frac12|K^\rmu_{21}|-\frac12|K^\rmu_{12}|}c^{|K^\rmu_{22}|}\max\left\{1,\tfrac{\ru^{|K^\rmu_{22}|}}{\qu^{|K^\rmu_{22}|}}\right\},\label{eq:Theta-user}
\end{align}
and
\begin{align}
    \Theta^\rma(S_1,S_2,T_1,T_2)\triangleq
    \ra^{|K^\rma_{11}|}\qa^{-\frac12|K^\rma_{21}|-\frac12|K^\rma_{12}|}c^{|K^\rma_{22}|}\max\left\{1,\tfrac{\ra^{|K^\rma_{22}|}}{\qa^{|K^\rma_{22}|}}\right\},\label{eq:Theta-attri}
\end{align}
% Furthermore, we notice that sets $K^\rmu_{11},K^\rmu_{21},K^\rmu_{12},K^\rmu_{20},K^\rmu_{02},K^\rmu_{22}$ partition $\eu(H)$ and $K^\rma_{11},K^\rma_{21},K^\rma_{12},K^\rma_{20},K^\rma_{02},K^\rma_{22}$ partition $\ea(H)$, we have 
% \begin{equation}
%     \label{eq:user-partition}
%     \eu(H)=K^\rmu_{11}+K^\rmu_{02}+K^\rmu_{20}+K^\rmu_{12}+K^\rmu_{21}+K^\rmu_{22},
% \end{equation}
% and
% \begin{equation}
%     \label{eq:attri-partition}
%     \ea(H)=K^\rma_{11}+K^\rma_{02}+K^\rma_{20}+K^\rma_{12}+K^\rma_{21}+K^\rma_{22}.
% \end{equation}
% Taking~\eqref{eq:user-sum}$-2\cdot$\eqref{eq:user-partition} yields
% \begin{equation}
% \label{eq:user-property}
%     4k-2\eu(H)=K^\rmu_{12}+K^\rmu_{21}+2K^\rmu_{22},
% \end{equation}
% and taking~\eqref{eq:attri-sum}$-2\cdot$\eqref{eq:attri-partition} yields
% \begin{equation}
% \label{eq:attri-property}
%     4k-2\ea(H)=K^\rma_{12}+K^\rma_{21}+2K^\rma_{22}.
% \end{equation}
% Because of~\eqref{eq:user-property} and~\eqref{eq:attri-property}, we can further rewrite~\eqref{eq:cross-moment-2} as
% \begin{align}
%     &\E[\omega_1(S_1)\omega_2(S_2)\omega_1(T_1)\omega_2(T_2)]\nonumber\\
%     &=\sigma_\rmu^{4k}\sigma_\rma^{4k}\ru^{|K^\rmu_{11}|}\ra^{|K^\rma_{11}|}\qu^{-2k+\eu(H)}\qa^{-2k+\ea(H)}.\label{eq:cross-moment-3}
% \end{align}
As a result, the variance expression~\eqref{eq:var-sum} can be rewritten as
\begin{align}
    &\var(\Phi_{ij})\nonumber\\
    &\le \sigma_\rmu^{4k}\sigma_\rma^{4k}\sum_{(S_1,S_2,T_1,T_2)\in \L_{ij}}\Theta^\rmu(S_1,S_2,T_1,T_2)\Theta^\rma(S_1,S_2,T_1,T_2)\triangleq \Gamma_{ij}.
    \label{eq:gamma}
\end{align}
To prove Propositions~\ref{prop:var-correct-pair} and~\ref{prop:var-wrong-pair}, we perform a type class enumeration analysis based on the different overlapping patterns of the four graphs $S_1,S_2,T_1,T_2$. For each overlapping pattern, we provide a tight-up-to-constant characterization for $\Theta^\rmu(S_1,S_2,T_1,T_2)$ and $\Theta^\rma(S_1,S_2,T_1,T_2)$, so that we can provide an upper bound for $\Gamma_{ij}$, which is tight up to constant.

\subsection{Proof of Proposition~\ref{prop:var-correct-pair}}
To prove Proposition~\ref{prop:var-correct-pair}, it suffices to show that $\frac{\Gamma_{ii}}{\E[\Phi_{ii}]^2}=o(1)$. The key idea in this proof is to partition the set of 4-tuples $\L_{ii}$ into smaller subsets according to the structure of the union graph $H=S_1\cup S_2\cup T_1\cup T_2$. Within each subset, because the union graph structure is fixed, we can manage to tightly bound $\Theta^\rmu(S_1,S_2,T_1,T_2)$ and $\Theta^\rma(S_1,S_2,T_1,T_2)$ and hence provide a tight upper bound for $\Gamma_{ii}$. Firstly, we partition $\L_{ii}$ into $k+1$ disjoint subsets according to the number of shared attributes:
\begin{equation}
    \label{eq:lambda-parition}
    \L_{ii}=\cup_{M=0}^k \L_{ii}^{(M)}, 
\end{equation}
where $\L_{ii}^{(M)}\triangleq \{(S_1,S_2,T_1,T_2)\in\L_{ii}:|\va(S_1)\cap\va(T_1)|=M\}$. Recall that $\va(S_1)=\va(S_2)$ and $\va(T_1)=\va(T_2)$ because $S_1,S_2\in\cg_{i,\cA}$ and $T_1,T_2\in \cg_{i,\cB}$. Thus, $\L_{ii}^{(M)}$ includes all the 4-tuples such that there are $M$ overlap attributes between $\cA$ and $\cB$. By the partition defined, we have
\begin{align}
    \Gamma_{ii}&=\sigma_\rmu^{4k}\sigma_\rma^{4k}\sum_{M=0}^k\sum_{(S_1,S_2,T_1,T_2)\in \L_{ii}^{(M)}}\Theta^\mathrm{a}(S_1,S_2,T_1,T_2)\Theta^\mathrm{u}(S_1,S_2,T_1,T_2)\nonumber\\
    &\le (k+1)\sigma_\rmu^{4k}\sigma_\rma^{4k}\max_{0\le M\le k}\sum_{(S_1,S_2,T_1,T_2)\in \L_{ii}^{(M)}}\Theta^\mathrm{a}(S_1,S_2,T_1,T_2)\Theta^\mathrm{u}(S_1,S_2,T_1,T_2).\label{eq:lambda-sepatate}
\end{align}
Because $k=\Theta(1)$, it suffices to show that for each $0\le M\le k$,
\begin{equation}
\label{eq:gamma-m}
    \sigma_\rmu^{4k}\sigma_\rma^{4k}\sum_{(S_1,S_2,T_1,T_2)\in \L_{ii}^{(M)}}\Theta^\mathrm{a}(S_1,S_2,T_1,T_2)\Theta^\mathrm{u}(S_1,S_2,T_1,T_2)\triangleq\Gamma_{ii}^{(M)}=o(\E[\Phi_{ii}]^2).
\end{equation}
To show this, we separately consider the cases of $M=0$ and $1\le M\le k$. 

    \emph{\textbf{Case 1: $M=0$.}} In this case, the two attribute subsets $\va(S_1)\cap \va(T_1)=\emptyset$. It turns out the fact that each edge in the union graph must appear in at least two of $S_1,S_2,T_1,T_2$ largely restrict the valid choices to the 4-tuple. In the rest of the paper, we call the set of non-root user vertices in an attributed tree as the port vertices of the tree.
    \begin{lem}
    \label{lem:correct-pair-1}
        For any $(S_1,S_2,T_1,T_2)\in \L_{ii}^{(0)}$, we have $S_1=S_2$ and $T_1=T_2$.
    \end{lem}
    \renewcommand*{\proofname}{Proof of Lemma~\ref{lem:correct-pair-1}.}
    \begin{proof}
        In this proof, we show that $S_1=S_2$. The proof of $T_1=T_2$ follows by exactly the same argument. Towards contradiction, we assume that $S_1\neq S_2$. Recall that $S_1,S_2\in \cg_{i,\mathcal{A}}$ for some $|\cA|=k$. We claim 
        that $S_1\neq S_2$ implies
        $\ea(S_1)\neq \ea(S_2)$. To see this, we note that if $\ea(S_1)= \ea(S_2)$, then $S_1$ and $S_2$ have the same set of port vertices, and it follows that $\eu(S_1)=\eu(S_2)$, which further implies that $S_1=S_2$. With the claim and the fact that $|\ea(S_1)|=|\ea(S_2)|$, we know that there exists some edge $e^*\in\ea(S_1)$, but $e^*\notin\ea(S_2)$. 
        Let $a^*$ denote the attribute vertex incident to $e^*$. Because $\va(S_1)\cap\va(T_1)=\emptyset$, we have that $a^*\notin\va(T_1)$. Therefore, it follows that $e^*\notin\ea(T_1)$ and $e^*\notin\ea(T_2)$. This leads to a contradiction because the edge $e^*$ only appears in graph $S_1$.
    \end{proof}

    One implication of Lemma~\ref{lem:correct-pair-1} is on the user-attribute edges in the union graph $H$. Consider an attribute $a\in \va(S_1)$. We know that $a$ is included in the two graphs $S_1$ and $S_2$, and in both graphs there is a single user-attribute edge connected to $a$. Therefore, in the union graph $H$, there must be exactly one edge incident to $a$ that is from both $S_1$ and $S_2$. By the similar argument, each attribute $a'\in \va(T_1)$ is incident to one single edge that is from both $T_1$ and $T_2$. This suggests all the user-attribute edges in $H$ are shared by $S_1,S_2$ or $T_1,T_2$, i.e., $K_{11}^\rma=\ea(H)$. Because $|\va(S_1)|=|\va(T_1)|=k$, we further have $|K_{11}^\rma|=2k$. Thus, we know that for any $(S_1,S_2,T_1,T_2)\in \L_{ii}^{(0)}$, we have
    \begin{equation}
    \label{eq:m0-attri}
        \Theta^{\rma}(S_1,S_2,T_1,T_2)=\ra^{2k}.
    \end{equation}
    
\begin{figure}[htbp]
\centering
    \def\svgwidth{0.5\columnwidth}
    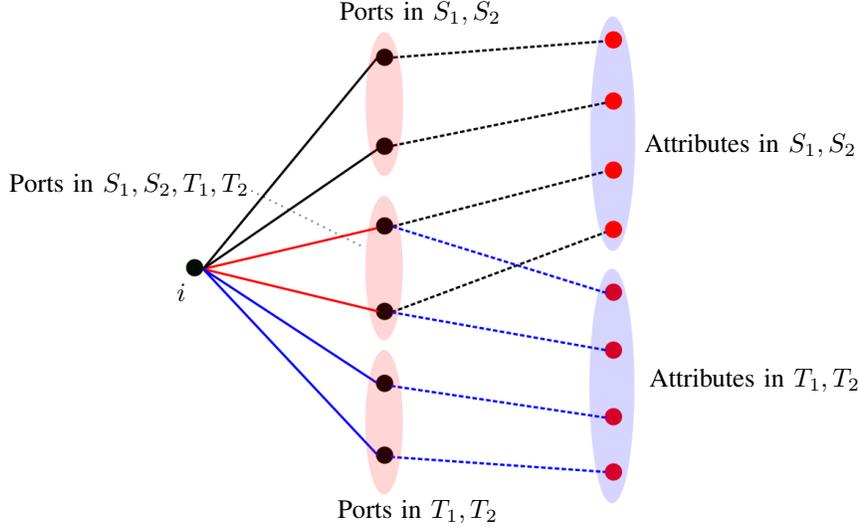
    \caption{Structure of union graph of $(S_1,S_2,T_1,T_2)\in \L_{ii}^{(0,2)}$. In this example, the number of attribute in each subgraph is $k=4$. The black solid lines represent user-user edges shared by $S_1,S_2$, the blue solid lines represent user-user edges shared by $T_1,T_2$, the red solid lines represent user-user edges shared by $S_1,S_2,T_1,T_2$, the black dotted lines represent user-attribute edges shared by $S_1,S_2$ and the blue dotted lines represent user-attribute edges shared by $T_1,T_2$.}
    \label{fig:correct-pair-no-overlap}
\end{figure}

    Now, we move on to consider the user-user edges in the union graph $H$. Notice that all the user-user edges in the four graphs $S_1,S_2,T_1,T_2$ are incident to the root user $i$. Therefore, in the union graph $H$, we also only have user-user edges incident to $i$. Since we have shown in Lemma~\ref{lem:correct-pair-1} that $S_1=S_2$ and $T_1=T_2$, all the user-user edge in $H$ must be from $S_1$ and $S_2$, or from $T_1$ and $T_2$, or from all four graphs $S_1,S_2,T_1,T_2$. Based on this fact, we further separate $\L_{ii}^{(0)}$ into small sets according to the number of user-user edges shared by $S_1,S_2,T_1,T_2$. Let 
    \begin{equation}
    \label{eq:partition-alpha}
    \L_{ii}^{(0,\alpha)}\triangleq\{(S_1,S_2,T_1,T_2)\in\L_{ii}^{(0)}:|\eu(S_1)|\cap|\eu(S_2)|\cap|\eu(T_1)|\cap|\eu(T_2)|=\alpha\}
    \end{equation}
    denote the set of all $(S_1,S_2,T_1,T_2)\in\L_{ii}^{(0)}$ such that there are $\alpha$ user-user edges in the union graph that are shared by all the four graphs. An example of $(S_1,S_2,T_1,T_2)\in \L_{ii}^{(0,2)}$ is illustrated in Figure~\ref{fig:correct-pair-no-overlap}.
    We claim that $ \L_{ii}^{(0,\alpha)}\neq \emptyset$ only if $1\le \alpha\le k$. The upper bound $k$ is easy to see because $S_1,S_2,T_1,T_2$ only have $4k$ user-user vertices in total. To see the lower bound $1$, we notice that if $\alpha=0$, then we are in the case when $S_1=S_2$, $T_1=T_2$, and $V(S_1)\cap V(S_2)=\{i\}$, i.e., $f_1(S_1,S_2,T_1,T_2)=0$, so $(S_1,S_2,T_1,T_2)\notin \L_{ii}$. As a result, $\L_{ii}^{(0)}=\cup_{\alpha=1}^k \L_{ii}^{(0,\alpha)}$.

    With the number of edges shared by all four graphs fixed to $\alpha$ for $(S_1,S_2,T_1,T_2)\in \L_{ii}^{(0,\alpha)}$, the structure of the union graph is clear to us.  By the fact that $|\eu(S_1)|=|\eu(T_1)|=k$, we know that for any $(S_1,S_2,T_1,T_2)\in\L_{ii}^{(0,\alpha)}$, out of the user-user edges in $H$, there are $k-\alpha$ edges shared by $S_1,S_2$, $k-\alpha$ edges shared by $T_1,T_2$, and $\alpha$ edges shared by $S_1,S_2,T_1,T_2$. This suggests that $|K_{11}^\rmu|=2k-2\alpha$, $|K_{02}^\rmu|=|K_{20}^\rmu|=0$, $|K_{12}^\rmu|=|K_{21}^\rmu|=0$ and $|K_{22}^\rmu|=\alpha$.
    Thus, we have
    \begin{equation}
    \label{eq:m0-user}
        \Theta^\rmu(S_1,S_2,T_2,T_2)=\ru^{2k-2\alpha}c^{\alpha}\max\{1,\tfrac{\ru^\alpha}{\qu^\alpha}\}
    \end{equation}
    % Thus, we have $|\eu(H)|=2k-\alpha$ and $|K_{11}^\rmu|=2k-2\alpha$, which implies that
    % \begin{equation}
    % \label{eq:m0-user}
    %     \ru^{|K_{11}^\rmu|}\qu^{-2k+\eu(H)}=\ru^{2k-2\alpha}\qu^{-\alpha}.
    % \end{equation}
    By equations
    Plugging~\eqref{eq:m0-attri} and~\eqref{eq:m0-user} into~\eqref{eq:gamma-m} yields
    \begin{align}
        \Gamma_{ii}^{(0)}&=\sigma_\rmu^{4k}\sigma_\rma^{4k}\sum_{\alpha=1}^k\sum_{(S_1,S_2,T_1,T_2)\in \L_{ii}^{(0,\alpha)}}
        \ra^{2k}\ru^{2k-2\alpha}c^{\alpha}\max\{1,\tfrac{\ru^\alpha}{\qu^\alpha}\}\nonumber\\
        &=\sigma_\rmu^{4k}\sigma_\rma^{4k}\sum_{\alpha=1}^k|\L_{ii}^{(0,\alpha)}|
        \ra^{2k}\ru^{2k-2\alpha}c^{\alpha}\max\{1,\tfrac{\ru^\alpha}{\qu^\alpha}\}.
    \end{align}
    Now, we consider the carnality of the set $\L_{ii}^{(0,\alpha)}$. To count the number of 4-tuples $(S_1,S_2,T_1,T_2)$, it suffices to count the possible union graph $H$ of the four graphs and the belonging of each edge in $H$ to the four graphs $S_1,S_2,T_1,T_2$. Note in the union graph $H$ there are $k$ attributes from $S_1,S_2$ and and $k$ attributes from $T_1,T_2$. The way of choosing those two disjoint attribute subset is given by $\binom{m}{k}\binom{m-k}{k}$. For the user vertices in $H$, recall that there are in $k-\alpha$ edges shared by $S_1,S_2$, $k-\alpha$ edges shared by $T_1,T_2$, and $\alpha$ edges shared by $S_1,S_2,T_1,T_2$, and all of them are incident to the root $i$. Therefore, we have the flexibility to choose $\alpha$ users shared by $S_1,S_2,T_1,T_2$, $k-\alpha$ users shared by $S_1,S_2$ and $k-\alpha$ users shared by $T_1,T_2$ from $n-1$ users. The number of choices is given by $\binom{n-1}{\alpha}\binom{n-1-\alpha}{k-\alpha}\binom{n-1-k}{k-\alpha}$. Finally, the number of ways wiring the edges between the $2k$ attributes and $2k-\alpha$ users is given by $(k!)^2$. This is because each attribute chosen for $S_1,S_2$ need to connect to one of the $k$ users chosen to be shared by $S_1,S_2$ or $S_1,S_2,T_1,T_2$, and each attribute chosen for $T_1,T_2$ need to connect to one of the $k$ users chosen to be shared by $T_1,T_2$ or $S_1,S_2,T_1,T_2$. As a result, we have 
    \begin{align}
        \left|\L_{ii}^{(0,\alpha)}\right|&=\binom{m}{k}\binom{m-k}{k}\binom{n-1}{\alpha}\binom{n-1-\alpha}{k-\alpha}\binom{n-1-k}{k-\alpha}(k!)^2.
        \label{eq:size-alpha}
    \end{align}
    Because $k=\Theta(1)$ and $1\le \alpha\le k$, we have $$\left|\L_{ii}^{(0,\alpha)}\right|\le C m^{2k}n^{2k-\alpha},$$
    for some constant $C$.
    By Proposition~\ref{prop:expectation}, we have 
    \begin{align*}
        \frac{\Gamma_{ii}^{(0)}}{\E[\Phi_{ii}]^2}&\le\frac{\sigma_\rmu^{4k}\sigma_\rma^{4k}\sum_{\alpha=1}^kC m^{2k}n^{2k-\alpha}\ra^{2k}\ru^{2k-2\alpha}c^{\alpha}\max\{1,\tfrac{\ru^\alpha}{\qu^\alpha}\}}{\left(\binom{m}{k}\binom{n-1}{k}k!(\sigma_\rmu^2\ru)^k(\sigma_\rma^2\ra)^k\right)^2}\\
        &\le C'\sum_{\alpha=1}^k\frac{\max\{1,\tfrac{\ru^\alpha}{\qu^\alpha}\}}{n^\alpha\ru^{2\alpha}},\\
        % &\le k \max_{1\le\alpha\le k}\frac{O(1)}{n^\alpha\qu^{\alpha}\ru^{2\alpha}}\\
        % &\stackrel{(c)}{=}o(1),
    \end{align*}
for some constant $C'$, because $k=\Theta(1)$ and $1\le \alpha\le k$. Finally, to see that $\frac{\Gamma_{ii}^{(0)}}{\E[\Phi_{ii}]^2}=o(1)$, we observe that $$\frac{\max\{1,\tfrac{\ru^\alpha}{\qu^\alpha}\}}{n^\alpha\ru^{2\alpha}}=\frac{1}{\min\{n^\alpha\ru^{2\alpha},n^\alpha\ru^\alpha\qu^\alpha\}}.$$ Because we assume $n\rho_\rmu^2=\omega(1)$ and $n\ru\qu=\omega(1)$, it follows that 
$$\frac{\Gamma_{ii}^{(0)}}{\E[\Phi_{ii}]^2}\le C'\sum_{\alpha=1}^k\frac{1}{\min\{n^\alpha\ru^{2\alpha},n^\alpha\ru^\alpha\qu^\alpha\}}=\frac{C'(1+o(1))}{\min\{n\ru^{2},n\ru\qu\}}=o(1),$$
which completes the proof for the case of $M=0$.
    
    % where (b) follows because $$k=\Theta(1),$$ 
    % $$\frac{\binom{n-1}{\alpha}\binom{n-1-\alpha}{k-\alpha}\binom{n-1-k}{k-\alpha}}{\binom{n-1}{k}^2}=n^\alpha(1+o(1)),$$
    % and
    % $$\frac{\binom{m}{k}\binom{m-k}{k}}{\binom{m}{k}^2}=1+o(1),$$
    % and (c) follows by the assumption that $n\qu\rho_\rmu^2=\omega(1)$. This completes the proof for the case of $M=0$. 
    \emph{\textbf{Case 2: $1\le M\le k$.}} In this case, there are some overlaps in the attribute sets of $S_1$ and $T_1$. In the following lemma, we restrict the edge connection configuration of those common attributes of $S_1$ and $T_1$ in the union graph $H$. 
    \begin{lem}
        \label{lem:correct-pair-2}
        Let $(S_1,S_2,T_1,T_2)\in \L_{ij}^{(M)}$ for some $1\le M\le k$. Let $H$ denote the union graph $S_1\cup S_2\cup T_1\cup T_2$. For any shared attribute $a\in \va(S_1)\cap\va(T_1)$, exactly one of the following claims holds true:
        \begin{enumerate}
            \item $a$ is incident to two edges in $H$. Moreover, one of the edges is shared by $S_1,S_2$ and the other one is shared by $T_1,T_2$;
            \item $a$ is incident to two edges in $H$. Moreover, one of the edges is shared by $S_1,T_1$ and the other one is shared by $S_2,T_2$;
            \item $a$ is incident to two edges in $H$. Moreover, one of the edges is shared by $S_1,T_2$ and the other one is shared by $S_2,T_1$;
            \item $a$ is incident to one edge in $H$, which is shared by all four graphs $S_1,S_2,T_1,T_2$.
        \end{enumerate}
    \end{lem}
    \begin{rem}
        Although we state Lemma~\ref{lem:correct-pair-2} in the proof of Proposition~\ref{prop:var-correct-pair}, where we are considering correct pairs $(i,i)$, the lemma actually also applies to the case of $(i,j)$ with $i\neq j$. The lemma will also be used in the proof of Proposition~\ref{prop:var-wrong-pair}.
    \end{rem}
    \renewcommand*{\proofname}{Proof of Lemma~\ref{lem:correct-pair-2}.}
    \begin{proof}
        Consider a shared attributed $a\in \va(S_1)\cap\va(T_1)$. We claim that $a$ is incident to one or two edges in $H$. Notice that in each of $S_1,S_2,T_1,T_2$, there exists exactly one edge incident to $a$. Therefore, there must be at least one edge incident to $a$ in $H$. Moreover, suppose there are three edge incident to $a$ in $H$. By the pigeon hole principle, there must be at least one of them that appear in only one graph out of $S_1,S_2,T_1,T_2$, which is not allowed. If $a$ is incident to one edge in $H$, the edge must be shared by all four graphs, which is exactly case 4 in the lemma. If $a$ is incident to two edges, one of the edges must be shared by two graphs out of $S_1,S_2,T_1,T_2$ and the other one must be shared by the left two graphs, otherwise there exists an edge in $H$ only appearing in one of $S_1,S_2,T_1,T_2$. This would lead to cases 1-3 in the lemma.
    \end{proof}
        \begin{figure}[htbp]
\centering
    \def\svgwidth{0.5\columnwidth}
    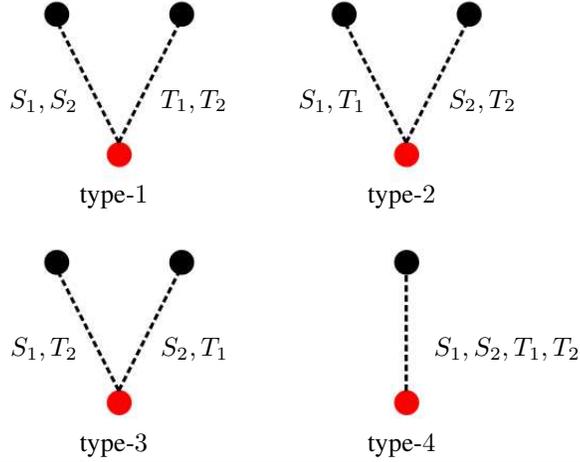
    \caption{Illustration of the four types of shared attributes in the union graph. In this figure, the red nodes represent the attributes and black nodes represent users. The labels beside the edges represent their belonging to graphs $S_1,S_2,T_1,T_2$.}
    \label{fig:4-types}
\end{figure}

    Given Lemma~\ref{lem:correct-pair-2}, we can partition those shared attributes in $H$ into $4$ types as visualized in Figure~\ref{fig:4-types}. A shared attribute satisfying $i$th claim in Lemma~\ref{lem:correct-pair-2} is called a type-$i$ attribute. Given a union graph $H$, we denote the number of type-$i$ attribute in $H$ as $Y_i(H)$. With the $4$ types of attributes defined, we further define the set of 4-tuples
\begin{align}
    &\L_{ii}^{(M_1,M_2,M_3,M_4)}\nonumber\\&\triangleq \{(S_1,S_2,S_3,S_4)\in\L_{ii}^{(M)}:Y_1(H)=M_1,Y_2(H)=M_2,Y_3(H)=M_3,Y_4(H)=M_4\},\label{eq:lambda-m1-4}
\end{align}
and
\begin{equation}
\label{eq:gamma-m1-4}
    \Gamma_{ii}^{(M_1,M_2,M_3,M_4)}\triangleq \sigma_\rmu^{4k}\sigma_\rma^{4k}\sum_{(S_1,S_2,T_1,T_2)\in \L_{ii}^{(M_1,M_2,M_3,M_4)}}\Theta^\rmu(S_1,S_2,T_1,T_2)\Theta^\rma(S_1,S_2,T_1,T_2)
\end{equation}
By these definitions, it follows that 
\begin{equation}
    \L_{ii}^{(M)}=\bigcup\limits_{M_1+M_2+M_3+M_4=M} \L_{ii}^{(M_1,M_2,M_3,M_4)},
\end{equation}
and
\begin{equation}
    \Gamma_{ii}^{(M)}=\sum_{M_1+M_2+M_3+M_4=M} \Gamma_{ii}^{(M_1,M_2,M_3,M_4)}.
\end{equation}
To show that $\Gamma_{ii}^{(M)}=o(\E[\Phi_{ii}]^2)$, it suffices to show that 
$$\max_{M_1+M_2+M_3+M_4=M}\Gamma_{ii}^{(M_1,M_2,M_3,M_4)}=o(\E[\Phi_{ii}]^2).$$
This is because $M\le k$ is a constant and hence, the number of ways of partitioning $M$ into four integers is also a constant.
\begin{sloppypar}
    Now, consider an arbitrary $1\le M\le k$, and fix some $M_1+M_2+M_3+M_4=M$. To bound $\Gamma_{ii}^{(M_1,M_2,M_3,M_4)}$, first consider the user-attribute edges in the union graph $H$ of some $(S_1,S_2,T_1,T_2)\in \L_{ii}^{(M_1,M_2,M_3,M_4)}$. By the definition of the four types of shared attributes, we know that each type-$1$ attributes is incident to two edges in $K_{11}^\rma$, each type-$2$ attributes is incident to one edge in $K_{20}^\rma$ and one edge in $K_{02}^\rma$, each type-$3$ attributes is incident to two edges in $K_{11}^\rma$ and each type-$4$ attributes is incident to one edge in $K_{22}^\rma$. Moreover, apart from the $M$ shared attributes, there are $2k-2M$ attributes in $\va(S_1)\triangle\va(T_1)$. Each of those attributes is incident to exactly one edge in $K_{11}^\rma$. Therefore, we get
    \begin{equation}
    \label{eq:true-pair-k11}
|K_{11}^\rma|=(2k-2M)+2M_1+2M_3=2k-2M_2-2M_4,
\end{equation}
\begin{equation}
    \label{eq:true-pair-k12}
|K_{12}^\rma|=|K_{21}^\rma|=0,
\end{equation}
and
\begin{equation}
    \label{eq:true-pair-k22}
|K_{22}^\rma|=M_4.
\end{equation}
It  follows that for any $(S_1,S_2,T_1,T_2)\in \L_{ii}^{(M_1,M_2,M_3,M_4)}$, we have 
\begin{equation}
    \label{eq:m1-4-attri}
    \Theta^\rma(S_1,S_2,T_1,T_2)=\ra^{2k-2M_2-2M_4}c^{M_4}\max\{1,\tfrac{\ra^{M_4}}{\qa^{M_4}}\}.
\end{equation}
\end{sloppypar}

Next, we move on to bound $\Theta^\rmu(S_1,S_2,T_1,T_2)$.
Similarity to the case of $M=0$, we separately consider the 4-tuples in the set $\L_{ii}^{(M_1,M_2,M_3,M_4)}$ according to the number of user-user edges shared by all 4 graphs:
\begin{align}
    &\L_{ii}^{(M_1,M_2,M_3,M_4,\alpha)}\nonumber\\
    &\triangleq\{(S_1,S_2,T_1,T_2)\in \L_{ii}^{(M_1,M_2,M_3,M_4)}:|\eu(S_1)|\cap|\eu(S_2)|\cap|\eu(T_1)|\cap|\eu(T_2)|=\alpha.\}
\end{align}
We claim that $\L_{ii}^{(M_1,M_2,M_3,M_4,\alpha)}\neq\emptyset$ only if $M_4\le \alpha\le k$. The upper bound $\alpha\le k$ is easy to see. To see the lower bound $\alpha\ge M_4$. We notice that each type-$4$ attribute in $H$ is connected to a port vertex that is shared by all four graphs. Therefore, there must be a user-user edge between root $i$ and this port vertex. As a result, we can rewrite $\Gamma_{ii}^{(M_1,M_2,M_3,M_4)}$ as
\begin{align}
    &\Gamma_{ii}^{(M_1,M_2,M_3,M_4)}\nonumber\\
    &=\sigma_\rmu^{4k}\sigma_\rma^{4k}\sum_{\alpha=M_4}^k\sum_{(S_1,S_2,T_1,T_2)\in \L_{ii}^{(M_1,M_2,M_3,M_4,\alpha)}}\Theta^\rmu(S_1,S_2,T_1,T_2)\Theta^\rma(S_1,S_2,T_1,T_2)\nonumber\\
    &\le\sigma_\rmu^{4k}\sigma_\rma^{4k} k\max_{M_4\le\alpha\le k}\sum_{(S_1,S_2,T_1,T_2)\in \L_{ii}^{(M_1,M_2,M_3,M_4,\alpha)}}\Theta^\rmu(S_1,S_2,T_1,T_2)\Theta^\rma(S_1,S_2,T_1,T_2).
\end{align}
Now, it suffice to show that
\begin{equation}
    \sigma_\rmu^{4k}\sigma_\rma^{4k}\max_{M_4\le\alpha\le k}\sum_{(S_1,S_2,T_1,T_2)\in \L_{ii}^{(M_1,M_2,M_3,M_4,\alpha)}}\Theta^\rmu(S_1,S_2,T_1,T_2)\Theta^\rma(S_1,S_2,T_1,T_2)=o(\E[\Phi_{ii}]^2).
    \label{eq:alpha-max}
\end{equation}
We have already found $\Theta^\rma(S_1,S_2,T_1,T_2)$ in~\eqref{eq:m1-4-attri} for any $(S_1,S_2,T_1,T_2)\in \L_{ii}^{(M_1,M_2,M_3,M_4)}$. Now, we focus the user-user part on a 4-tuple $(S_1,S_2,T_1,T_2)\in \L_{ii}^{(M_1,M_2,M_3,M_4,\alpha)}$. By our definition of $\L_{ii}^{(M_1,M_2,M_3,M_4,\alpha)}$, we know that $K_{22}^\rmu=\alpha$. Here we show that $|K_{12}^\rmu|=|K_{21}^\rmu|=0$ and that $|K_{11}^\rmu|\ge 2k-2\alpha-2M_2$. To see $|K_{12}^\rmu|=|K_{21}^\rmu|=0$, towards contradiction, suppose there exists such a user-user edge $(i,u^*)$. Then $u^*$ is a user shared by three graphs. Thus, in $H$, $u^*$ must be incident to an user-attribute edges that is shared by three graphs, otherwise it is connected to  an edge that only appear in one of $S_1,S_2,T_1,T_2$, which is not allowed. However, this leads to a contradiction, because such a user-attribute edge shared by three graphs does not exist by Lemma~\ref{lem:correct-pair-2}. This shows that $|K_{12}^\rmu|=|K_{21}^\rmu|=0$. To see that $|K_{11}^\rmu|\ge 2k-2\alpha-2M_2$, we notice there are $2k-2\alpha$ user-user edges shared by two graphs out of $S_1,S_2,T_1,T_2$ because there are $\alpha$ edges shared by $S_1,S_2,T_1,T_2$. Out of these edges shared by two graphs, only edges shared by $S_1,T_1$ or $S_2,T_2$ do not belong to $K_{11}^\rmu$. We claim that the number of such edges is upper bounded by $2M_2$. To see the claim, we note that each user-user edge shared by $S_1,T_1$ or $S_2,T_2$ is connected between root $i$ and a port vertex shared by $S_1,T_1$ or $S_2,T_2$. In the union graph $H$, each of these port vertex must be incident to a user-attribute edge that is shared by $S_1,T_1$ or $S_2,T_2$. However, by the definition of our 5 types of attributes in $H$, there are at most $2M_3$ user-attribute edges shared by $S_1,T_1$ or $S_2,T_2$. This implies that we have at most $2M_3$ user-user edges shared by $S_1,T_1$ or $S_2,T_2$.

% We are going to show that the user-user edges in the union graph $H$ satisfies $|\eu(H)|=2k-\alpha$ and $|K_{11}^\rmu|\ge 2k-2\alpha-2M_2$. To see that $|\eu(H)|=2k-\alpha$, we notice that each graphs of $S_1,S_2,T_1,T_2$ has $k$ user-user vertices, and that in $\eu(H)$, $\alpha$ of the edges are shared by all $S_1,S_2,T_1,T_2$ and the rest $|\eu(H)|-\alpha$ are shared by two graphs. These implies $4\alpha+2(|\eu(H)|-\alpha)=4k$, which gives $|\eu(H)|=2k-\alpha$. To see that $|K_{11}^\rmu|\ge 2k-2\alpha-2M_2$, we notice there are $2k-2\alpha$ user-user edges shared by two graphs out of $S_1,S_2,T_1,T_2$. Out of these edges, only edges shared by $S_1,T_1$ or $S_2,T_2$ do not belong to $K_{11}^\rmu$. We claim that the number of such edges is upper bounded by $2M_2$. To see the claim, we note that each user-user edge shared by $S_1,T_1$ or $S_2,T_2$ is connected between root $i$ and a port vertex shared by $S_1,T_1$ or $S_2,T_2$. In the union graph $H$, each of these port vertex must be incident to a user-attribute edge that is shared by $S_1,T_1$ or $S_2,T_2$. However, by the definition of our 5 types of attributes in $H$, there are at only $2M_3$ user-attribute edges shared by $S_1,T_1$ or $S_2,T_2$. This implies that we have at most $2M_3$ user-user edges shared by $S_1,T_1$ or $S_2,T_2$.

Now, we have that for any $(S_1,S_2,T_1,T_2)\in \L_{ii}^{(M_1,M_2,M_3,M_4,\alpha)}$
\begin{equation}
    \label{eq:m1-4-user}
    \Theta^\rmu(S_1,S_2,T_1,T_2)\le \ru^{2k-2\alpha-2M_2}c^{\alpha}\max\{1,\tfrac{\ru^\alpha}{\qu\alpha}\} .
\end{equation}
Plugging~\eqref{eq:m1-4-attri} and~\eqref{eq:m1-4-user} into the left-hand side of~\eqref{eq:alpha-max} gives
\begin{align}
     \sigma_\rmu^{4k}\sigma_\rma^{4k}\max_{M_4\le\alpha\le k}|\L_{ii}^{(M_1,M_2,M_3,M_4,\alpha)}|\ra^{2k-2M_2-2M_4}c^{M_4}\max\{1,\tfrac{\ra^{M_4}}{\qa^{M_4}}\} \ru^{2k-2\alpha-2M_2}c^{\alpha}\max\{1,\tfrac{\ru^\alpha}{\qu\alpha}\} .\label{eq:alpha-max-1}
\end{align}
Now, we want to bound the size of the set $\L_{ii}^{(M_1,M_2,M_3,M_4,\alpha)}$. To count the number of $4$-tuples inside, it suffices to count the ways of forming union graphs $H$ and the belonging of each edge in $H$ to $S_1,S_2,T_1,T_2$. The way of forming the user-user part of $H$ is upper bounded by
\begin{equation}
\label{eq:overlap-user-count}
    \binom{n-1}{\alpha}\binom{n-1-\alpha}{2k-2\alpha}6^{2k-2\alpha}.
\end{equation}
To see this, recall that there are $\alpha$ user-user edges shared by all four graphs and $2k-2\alpha$ edges shared by two out of $S_1,S_2,T_1,T_2$. In~\eqref{eq:overlap-user-count}, $ \binom{n-1}{\alpha}$ represents the number of ways to choose $\alpha$ port vertices shared by four graphs and $\binom{n-1-\alpha}{2k-2\alpha}$ represents the number of ways to choose the other $2k-2\alpha$ port vertices. But because each port vertices shared by two graphs could belong to $S_1,S_2$ or $T_1,T_2$ or $S_1,T_2$ or $S_2,T_1$ or $S_1,T_1$ or $S_2,T_2$, we have the additional term $6^{2k-2\alpha}$ bounding the total number of configurations.

The way of forming the user-attribute part of $H$ is upper bounded by 
\begin{equation}
\label{eq:overlap-attribute-count}
    \binom{m}{M_1}\binom{m}{M_2}\binom{m}{M_3}\binom{m}{M_4}\binom{m}{2k-2M}(2k-\alpha)^{2(2k-M)}.
\end{equation}
To see this, we note that $\binom{m}{M_1}\binom{m}{M_2}\binom{m}{M_3}\binom{m}{M_4}\binom{m}{2k-2M}$ upper bounds the way of choosing the five types of attributes from a total of $m$ attribute. The term $(2k-\alpha)^{2(2k-M)}$ upper bounds the ways of wiring between the $2k-\alpha$ ports and $M_1+M_2+M_3+M_4+2k-2M=2k-M$ attributes.

As a result, we have 
\begin{align}
    &|\L_{ii}^{(M_1,M_2,M_3,M_4,\alpha)}|\nonumber\\
    &\le \binom{n-1}{\alpha}\binom{n-1-\alpha}{2k-2\alpha}6^{2k-2\alpha}\binom{m}{M_1}\binom{m}{M_2}\binom{m}{M_3}\binom{m}{M_4}\binom{m}{2k-2M}(2k-\alpha)^{2(2k-M)}\nonumber\\
    &=Cn^{2k-\alpha}m^{2k-M},\label{eq:lambda-size-bound}
\end{align}
for some large enough constant $C$ because of the fact that $k,\alpha,M$ are all constants.

Plugging~\eqref{eq:lambda-size-bound} into~\eqref{eq:alpha-max-1} gives
\begin{align}
     &\sigma_\rmu^{4k}\sigma_\rma^{4k}\max_{M_4\le\alpha\le k}Cn^{2k-\alpha}m^{2k-M}\ra^{2k-2M_2-2M_4}c^{M_4}\max\{1,\tfrac{\ra^{M_4}}{\qa^{M_4}}\} \ru^{2k-2\alpha-2M_2}c^{\alpha}\max\{1,\tfrac{\ru^\alpha}{\qu^\alpha}\}\nonumber\\
     &=\sigma_\rmu^{4k}\sigma_\rma^{4k}Cn^{2k}m^{2k-M}\ra^{2k-2M_2-2M_4}c^{M_4}\max\{1,\tfrac{\ra^{M_4}}{\qa^{M_4}}\}\ru^{2k-2M_2}\nonumber\\
     &\;\;\;\cdot\max_{M_4\le\alpha\le k}n^{-\alpha}\ru^{-2\alpha}c^\alpha \max\{1,\tfrac{\ru^\alpha}{\qu^\alpha}\}
     \label{eq:alpha-max-2}
\end{align}
Because it is assumed that $n\ru\qu=\omega(1)$ and $n\rho_\rmu^2=\omega(1)$, the maximum $$\max_{M_4\le\alpha\le k}n^{-\alpha}\ru^{-2\alpha}c^\alpha \max\{1,\tfrac{\ru^\alpha}{\qu^\alpha}\}$$ is attained at $\alpha=M_4$. Therefore, it suffices to show that 
\begin{align}
    \label{eq:alpha-max-3}
    \sigma_\rmu^{4k}\sigma_\rma^{4k}n^{2k-M_4}m^{2k-M}\ra^{2k-2M_2-2M_4}\max\{1,\tfrac{\ra^{M_4}}{\qa^{M_4}}\} \ru^{2k-2M_2-2M_4}\max\{1,\tfrac{\ru^{M_4}}{\qu^{M_4}}\}=o(\E[\Phi_{ii}]^2),
\end{align}
since $C,c$ and $M_4$ are all constants.
Notice that the left-hand side of~\eqref{eq:alpha-max-3} involves two maximum. We prove~\eqref{eq:alpha-max-3} by considering all $4$ combinations of the two maximum values.

    \emph{\textbf{Suppose $\ru<\qu$ and $\ra<\qa$.}} Then we have $\max\{1,\tfrac{\ru^{M_4}}{\qu^{M_4}}\}=1$ and $\max\{1,\tfrac{\ra^{M_4}}{\qa^{M_4}}\}=1$. Dividing left-hand side of~\eqref{eq:alpha-max-3} by $\E[\Phi_{ii}]^2$ yields
    \begin{align*}
        &\frac{\sigma_\rmu^{4k}\sigma_\rma^{4k}n^{2k-M_4}m^{2k-M}\ra^{2k-2M_2-2M_4} \ru^{2k-2M_2-2M_4}}{\E[\Phi_{ii}]^2}\\
        &=\frac{\sigma_\rmu^{4k}\sigma_\rma^{4k}n^{2k-M_4}m^{2k-M}\ra^{2k-2M_2-2M_4} \ru^{2k-2M_2-2M_4}}{\left(\binom{m}{k}(\ru\sigma_\rmu^2)^k(\ra\sigma_\rma^2)^k\binom{n-1}{k}k!\right)^2}\\
        &\le \frac{C'}{m^{M}n^{M_4}\ra^{2M_2+2M_4}\ru^{2M_2+2M_4}},
    \end{align*}
    for some constant $C'$. To see~\eqref{eq:alpha-max-3} in this case, we notice
    \begin{align*}
        m^{M}n^{M_4}\ra^{2M_2+2M_4}\ru^{2M_2+2M_4}&\stackrel{(a)}{\ge} m^{M}n^{M_4}\ra^{2M}\ru^{2M}\\
        &=n^{M_4}(m\rho_\rma^2\rho_\rmu^2)^M\\
        &\stackrel{(b)}{=}\omega(1),
    \end{align*}
    where (a) follows because $M_2+M_4\le M$ and (b) follows by the assumption $m\rho_\rma^2\rho_\rmu^2=\omega(1)$ and $M\ge 1$.
    
    \emph{\textbf{Suppose $\ru\ge\qu$ and $\ra\ge\qa$.}} Then we have $\max\{1,\tfrac{\ru^{M_4}}{\qu^{M_4}}\}=\tfrac{\ru^{M_4}}{\qu^{M_4}}$ and $\max\{1,\tfrac{\ra^{M_4}}{\qa^{M_4}}\}=\tfrac{\ra^{M_4}}{\qa^{M_4}}$.
    Dividing left-hand side of~\eqref{eq:alpha-max-3} by $\E[\Phi_{ii}]^2$ yields
    \begin{align*}
        &\frac{\sigma_\rmu^{4k}\sigma_\rma^{4k}n^{2k-M_4}m^{2k-M}\ra^{2k-2M_2-M_4} \ru^{2k-2M_2-M_4}\qa^{-M_4}\qu^{-M_4}}{\E[\Phi_{ii}]^2}\\
        &=\frac{\sigma_\rmu^{4k}\sigma_\rma^{4k}n^{2k-M_4}m^{2k-M}\ra^{2k-2M_2-M_4} \ru^{2k-2M_2-M_4}\qa^{-M_4}\qu^{-M_4}}{\left(\binom{m}{k}(\ru\sigma_\rmu^2)^k(\ra\sigma_\rma^2)^k\binom{n-1}{k}k!\right)^2}\\
        &\le \frac{C'}{m^{M}n^{M_4}\ra^{2M_2+M_4}\ru^{2M_2+M_4}\qa^{M_4}\qu^{M_4}},
    \end{align*}
    for some constant $C'$. To see~\eqref{eq:alpha-max-3} in this case, we notice
    \begin{align*}
        m^{M}n^{M_4}\ra^{2M_2+M_4}\ru^{2M_2+M_4}\qa^{M_4}\qu^{M_4}&=(m\qa\ra n\qu\ru)^{M_4}(m\rho_\rmu^2\rho_\rma^2)^{M_2}m^{M-M_2-M_4}=\omega(1),
    \end{align*}
    where the last equality follows because $m\qa\ra n\qu\ru=\omega(1)$, $m\rho_\rmu^2\rho_\rma^2=\omega(1)$, and at least one of $M_2,M_4$ and $M-M_2-M_4$ is strictly positive.
    
    \emph{\textbf{Suppose $\ru<\qu$ and $\ra\ge \qa$.}} Then we have $\max\{1,\tfrac{\ru^{M_4}}{\qu^{M_4}}\}=1$ and $\max\{1,\tfrac{\ra^{M_4}}{\qa^{M_4}}\}=\tfrac{\ra^{M_4}}{\qa^{M_4}}$.
    Dividing left-hand side of~\eqref{eq:alpha-max-3} by $\E[\Phi_{ii}]^2$ yields
    \begin{align*}
        &\frac{\sigma_\rmu^{4k}\sigma_\rma^{4k}n^{2k-M_4}m^{2k-M}\ra^{2k-2M_2-M_4} \ru^{2k-2M_2-2M_4}\qa^{-M_4}}{\E[\Phi_{ii}]^2}\\
        &=\frac{\sigma_\rmu^{4k}\sigma_\rma^{4k}n^{2k-M_4}m^{2k-M}\ra^{2k-2M_2-M_4} \ru^{2k-2M_2-2M_4}\qa^{-M_4}}{\left(\binom{m}{k}(\ru\sigma_\rmu^2)^k(\ra\sigma_\rma^2)^k\binom{n-1}{k}k!\right)^2}\\
        &\le \frac{C'}{m^{M}n^{M_4}\ra^{2M_2+M_4}\ru^{2M_2+2M_4}\qa^{M_4}},
    \end{align*}
    for some constant $C'$. To see~\eqref{eq:alpha-max-3} in this case, we notice
    \begin{align*}
        m^{M}n^{M_4}\ra^{2M_2+M_4}\ru^{2M_2+2M_4}\qa^{M_4}=(n\rho_\rmu^2 m\qa\ra)^{M_4}(m\rho_\rma^2 \rho_\rmu^2)^{M_2}m^{M-M_2-M_4}=\omega(1),
    \end{align*}
    where the last equality follows because $m\qa\ra n\rho_\rmu^2=\omega(1)$, $m\rho_\rmu^2\rho_\rma^2=\omega(1)$, and at least one of $M_2,M_4$ and $M-M_2-M_4$ is strictly positive.
    
     \emph{\textbf{Suppose $\ru\ge\qu$ and $\ra< \qa$.}} Then we have $\max\{1,\tfrac{\ru^{M_4}}{\qu^{M_4}}\}=\tfrac{\ru^{M_4}}{\qu^{M_4}}$ and $\max\{1,\tfrac{\ra^{M_4}}{\qa^{M_4}}\}=1$.
    Dividing left-hand side of~\eqref{eq:alpha-max-3} by $\E[\Phi_{ii}]^2$ yields
    \begin{align*}
        &\frac{\sigma_\rmu^{4k}\sigma_\rma^{4k}n^{2k-M_4}m^{2k-M}\ra^{2k-2M_2-2M_4} \ru^{2k-2M_2-M_4}\qu^{-M_4}}{\E[\Phi_{ii}]^2}\\
        &=\frac{\sigma_\rmu^{4k}\sigma_\rma^{4k}n^{2k-M_4}m^{2k-M}\ra^{2k-2M_2-2M_4} \ru^{2k-2M_2-M_4}\qu^{-M_4}}{\left(\binom{m}{k}(\ru\sigma_\rmu^2)^k(\ra\sigma_\rma^2)^k\binom{n-1}{k}k!\right)^2}\\
        &\le \frac{C'}{m^{M}n^{M_4}\ra^{2M_2+2M_4}\ru^{2M_2+M_4}\qu^{M_4}},
    \end{align*}
    for some constant $C'$. To see~\eqref{eq:alpha-max-3} in this case, we notice
        \begin{align*}
        m^{M}n^{M_4}\ra^{2M_2+2M_4}\ru^{2M_2+M_4}\qu^{M_4}=(n\ru\qu m\rho_\rma^2)^{M_4}(m\rho_\rma^2 \rho_\rmu^2)^{M_2}m^{M-M_2-M_4}=\omega(1),
    \end{align*}
    where the last equality follows because $n\ru\qu=\omega(1)$, $m\rho_\rmu^2\rho_\rma^2=\omega(1)$, and at least one of $M_2,M_4$ and $M-M_2-M_4$ is strictly positive.

    \subsection{Proof of Proposition~\ref{prop:var-wrong-pair}}
In this section, we prove Proposition~\ref{prop:var-wrong-pair} by showing that for any $i\neq j$,
\begin{equation}
    \label{eq:prop3-wts}
    \Gamma_{ij}=\sigma_\rmu^{4k}\sigma_\rma^{4k}\sum_{(S_1,S_2,T_1,T_2)\in \L_{ij}}\ru^{|K^\rmu_{11}|}\Theta^\rmu(S_1,S_2,T_1,T_2)\Theta^\rma(S_1,S_2,T_1,T_2)=o(\E[\Phi_{ij}]^2n^{-2}).
\end{equation}
In this proof, we exploit the property that every edge in the union graph $H$ must appear in at least two subgraphs of $S_1,S_2,T_1,T_2$ to restrict the structure of union graph. In the following, we provide a lemma capturing a key property of the 4-tuples in $\L_{ij}$.
\begin{lem}
    \label{lem:wrong-pair-1}
    For any $(S_1,S_2,T_1,T_2)\in\L_{ij}$, we have $\eu(S_1)=\eu(T_1)$ and $\eu(S_2)=\eu(T_2)$.
\end{lem}
\renewcommand*{\proofname}{Proof of Lemma~\ref{lem:wrong-pair-1}.}
\begin{proof}
    Recall that for any 4-tuple $(S_1,S_2,T_1,T_2)\in\L_{ij}$, we have $\va(S_1)=\va(S_2)$, $\va(T_1)=\va(T_2)$. Moreover, $S_1,T_1$ are rooted at $i$ and $S_2,T_2$ are rooted at $j$. In this proof, we argue that  $\eu(S_1)=\eu(T_1)$. The proof of $\eu(S_2)=\eu(T_2)$ would follow by the same argument. Towards contradiction, suppose that $\eu(S_1)\neq\eu(T_1)$. Recall that the user-user part of both $S_1$ and $T_1$ are $k$ user-user edges incident to the root $i$. Thus, it follows that there must exist a user-user edge $(i,u^*)\in\eu(S_1)$, but $(i,u^*)\notin\eu(T_1)$, and also another user-user edge $(i,v^*)\in\eu(T_1)$, but $(i,v^*)\notin\eu(S_1)$. We observe that $u^*\neq v^*$, and hence, at least one of the following holds true:
    \begin{enumerate}
        \item $u^*\neq j$;
        \item $v^*\neq j$.
    \end{enumerate}
    We assume without loss of generality that $u^*\neq j$ holds true. This would imply that $(i,u^*)\notin \eu(S_2)$ and $(i,u^*)\notin \eu(T_2)$ because all the user-user edges in $S_2$ and $T_2$ are incident to their root $j$. This leads to a contradiction because we then have edge $(i,u^*)$ appear only in the graph $S_1$.
\end{proof}
To prove~\eqref{eq:prop3-wts}, we separately consider two subcases of $\L_{ij}$. We define:
\begin{align}
    \bar{\L}_{ij}&\triangleq\{(S_1,S_2,T_1,T_2)\in \L_{ij}:(i,j)\notin \eu(H)\}\label{eq:lambda-bar},\\
    \tilde{\L}_{ij}&\triangleq\{(S_1,S_2,T_1,T_2)\in \L_{ij}:(i,j)\in \eu(H)\}\label{eq:lambda-td}.
\end{align}
In other words, $\bar{\L}_{ij}$ includes the cases when the edge $(i,j)$ is excluded from the union graph $H$, while $\tilde{\L}_{ij}$ includes the cases when the edge $(i,j)$ is included in the union graph $H$. Following the above definitions, we further define
\begin{align}
    \bar{\Gamma}_{ij}&\triangleq\sigma_\rmu^{4k}\sigma_\rma^{4k}\sum_{(S_1,S_2,T_1,T_2)\in \bar{\L}_{ij}}\Theta^\rmu(S_1,S_2,T_1,T_2)\Theta^\rma(S_1,S_2,T_1,T_2),\label{eq:gamma-bar}\\
    \tilde{\Gamma}_{ij}&\triangleq\sigma_\rmu^{4k}\sigma_\rma^{4k}\sum_{(S_1,S_2,T_1,T_2)\in \tilde{\L}_{ij}}\Theta^\rmu(S_1,S_2,T_1,T_2)\Theta^\rma(S_1,S_2,T_1,T_2).\label{eq:gamma-td}
\end{align}
Because the two sets $\bar{\L}_{ij}$ and $\tilde{\L}_{ij}$ partition the set $\L_{ij}$, it follows that $\Gamma_{ij}=\bar{\Gamma}_{ij}+\tilde{\Gamma}_{ij}$. Therefore, it suffices to show that 
\begin{align}
    \bar{\Gamma}_{ij}=o(\E[\Phi_{ij}]^2n^{-2})
    \label{eq:prop3-wts-1},
\end{align}
and
\begin{align}
    \tilde{\Gamma}_{ij}=o(\E[\Phi_{ij}]^2n^{-2})
    \label{eq:prop3-wts-2}.
\end{align}
We first prove~\eqref{eq:prop3-wts-1}, and then prove~\eqref{eq:prop3-wts-2} by showing that $\tilde{\Gamma}_{ij}=O(\bar{\Gamma}_{ij})$.

Consider the user-user edges in $H$ for some $(S_1,S_2,T_1,T_2)\in \bar{\L}_{ij}$. We note that all the user-user edges in $H$ are shared either by $S_1,T_1$ or $S_2,T_2$. This follows by the fact that $\eu(S_1)=\eu(T_1)$, $\eu(S_2)=\eu(T_2)$, and $\eu(S_1)\cap\eu(T_1)=\emptyset$ since $(i,j)\notin\eu(H)$. As a result, we have $|K_{11}^\rmu|=|K_{21}^\rmu|=|K_{12}^\rmu|=|K_{22}^\rmu|=0$. Therefore, for any $(S_1,S_2,T_1,T_2)\in \bar{\L}_{ij}$, 
\begin{equation}
    \label{eq:bar-user}
    \Theta^\rmu(S_1,S_2,T_1,T_2)=1.
\end{equation}
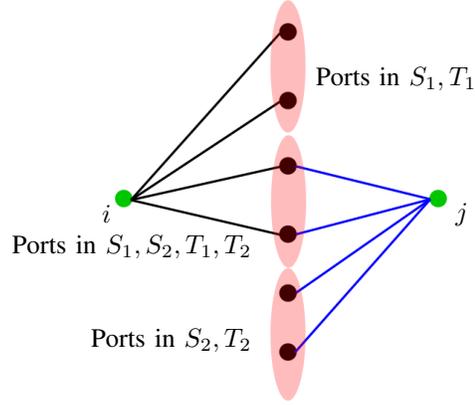
\begin{figure}
\centering
    \def\svgwidth{0.3\columnwidth}
    %% Creator: Inkscape 1.2.2 (b0a84865, 2022-12-01), www.inkscape.org
%% Accompanies image file 'wrong_pair_user_COLT.eps' (pdf, eps, ps)
\begingroup%
  \makeatletter%
  \providecommand\color[2][]{%
    \errmessage{(Inkscape) Color is used for the text in Inkscape, but the package 'color.sty' is not loaded}%
    \renewcommand\color[2][]{}%
  }%
  \providecommand\transparent[1]{%
    \errmessage{(Inkscape) Transparency is used (non-zero) for the text in Inkscape, but the package 'transparent.sty' is not loaded}%
    \renewcommand\transparent[1]{}%
  }%
  \providecommand\rotatebox[2]{#2}%
  \newcommand*\fsize{\dimexpr\f@size pt\relax}%
  \newcommand*\lineheight[1]{\fontsize{\fsize}{#1\fsize}\selectfont}%
  \ifx\svgwidth\undefined%
    \setlength{\unitlength}{120.93419113bp}%
    \ifx\svgscale\undefined%
      \relax%
    \else%
      \setlength{\unitlength}{\unitlength * \real{\svgscale}}%
    \fi%
  \else%
    \setlength{\unitlength}{\svgwidth}%
  \fi%
  \global\let\svgwidth\undefined%
  \global\let\svgscale\undefined%
  \makeatother%
  \begin{picture}(1,1.17306174)%
    \lineheight{1}%
    \setlength\tabcolsep{0pt}%
    \put(0,0){\includegraphics[width=\unitlength]{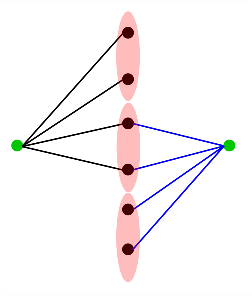}}%
    \put(0.00939697,0.5324979){\color[rgb]{0,0,0}\makebox(0,0)[lt]{\lineheight{1.25}\smash{\begin{tabular}[t]{l}$i$\end{tabular}}}}%
    \put(0.95849657,0.53876267){\color[rgb]{0,0,0}\makebox(0,0)[lt]{\lineheight{1.25}\smash{\begin{tabular}[t]{l}$j$\end{tabular}}}}%
    \put(0.58261541,0.9021143){\color[rgb]{0,0,0}\makebox(0,0)[lt]{\lineheight{1.25}\smash{\begin{tabular}[t]{l}Ports in $S_1,T_1$\end{tabular}}}}%
    \put(0.41033654,0.44949086){\color[rgb]{0,0,0}\makebox(0,0)[rt]{\lineheight{1.25}\smash{\begin{tabular}[t]{r}Ports in $S_1,S_2,T_1,T_2$\end{tabular}}}}%
    \put(0.41033655,0.19890389){\color[rgb]{0,0,0}\makebox(0,0)[rt]{\lineheight{1.25}\smash{\begin{tabular}[t]{r}Ports in $S_2,T_2$\end{tabular}}}}%
  \end{picture}%
\endgroup%

    \caption{Illustration of the user-user part structure of the union graph of $(S_1,S_2,T_1,T_2)\in\Bar{\L}_{ij}$. In this example, we set $k=4$. The two green nodes are the two roots $i$ and $j$. The black lines represents user-user edges from $S_1,T_1$, and the blue lines represents user-user edges from $S_2,T_2$.}
    \label{fig:wrong-pair-user}
\end{figure}
Now, we consider the user-attribute part of the union graph $H$. By Lemma~\ref{lem:correct-pair-2}, the attributes in $\va(S_1)\cap\va(T_1)$ can be separated in four types in the same way as we have done in the proof of Proposition~\ref{prop:var-correct-pair}. We partition the set $\bar{\L}_{ij}$ according the number of each type of shared attributes:
\begin{align}
    &\bar{\L}_{ij}^{(M_1,M_2,M_3,M_4)}\nonumber\\
    &\triangleq \{(S_1,S_2,T_1,T_2)\in \bar{\L}_{ij}: Y_1(H)=M_1,Y_2(H)=M_2,Y_3(H)=M_3,Y_4(H)=M_4\},\label{eq:bar-m1-4}
\end{align}
where $Y_t$ denotes the number of type-$t$ attributes in $H$.
Let
\begin{equation}
    \label{eq:bar-gamma-m1-4}
    \bar{\Gamma}_{ij}^{(M_1,M_2,M_3,M_4)}\triangleq\sigma_\rmu^{4k}\sigma_\rma^{4k}\sum_{(S_1,S_2,T_1,T_2)\in \bar{\L}_{ij}^{(M_1,M_2,M_3,M_4)}}\ru^{|K^\rmu_{11}|}\ra^{|K^\rma_{11}|}\qu^{-2k+\eu(H)}\qa^{-2k+\ea(H)}.
\end{equation}
Therefore, 
\begin{equation}
    \bar{\Gamma}_{ij}=\sum_{0\le M_1+M_2+M_3+M_4\le k}\bar{\Gamma}_{ij}^{(M_1,M_2,M_3,M_4)}.
\end{equation}
Because $k=\Theta(1)$, it suffices to show that $\bar{\Gamma}_{ij}^{(M_1,M_2,M_3,M_4)}=o(\E[\Phi_{ij}]^2n^{-2})$ for any $0\le M_1+M_2+M_3+M_4\le k$.
\begin{sloppypar}
Now, we consider the user-attribute edge connections in $H$ of some $(S_1,S_2,T_1,T_2)\in \bar{\L}_{ij}^{(M_1,M_2,M_3,M_4)}$. By a similar argument as for~\eqref{eq:true-pair-k11}, ~\eqref{eq:true-pair-k12} and~\eqref{eq:true-pair-k22} in the proof of Proposition~\ref{prop:var-correct-pair}, we still have $|K_{11}^\rma|=2k-2M_2-2M_4$, and $|K_{12}^\rma|=|K_{21}^\rma|=0$ and $|K_{22}^\rma|=M_4$. Therefore, for any $(S_1,S_2,T_1,T_2)\in \bar{\L}_{ij}^{(M_1,M_2,M_3,M_4)}$, 
\begin{equation}
    \label{eq:bar-attri}
    \Theta^\rma(S_1,S_2,T_1,T_2)=\ra^{2k-2M_2-2M_4}c^{M_4}\max\{1,\tfrac{\ra^{M_4}}{\qa^{M_4}}\}.
\end{equation}
Plugging~\eqref{eq:bar-user} and~\eqref{eq:bar-attri} into~\eqref{eq:bar-gamma-m1-4} yields
\begin{equation}
    \label{eq:gamma-bar-bound-1}
    \bar{\Gamma}_{ij}^{(M_1,M_2,M_3,M_4)}=\sigma_\rmu^{4k}\sigma_\rma^{4k}|\bar{\L}_{ij}^{(M_1,M_2,M_3,M_4)}|\ra^{2k-2M_2-2M_4}c^{M_4}\max\{1,\tfrac{\ra^{M_4}}{\qa^{M_4}}\}.
\end{equation}
Now, we need to bound $|\bar{\L}_{ij}^{(M_1,M_2,M_3,M_4)}|$. To bound the size of this set, we bound the number of possible ways of forming union graph $H$ of four graphs $S_1,S_2,T_1,T_2$, and the belonging of each edge in $H$. We first consider the user-user part of the union graph $H$. Remember that $\eu(S_1)=\eu(T_1)$ and $\eu(S_2)=\eu(T_2)$. It follows that all user vertices in $H$ are shared by $S_1,T_1$ or $S_2,T_2$ or all four graphs $S_1,T_1,S_2,T_2$. Let $\beta$ denote the number of users shared by all graphs $S_1,T_1,S_2,T_2$. Then it follows that except for $i$ and $j$, there are $k-\beta$ users in $H$ that are shared by $S_1,T_1$ and $k-\beta$ users in $H$ that are shared by $S_2,T_2$. Moreover, we claim that $k-M_2\le \beta\le  k$. To see the claim, we note that each port vertex in $H$ shared by $S_1,T_1$ or $S_2,T_2$ is incident to a user-attribute edge shared by $S_1,T_1$ or $S_2,T_2$ respectively. However, by the definition of attribute types, we know that there are in total $2M_2$ such user-attribute edges in $H$. This implies that $2(k-\beta)\le 2M_2 $, which gives the desired bound. 
\end{sloppypar}
Then, we can upper bound the number of ways of forming the user-user part of $H$ by
\begin{equation}
\label{eq:bar-user-count}
    \sum_{\beta=k-M_2}^k\binom{n}{\beta}\binom{n}{k-\beta}\binom{n}{k-\beta}.
\end{equation}
In this upper bound, the term in the summation upper bounds the number of the ways of choosing $\beta$ users shared by $S_1,S_2,T_1,T_2$, $k-\beta$ users shared by $S_1,T_1$ and $k-\beta$ users shared by $S_2,T_2$ from $n-2$ users.

For the user-attribute connections, the number of choices is upper bounded by
\begin{equation}
     \binom{m}{M_1}\binom{m}{M_2}\binom{m}{M_3}\binom{m}{M_4}\binom{m}{2k-2M}(2k)^{2(2k-M)},
\end{equation}
where $M\triangleq M_1+M_2+M_3+M_4$.
The term $\binom{m}{M_1}\binom{m}{M_2}\binom{m}{M_3}\binom{m}{M_4}\binom{m}{2k-2M}$ upper bounds the number of ways of choosing the attributes in $H$ from $m$ attributes, and the term $(2k)^{2(2k-M)}$ upper bounds the number of ways of edge connections between those $2k-M$ chosen attributes and at most $2k$ users.

Because $k,M,\beta=\Theta(1)$, we have 
\begin{align}
    &|\bar{\L}_{ij}^{(M_1,M_2,M_3,M_4)}|\nonumber\\&\le \binom{m}{M_1}\binom{m}{M_2}\binom{m}{M_3}\binom{m}{M_4}\binom{m}{2k-2M}(2k)^{2(2k-M)}\sum_{\beta=k-M_2}^k\binom{n}{\beta}\binom{n}{k-\beta}\binom{n}{k-\beta}\nonumber\\
    &\le \bar{C}m^{2k-M}\sum_{\beta=k-M_2}^k n^{2k-\beta}\nonumber\\
    &=(1+o(1))\bar{C}m^{2k-M}n^{k+M_2},\label{eq:lambda-bar-size}
\end{align}
for some large enough constant $\bar{C}$. Finally, to see $\bar{\Gamma}_{ij}^{(M_1,M_2,M_3,M_4)}=o(\E[\Phi_{ij}]^2n^{-2})$, we divide~\eqref{eq:gamma-bar-bound-1} by $\E[\Phi_{ii}]^2$:
\begin{align}
    \frac{\bar{\Gamma}_{ij}^{(M_1,M_2,M_3,M_4)}}{\E[\Phi_{ii}]^2}&\le \frac{\sigma_\rmu^{4k}\sigma_\rma^{4k}(1+o(1))\bar{C}m^{2k-M}n^{k+M_2}\ra^{2k-2M_2-2M_4}c^{M_4}\max\{1,\tfrac{\ra^{M_4}}{\qa^{M_4}}\}}{\left(\binom{m}{k}(\ru\sigma_\rmu^2)^k(\ra\sigma_\rma^2)^k\binom{n-1}{k}k!\right)^2}\nonumber\\
    &\le\frac{\bar{C}'\max\{1,\tfrac{\ra^{M_4}}{\qa^{M_4}}\}}{m^M n^{k-M_2}\ra^{2M_2+2M_4}\ru^{2k}},
\end{align}
for some constant $C'$. To show that $\frac{\bar{\Gamma}_{ij}^{(M_1,M_2,M_3,M_4)}}{\E[\Phi_{ii}]^2}=o(1/n^2)$, it suffices to show that 
\begin{equation}
    m^M n^{k-M_2}\ra^{2M_2+2M_4}\ru^{2k}=\omega(n^2),
    \label{eq:wrong-pair-final1}
\end{equation}
and 
\begin{equation}
    m^M n^{k-M_2}\ra^{2M_2+M_4}\qa^{M_4}\ru^{2k}=\omega(n^2).
    \label{eq:wrong-pair-final2}
\end{equation}
To see~\eqref{eq:wrong-pair-final1}, we notice that
\begin{align*}
    m^M n^{k-M_2}\ra^{2M_2+2M_4}\ru^{2k}&\stackrel{(a)}{\ge}m^M n^{k-M}\ra^{2M}\ru^{2k}\\
    &=(m\rho_\rma^2\rho_\rmu^2)^M(n\rho_\rmu^2)^{k-M}\\
    &\ge (\min\{m\rho_\rma^2\rho_\rmu^2,n\rho_\rmu^2\})^k\\
    &\stackrel{(b)}{=}\omega(n^2),
\end{align*}
where (a) follows because $M_2+M_4\le M$ and (b) follows by assumptions $m\rho_\rma^2\rho_\rmu^2=\omega(n^{2/k})$ and $n\rho_\rmu^2=\omega(n^{2/k})$.

To see~\eqref{eq:wrong-pair-final2}, we observe that
\begin{align*}
    m^M n^{k-M_2}\ra^{2M_2+M_4}\qa^{M_4}\ru^{2k}&=m^{M-M_2-M_4}m^{M_2+M_4} n^{k-M_2}\ra^{2M_2+M_4}\qa^{M_4}\ru^{2k}\\
    &=m^{M-M_2-M_4}(m\rho_\rma^2\rho_\rmu^2)^{M_2}(m\qa\ra)^{M_4}(n\rho_\rmu^2)^{k-M_2}.
\end{align*}

Suppose $m\qa\ra\ge 1$, then we have 
\begin{align*}
    m^{M-M_2-M_4}(m\rho_\rma^2\rho_\rmu^2)^{M_2}(m\qa\ra)^{M_4}(n\rho_\rmu^2)^{k-M_2}&\ge m^{M-M_2-M_4}(m\rho_\rma^2\rho_\rmu^2)^{M_2}(n\rho_\rmu^2)^{k-M_2}\\
    &\ge m^{M-M_2-M_4}(\min\{m\rho_\rma^2\rho_\rmu^2,n\rho_\rmu^2\})^k\\
    &\stackrel{(c)}{=}\omega(n^2),
\end{align*}
where (c) follows because  $m\rho_\rma^2\rho_\rmu^2=\omega(n^{2/k})$ and $n\rho_\rmu^2=\omega(n^{2/k})$.

Suppose $m\qa\ra< 1$. Because $M_4\le k-M_2$, we have
\begin{align*}
    m^{M-M_2-M_4}(m\rho_\rma^2\rho_\rmu^2)^{M_2}(m\qa\ra)^{M_4}(n\rho_\rmu^2)^{k-M_2}&\ge m^{M-M_2-M_4}(m\rho_\rma^2\rho_\rmu^2)^{M_2}(m\qa\ra n\rho_\rmu^2)^{k-M_2}\\
    &\ge m^{M-M_2-M_4}(\min\{m\rho_\rma^2\rho_\rmu^2,m\qa\ra n\rho_\rmu^2\})^k\\
    &\stackrel{(d)}{=}\omega(n^2),
\end{align*}
where (d) follows by assumptions $m\rho_\rma^2\rho_\rmu^2=\omega(n^{2/k})$ and $m\qa\ra n\rho_\rmu^2=\omega(n^{2/k})$.

Next, we show that $\tilde{\Gamma}_{ij}=O(\bar{\Gamma}_{ij})$. The key idea in this part of the proof is that since we already fixed an edge $(i,j)$ in the union graph of the four tuples in $\tilde{\L}_{ij}$, the choices of the 4-tuples is greatly restricted, and hence, it become a dominated case comparing to $\bar{\L}_{ij}$. 

Firstly, we further partition $\tilde{\L}_{ij}$ into three subsets
\begin{equation}
    \label{eq:tilde-partiion}
    \tilde{\L}_{ij}=\tilde{\L}_{ij}(1)\cup\tilde{\L}_{ij}(2)\cup\tilde{\L}_{ij}(3), 
\end{equation}
where
\begin{align*}
    \tilde{\L}_{ij}(1)&\triangleq\{(S_1,S_2,T_1,T_2)\in\tilde{\L}_{ij}:(i,j)\in (\eu(S_1)\cap\eu(T_1))\cap (\eu(S_2)\cup\eu(T_2))^c \}\\
    \tilde{\L}_{ij}(2)&\triangleq\{(S_1,S_2,T_1,T_2)\in\tilde{\L}_{ij}:(i,j)\in (\eu(S_1)\cup\eu(T_1))^c\cap (\eu(S_2)\cap\eu(T_2)) \}\\
    \tilde{\L}_{ij}(3)&\triangleq\{(S_1,S_2,T_1,T_2)\in\tilde{\L}_{ij}:(i,j)\in \eu(S_1)\cap\eu(T_2)\cap \eu(S_2)\cap\eu(T_2) \}.
\end{align*}
In other words, $\tilde{\L}_{ij}(1)$ includes the cases that $(i,j)$ is shared only by $S_1,T_1$, $\tilde{\L}_{ij}(2)$ includes the cases that $(i,j)$ is shared only by $S_2,T_2$ and $\tilde{\L}_{ij}(3)$ includes the cases that $(i,j)$ is shared all four graphs. We notice that $\tilde{\L}_{ij}(1),\tilde{\L}_{ij}(2),\tilde{\L}_{ij}(3)$ indeed partition the set $\tilde{\L}_{ij}$ because of $\eu(S_1)=\eu(T_1)$ and $\eu(S_2)=\eu(T_2)$. Following this definition, we define 
\begin{align*}
    \tilde{\Gamma}_{ij}(t)&\triangleq\sigma_\rmu^{4k}\sigma_\rma^{4k}\sum_{(S_1,S_2,T_1,T_2)\in \tilde{\L}_{ij}(t)}\Theta^\rmu(S_1,S_2,T_1,T_2)\Theta^\rma(S_1,S_2,T_1,T_2).
\end{align*}
for each $t\in [3]$. Thus, to show that $\tilde{\Gamma}_{ij}=O(\bar{\Gamma}_{ij})$, it suffices to show that $\tilde{\Gamma}_{ij}(t)=O(\bar{\Gamma}_{ij})$ for each $t\in [3]$.

Firstly, we show that $\tilde{\Gamma}_{ij}(1)=o(\bar{\Gamma}_{ij})$. Similarly to our analysis for  $\bar{\Gamma}_{ij}$, we can partition $\tilde{\L}_{ij}(1)$ as $$\tilde{\L}_{ij}(1)=\bigcup_{0\le M_1+M_2+M_3+M_4\le k}\tilde{\L}_{ij}^{(M_1,M_2,M_3,M_4)}(1),$$
according to the number of the four types of shared attributes in the union graph $H$, and define
$$\tilde{\Gamma}_{ij}^{(M_1,M_2,M_3,M_4)}(1)\triangleq \sigma_\rmu^{4k}\sigma_\rma^{4k}\sum_{(S_1,S_2,T_1,T_2)\in \tilde{\L}_{ij}^{(M_1,M_2,M_3,M_4)}(1)}\Theta^\rmu(S_1,S_2,T_1,T_2)\Theta^\rma(S_1,S_2,T_1,T_2).$$
For each $(S_1,S_2,T_1,T_2)\in \tilde{\L}_{ij}^{(M_1,M_2,M_3,M_4)}(1)$, it is easy to check that 
$$\Theta^\rmu(S_1,S_2,T_1,T_2)=1,$$
and 
$$\Theta^\rma(S_1,S_2,T_1,T_2)=\ra^{2k-2M_2-2M_4}c^{M_4}\max\{1,\tfrac{\ra^{M_4}}{\qa^{M_4}}\},$$
which are the same as a 4-tuple $(S_1,S_2,T_1,T_2)\in \bar{\L}_{ij}^{(M_1,M_2,M_3,M_4)}$. However, the cardinality of $\tilde{\L}_{ij}^{(M_1,M_2,M_3,M_4)}(1)$ is much smaller than the cardinality of $\bar{\L}_{ij}^{(M_1,M_2,M_3,M_4)}$. To see that, we observe that number of choices of the user-attribute part of the union graph in $\tilde{\L}_{ij}^{(M_1,M_2,M_3,M_4)}(1)$ is the same as the number of choices in $\bar{\L}_{ij}^{(M_1,M_2,M_3,M_4)}$. However, the number of ways of choosing the user-user part in $\tilde{\L}_{ij}^{(M_1,M_2,M_3,M_4)}(1)$ is upper bounded by
\begin{equation}
\label{eq:tilde1-user}
    \sum_{\beta=k-M_2}^k\binom{n}{\beta}\binom{n}{k-1-\beta}\binom{n}{k-\beta}.
\end{equation}
This is because we can choose $\beta$ users shared by $S_1,S_2,T_1,T_2$, $k-\beta$ users shared by $T_1,T_2$ and $k-\beta-1$ \emph{users other than $j$ that are shared by $S_1,T_1$}. If we compare~\eqref{eq:tilde1-user} to~\eqref{eq:bar-user-count}, we can observe that there is a factor $n$ missing in~\eqref{eq:tilde1-user}. This is because one of the users in $S_1$ and $T_1$ are already fixed to $j$, so we have one less user to choose. This implies that $\tilde{\Gamma}_{ij}^{(M_1,M_2,M_3,M_4)}(1)=O(\bar{\Gamma}_{ij}^{(M_1,M_2,M_3,M_4)}/n)$ for any $0\le M_1+M_2+M_3+M_4\le k$, and it follows that $\tilde{\Gamma}_{ij}(1)=o(\bar{\Gamma}_{ij})$. The proof of $\tilde{\Gamma}_{ij}(2)=o(\bar{\Gamma}_{ij})$ follows similarly as the proof of $\tilde{\Gamma}_{ij}(1)=o(\bar{\Gamma}_{ij})$.

Finally, we show that $\tilde{\Gamma}_{ij}(3)=O(\bar{\Gamma}_{ij})$. We perform the similar partition again: 
$$\tilde{\L}_{ij}(3)=\bigcup_{0\le M_1+M_2+M_3+M_4\le k}\tilde{\L}_{ij}^{(M_1,M_2,M_3,M_4)}(3),$$
according to the number of the four types of shared attributes in the union graph $H$, and define
$$\tilde{\Gamma}_{ij}^{(M_1,M_2,M_3,M_4)}(3)\triangleq \sigma_\rmu^{4k}\sigma_\rma^{4k}\sum_{(S_1,S_2,T_1,T_2)\in \tilde{\L}_{ij}^{(M_1,M_2,M_3,M_4)}(3)}\Theta^\rmu(S_1,S_2,T_1,T_2)\Theta^\rma(S_1,S_2,T_1,T_2).$$ 
For each $(S_1,S_2,T_1,T_2)\in \tilde{\L}_{ij}^{(M_1,M_2,M_3,M_4)}(3)$, in the user-attribute part of the union graph, we still have 
$$\Theta^\rma(S_1,S_2,T_1,T_2)=\ra^{2k-2M_2-2M_4}c^{M_4}\max\{1,\tfrac{\ra^{M_4}}{\qa^{M_4}}\}.$$

However, in the user-attribute part, we have $|K_{22}^\rmu|=1$ instead of $|K_{22}^\rmu|=0$ in the previous case. This is because we have the edge $(i,j)$ shared by all four graphs. It is easy to check that we still have $|K_{11}^\rmu|=|K_{12}^\rmu|=|K_{21}^\rmu|=0$. Thus, we have
$$\Theta^\rmu(S_1,S_2,T_1,T_2)=c\max\{1,\tfrac{\ru}{q_u}\}\le \tfrac{c}{\qu}.$$

% However, in the user-user part we have one less $\eu(H)=2k-1$ instead of $2k$, because we now have a user-user edge $(i,j)$ shared by all four graphs $S_1,S_2,T_1,T_2$. Hence, we have 
% $$\ru^{|K_{11}^\rmu|}\qu^{-2k+\eu(H)}=\qu^{-1}.$$ 
Moreover, if we compare $|\tilde{\L}_{ij}^{(M_1,M_2,M_3,M_4)}(3)|$ to $|\bar{\L}_{ij}^{(M_1,M_2,M_3,M_4)}|$, we observe that 
$$\frac{|\tilde{\L}_{ij}^{(M_1,M_2,M_3,M_4)}(3)|}{|\bar{\L}_{ij}^{(M_1,M_2,M_3,M_4)}|}=O(1/n^2).$$
This is because in the union graph of a 4-tuple $(S_1,S_2,T_1,T_2)\in \tilde{\L}_{ij}^{(M_1,M_2,M_3,M_4)}(3)$, we have two less users to choose, since one of the ports in $S_1,T_1$ is fixed to $j$ and one of the ports in $S_2,T_2$ is fixed to $i$. As a result, for any $0\le M_1+M_2+M_3+M_4\le k$, we have 
$$\frac{\tilde{\Gamma}_{ij}^{(M_1,M_2,M_3,M_4)}(3)}{\bar{\Gamma}_{ij}^{(M_1,M_2,M_3,M_4)}}=O\left(\frac{c}{n^2\qu}\right)=O(1),$$
because of assumption $n^2\qu=\Omega(1)$. This further implies that $\tilde{\Gamma}_{ij}(3)=O(\bar{\Gamma}_{ij})$, which completes the proof.

\section{Proof of Theorem~\ref{thm:exact}}\label{appd:exact}
In this section, we prove the 
theoretical performance guarantee for exact alignment. Recall that Theorem~\ref{thm:almost-exact} ensures that Algorithm~\ref{alg:subgraph-count} outputs a set of users $I$ with $|I|=n-o(n)$ and a mapping $\hat\Pi$ with $\hat\Pi(i)=\hat\Pi^*(i),\forall i\in I$. Therefore, to prove Theorem~\ref{thm:exact}, it suffices to show that Algorithms~\ref{alg:refine-sparse} and~\ref{alg:refine-rich} can exactly recover $\Pi^*$ with input $I,\hat\Pi$ under the \AttrSparse{} regime and \AttrRich{} regime respectively. In the following two propositions, we provide performance guarantee for the two refinement algorithms respectively.
To state the two propositions, we define function 
\begin{equation*}
    f(x)\triangleq x\log x-x+1.
\end{equation*}
Note that $f(x)$ is a function monotonically decreasing in $(0,1)$, and monotonically increasing in $(1,\infty)$.
\begin{prop}[Feasible region of Algorithm~\ref{alg:refine-sparse} in \AttrSparse{} regime]
    \label{prop:refine-sparse}
    Suppose $(G_1,G_2')\sim\cg(n,\qu,\ru;m,\qa,\ra)$, where 
    \begin{equation}
    \label{eq:sparse-cond-1}
        n\rho_\rmu^2=\omega(\log n),
    \end{equation}
    \begin{equation}
    \label{eq:sparse-cond-2}
        n\qu(\qu+\ru(1-\qu))\ge (1+\epsilon)\log n
    \end{equation}
    and
    \begin{equation}
    \label{eq:sparse-cond-3}
        \frac{\ru(1-\qu)}{\qu}\ge \epsilon
    \end{equation}   
    for some arbitrarily small constant $\epsilon$.
    Let $\hat{\pi}\equiv \hat{\pi}(G_1,G_2')$ be a mapping $I\rightarrow[n]$ such that $\hat{\pi}=\pi^*|_I$ and $|I|\ge (1-\epsilon/16)n$. Let $\gamma_1$ denote the unique solution in $(1,\infty)$ to $f(\gamma_1)=\frac{3\log n}{(n-2)q_\rmu^2}$. Then there exists an algorithm, namely Algorithm~\ref{alg:refine-sparse} with inputs $\gamma_1,I$ and $\hat{\pi}$, that outputs $\tilde{\pi}=\pi^*$ with high probability.
\end{prop}

\begin{prop}[Feasible region of Algorithm~\ref{alg:refine-rich} in \AttrRich{} regime]
    \label{prop:refine-rich}
    Suppose $(G_1,G_2')\sim\cg$ $(n,\qu,\ru;m,\qa,\ra)$, where 
    \begin{equation}
    \label{eq:rich-cond-1}
        n\rho_\rmu^2=\omega(\log n),
    \end{equation}
    \begin{equation}
    \label{eq:rich-cond-2}
        m\rho_\rma^2=\omega(\log n),
    \end{equation}
    \begin{equation}
    \label{eq:rich-cond-3}
        n\qu(\qu+\ru(1-\qu))+m\qa(\qa+\ra(1-\qu))\ge (1+\epsilon)\log n
    \end{equation}
    \begin{equation}
    \label{eq:rich-cond-4}
        n\qu(\qu+\ru(1-\qu))=\Omega(\log n),
    \end{equation}
    \begin{equation}
    \label{eq:rich-cond-5}
        m\qa(\qa+\ra(1-\qu))=\Omega(\log n),
    \end{equation}
    \begin{equation}
    \label{eq:rich-cond-6}
        \frac{\ru(1-\qu)}{\qu}\ge \epsilon,
    \end{equation}   
    and
    \begin{equation}
    \label{eq:rich-cond-7}
        \frac{\ra(1-\qa)}{\qa}\ge \epsilon,
    \end{equation}      
    for some arbitrarily small constant $\epsilon$.
    Let $\hat{\pi}\equiv \hat{\pi}(G_1,G_2')$ be a mapping $I\rightarrow[n]$ such that $\hat{\pi}=\pi^*|_I$ and $|I|\ge (1-\epsilon/16)n$. Let $\gamma_2$ denote the unique solution in $(1,\infty)$ to $f(\gamma_2)=\frac{3\log n}{(n-2)q_\rmu^2}$, and let $\gamma_3$ denote the unique solution in $(1,\infty)$ to $f(\gamma_3)=\frac{3\log n}{mq_\rma^2}$. Then there exists an algorithm, namely Algorithm~\ref{alg:refine-rich} with inputs $\gamma_2,\gamma_3,I$ and $\hat{\pi}$, that outputs $\tilde{\pi}=\pi^*$ with high probability.
\end{prop}
With Propositions~\ref{prop:refine-sparse} and~\ref{prop:refine-rich}, we are now ready to prove Theorem~\ref{thm:exact}.
\renewcommand*{\proofname}{Proof of Theorem~\ref{thm:exact}.}
\begin{proof}
    In this proof, we partition the information theoretic feasible region 
    \begin{equation*}
        n\qu(\qu+\ru(1-\qu))+m\qa(\qa+\ra(1-\qa))\ge (1+\epsilon)\log n
    \end{equation*}
    into three regimes:
\begin{enumerate}
    \item $m\qa(\qa+\ra(1-\qa))=o(\log n)$, and $ n\qu(\qu+\ru(1-\qu))\ge (1+\epsilon)\log n$;
    \item $m\qa(\qa+\ra(1-\qa))=\Omega(\log n)$, and $ n\qu(\qu+\ru(1-\qu))= \Omega(\log n)$;
    \item $m\qa(\qa+\ra(1-\qa))\ge(1+\epsilon)\log n$, and $ n\qu(\qu+\ru(1-\qu))= o(\log n)$,
\end{enumerate}
and separately prove the feasibility of exact alignment in each regime.

In the first regime, by the assumptions in Theorem~\ref{thm:exact}, all the conditions in Theorem~\ref{thm:almost-exact} are satisfied, so the outputs $I$ and $\hat{\pi}$ from Algorithm~\ref{alg:subgraph-count} satisfy that $\hat{\pi}=\pi^*|_I$, and $|I|\ge n-o(n)$ with high probability. Moreover, because condition~\eqref{eq:almost-cond-5} implies condition~\eqref{eq:sparse-cond-1} and we have $ n\qu(\qu+\ru(1-\qu))\ge (1+\epsilon)\log n$ in this regime, all the conditions in Proposition~\ref{prop:refine-sparse} are satisfied. Therefore, on the event that $\{\hat{\pi}=\pi^*|_I\}\cap\{|I|\ge n-o(n)\}$, Algorithm~\ref{alg:refine-sparse} outputs $\tilde{\pi}=\pi^*$ with high probability. This completes the proof for the regime.

In the second regime, similarly, we know that the outputs from Algorithm~\ref{alg:subgraph-count} satisfy $\hat{\pi}=\pi^*|_I$, and $|I|\ge n-o(n)$ with high probability. Notices that condition~\eqref{eq:almost-cond-4} implies condition~\eqref{eq:rich-cond-2}, and condition~\eqref{eq:almost-cond-5} implies condition~\eqref{eq:rich-cond-1}. By assumptions $m\qa(\qa+\ra(1-\qu))=\Omega(\log n)$, and $ n\qu(\qu+\ru(1-\qu))= \Omega(\log n)$ for the \AttrRich{} regime, we can see that all the conditions in Proposition~\ref{prop:refine-rich} are satisfied. Thus, on the event $\{\hat{\pi}=\pi^*|_I\}\cap\{|I|\ge n-o(n)\}$, Algorithm~\ref{alg:refine-rich} outputs $\tilde{\pi}=\pi^*$ with high probability. Therefore, exact recovery is achieved with high probability.

Finally, consider the third regime. In this regime, we consider a strictly harder problem where all user-user edges are removed from the two graphs $G_1$ and $G_2'$. Then both graphs reduce to bipartite graphs with edges only between the users and the attributes. For this bipartite graph alignment problem, it is well-known that the optimal MAP estimator can be implemented efficiently within $O(n^3)$ complexity (See for example~\cite{Cullina-database}). In Theorem 1 of~\cite{zhang2024TIT}, it is shown that the MAP estimator achieves exact recovery if
% \begin{equation}
%     m(\sqrt{\qa(\qa+\ra(1-\qa))(1-\qa(\qa+\ra(1-\qa))-2\qa(1-\qa-\ra(1-\qa)))}-\qa(1-\qa-\ra(1-\qa)))^2
% \end{equation}
\begin{equation}
\label{eq:bipa-IT}
    m(\sqrt{q_{11}q_{00}}-\sqrt{q_{01}q_{10}})^2\ge \log n + \omega(1),
\end{equation}
where \[q_{11}\triangleq \qa(\qa+\ra(1-\qa)),\]
\[q_{00}\triangleq1-\qa(\qa+\ra(1-\qa))-2\qa(1-\qa-\ra(1-\qa))\]
and
\[q_{10}=q_{01}\triangleq\qa(1-\qa-\ra(1-\qa)).\]
To complete the proof, it suffices to show that~\eqref{eq:bipa-IT} is satisfied under $m\qa(\qa+\ra(1-\qa))\ge(1+\epsilon)\log n$ and other conditions of Theorem~\ref{thm:exact}. Towards the proof, we define two more quantities $s_\rma\triangleq \qa+\ra(1-\ra)$ and $q\triangleq \qa/s_\rma$. In view of condition~\eqref{eq:exact-cond-2}, we know that $q$ is bounded away from $1$. We separately consider two regimes of $q$ to complete the proof. 

Firstly, consider the regime $q=o(1)$. Notice that $\frac{1}{q}=1+\frac{\ra(1-\qa)}{\qa}$, we then have $\ra=\omega(\qa)$, which further implies that $s_\rma=(1-o(1))\ra$. It follows that
\begin{align*}
    m(\sqrt{q_{11}q_{00}}-\sqrt{q_{01}q_{10}})^2&\ge m(\sqrt{q_{11}q_{00}}-\qa)^2\\
    &=m(\sqrt{\qa s_\rma}-\qa)^2(1-o(1))\\
    &=(1-o(1))m\qa s_\rma,
\end{align*}
where the penultimate equality follows because $q=o(1)$ and the last equality follows because $s_\rma=(1-o(1))\ra=\omega(\qa)$. Finally,~\eqref{eq:bipa-IT} follows by the assumption that $m\qa s_\rma\ge(1+\epsilon)\log n$.

Next, we consider the regime $q=\Theta(1)$. In this regime, by the definitions of $q$ and $s_\rma$, we have $\ra=\Theta(\qa)$ and $\ra=\Theta(s_\rma)$. Moreover, we have
\begin{align*}
    m(\sqrt{q_{11}q_{00}}-\sqrt{q_{01}q_{10}})^2&=(1-q)m\qa s_\rma\frac{1}{\left(\sqrt{1+\frac{q(1-s_\rma)^2}{1-q}}+\sqrt{\frac{q(1-s_\rma)^2}{1-q}}\right)^2}=\Theta(m\qa s_\rma),
\end{align*}
because $q$ is bounded away from 1. The proof is then completed by realizing
\begin{align*}
    m\qa s_\rma=mqs_\rma^2=\Omega(mq\rho_\rma^2)=\Omega(n^{\Theta(1)}),
\end{align*}
where the last equality follows by condition~\eqref{eq:almost-cond-4} and the assumption that $q=\Theta(1)$.
\end{proof}

\subsection{Proof of Proposition~\ref{prop:refine-sparse}}
In this section, we show that Algorithm~\ref{alg:refine-sparse} achieves exact recovery under the conditions specified by Proposition~\ref{prop:refine-sparse}. Towards that goal, we first present a lemma in~\cite{mao2023} that shows an \erdos--\renyi graph $G\sim\cg(n,p)$ has good expansion property with high probability.
\begin{lem}[Lemma 13 in~\cite{mao2023}]
\label{lem:er-expander}
    Suppose $G\sim\cg(n,p)$, where $np\ge (1+\epsilon)\log n$ for an arbitrarily small constant $\epsilon$. Let $\kappa_G(I,I^c)$ denote the number of edges between $I\subseteq [n]$ and $I^c$, where $I^c\triangleq [n]\setminus I$. Then with probability at least $1-O(n^{-\epsilon/8})$, for all subsets $I\subseteq [n]$ with $|I|\le \frac{\epsilon}{16}n$, we have $\kappa_G(I,I^c)\ge \eta |I||I^c|p$, where $\eta$ is the unique solution in $(0,1)$ to $f(\eta)=\frac{(1+\epsilon/8)\log n}{(1-\epsilon/16)np}$. Moreover, $\eta$ satisfies lower bound
    \begin{equation}
    \label{eq:eta-lower}
        \eta\ge \max\left\{\frac{\epsilon}{16},1-\sqrt{\frac{2(1+\epsilon/8)\log n}{(1-\epsilon/16)np}}\right\}.
    \end{equation}
\end{lem}
Now, we are ready to prove Proposition~\ref{prop:refine-sparse}. We assume without generality that the true permutation $\pi^*$ is the identity permutation. For a pair of user vertices $i\in \vuone$ and $j\in \vutwo$, let $N^\rmu(i,j)$ denote the number of user vertices $u\in [n]$ such that $u$ is connected to $i$ in $G_1$ and $u$ is connected to $j$ in $G_2'$. By the definition of the attributed \erdos--\renyi graph pair model, we have $N^\rmu(i,j)\sim\mathrm{Binom}(n-2,q_\rmu^2)$ for any $i\neq j$. By the multiplicative Chernoff bound (see for example~\citet[Theorem~4.4]{mitzenmacher_upfal_2005}), we have
\begin{equation*}
    \P(N^\rmu(i,j)\ge \gamma_1 (n-2)q_\rmu^2)\le \exp(-(n-2)q_\rmu^2h(\gamma_1))=n^{-3}.
\end{equation*}
By a union bound over all the $\binom{n}{2}$ pairs of users, we have that
\begin{equation}
\label{eq:sparse-no-wrong}
    \P(\exists i,j\in [n], i\neq j:N^\rmu(i,j)\ge \gamma_1(n-2)q_\rmu^2)\le \binom{n}{2}n^{-3}\le n^{-1}.
\end{equation}
Now, we assume that $\{\nexists i,j\in [n], i\neq j:N^\rmu(i,j)\ge \gamma_1(n-2)q_\rmu^2\}$. From this, we are going prove by induction that $\tilde{\pi}=\pi^*|_J$ throughout out the process of the algorithm. At the beginning of the algorithm, this certainly holds because we set $J=I$ and $\tilde{\pi}=\hat{\pi}$. Now we suppose $\tilde{\pi}=\pi^*|_J$ up to $l$ calls of the while loop in Algorithm~\ref{alg:refine-sparse}. Consider the $(l+1)$-th call of the while loop. By the induction hypothesis, we have that
\begin{equation*}
    N^\rmu_{\tilde{\pi}}(i,j)\le N^\rmu(i,j)<\gamma_1(n-2)q_\rmu^2,
\end{equation*}
for all $i,j\in[n]$, $i\neq j$. As a result, at $(l+1)$-th call of the while loop, we either have that Algorithm~\ref{alg:refine-sparse} terminates or some $i$ is added to $J$ and $\tilde{\pi}(i)=i$. This proves that $\tilde{\pi}=\pi^*|_J$ after $l+1$ calls of the while loop. 

To complete the proof, next we show that when the algorithm terminates, we have $J=[n]$ and $\tilde{\pi}$ equals to the identity permutation. Towards contradiction, we suppose that the algorithm ends with $|J|<n$. This suggests that $N^\rmu_{\tilde{\pi}}(i,i)<\gamma_1(n-2)q_\rmu^2$ for each $i\in J^c$, which further implies that
\begin{equation}
    \kappa_{G_1^\rmu\cap G_2^\rmu}(J,J^c)=\sum_{i\in J^c}N^\rmu_{\tilde{\pi}}(i,i)<\gamma_1(n-2)q_\rmu^2|J^c|.
\end{equation}
From the attributed \erdos--\renyi graph pair model, it is not hard to see that $G_1^\rmu\cap G_2'^\rmu\sim\cg(n,\qu s_\rmu)$, where $s_\rmu\triangleq \qu+\ru(1-\qu)$. Because condition~\eqref{eq:sparse-cond-2} holds, from Lemma~\ref{lem:er-expander}, we know that with probability at least $1-O(n^{-\epsilon/8})$, 
\begin{equation}
    \kappa_{G_1^\rmu\cap G_2^\rmu}(J,J^c)\ge \eta|J||J^c|\qu s_\rmu\stackrel{(a)}{\ge}\eta(1-\epsilon/16)n|J^c|\qu s_\rmu,
\end{equation}
where $\eta$ is the unique solution in $(0,1)$ to $h(\eta)=\frac{(1+\epsilon/8)\log n}{(1-\epsilon/16)n\qu s_\rmu}$ and (a) follows because we assume $|J|\ge (1-\epsilon/16)n$ at the very beginning of the algorithms. To complete the proof, it suffices to show that $\gamma_1(n-2)q_\rmu^2|J^c|\le \eta(1-\epsilon/16)n|J^c|\qu s_\rmu$, i.e., we want to show that
\begin{equation*}
    \gamma_1\le \eta(1-\epsilon/16)\frac{s_\rmu}{\qu}\triangleq \bar{\gamma}_1.
\end{equation*}
By taking derivative, one can check that $f(x)$ is monotone increasing on $(1,\infty)$. Thus, it suffices to show that $\bar{\gamma}_1>1$ and $f(\bar{\gamma}_1)\ge f(\gamma_1)$.
We show these by considering two different cases. 

\underline{\textbf{Case 1: $\qu=o(\ru):$}} In view of $\eta\ge \epsilon/16$, we have
\begin{align*}
    \bar{\gamma}_1&\ge (1-\epsilon/16)\frac{\epsilon s_\rmu}{16\qu}\\
    &=(1-\epsilon/16)\frac{\epsilon}{16}\left(1+\frac{\ru(1-\qu)}{\qu}\right)\\
    &=\omega(1),
\end{align*}
where the last equality follows because $\qu=o(\ru)$. Therefore, it follows that
\begin{align*}
    f(\bar{\gamma}_1)>\bar{\gamma}_1(\log \bar{\gamma}_1 - 1)
    \stackrel{(b)}{=}\omega(\bar{\gamma}_1)
    \stackrel{(c)}{=}\omega\left(\frac{s_\rmu}{\qu}\right),
\end{align*}
where (b) follows because $\bar{\gamma}_1=\omega(1)$ and (c) follows because $\eta\ge \epsilon/16$ and $\epsilon=\Theta(1)$.
Finally, the proof of Case 1 is completed by realizing $f(\bar{\gamma}_1)\ge f(\gamma_1)$ because 
\begin{equation*}
\frac{s_\rmu/\qu}{f(\bar{\gamma}_1)}=\frac{s_\rmu/\qu}{3\log n/((n-2)q_\rmu^2)}=\frac{(n-2)\qu s_\rmu}{3\log n}=\Omega(1),
\end{equation*}
where the last equality follows by the assumption that $n\qu s_\rmu\ge (1+\epsilon)\log n$.

\underline{\textbf{Case 2: $\qu=\Omega(\ru):$}} Recall that we have $\eta\ge 1-\sqrt{\frac{2(1+\epsilon/8)\log n}{(1-\epsilon/16)n\qu s_\rmu}} $. Notice that 
\begin{equation*}
    n\qu s_\rmu\ge nq_\rmu^2=\Omega(n\rho_\rmu^2)=\omega(\log n), 
\end{equation*}
where the last equality follows by condition~\eqref{eq:sparse-cond-1} in the proposition. Then it follows that $\eta\ge 1-\epsilon/16$. Therefore, we have 
\begin{align*}
    \bar{\gamma}_1&\ge (1-\epsilon/16)^2\frac{s_\rmu}{\qu}\\
    &\stackrel{(d)}{\ge} (1-\epsilon/16)^2(1+\epsilon)\\
    &\ge 1+\epsilon/2,
\end{align*}
where (d) follows by condition~\eqref{eq:sparse-cond-3} in the proposition. Finally, the proof is completed by 
\begin{align*}
    f(\bar{\gamma}_1)\ge f(1+\epsilon/2)\stackrel{(e)}{>}\frac{3\log n}{(n-2)q_\rmu^2}=f(\gamma_1),
\end{align*}
where (e) follows because $f(1+\epsilon/2)=\Omega(1)$, while $\frac{3\log n}{(n-2)q_\rmu^2}=O(\frac{3\log n}{(n-2)\rho_\rmu^2})=o(1)$ by condition~\eqref{eq:sparse-cond-1}.
\subsection{Proof of Proposition~\ref{prop:refine-rich}}
To show Proposition~\ref{prop:refine-rich}, we firstly study some expansion property of random attributed graphs. To state the result, we need to define the \emph{attributed \erdos--\renyi graph model} $\cg(n,p;m,q)$, which generates random attributed graphs. In this model, we generate an attributed graph $G$ with $n+m$ vertices, where $\vu(G)=[n]$ and $\va(G)=\{a_1,\ldots,a_m\}$. Between each pair $(i,j)\in \vu(G)\times\vu(G), i\neq j$, a user-user edge is generated i.i.d. with probability $p$, and between each pair $(i,a_j)\in \vu(G)\times \va(G)$, a user-attribute edge is generated i.i.d. with probability $q$. We show that attributed graphs generated from the attributed \erdos--\renyi graph model satisfies good expansion property with high probability. 
% Before we state the expansion result, recall that we assume $$
\begin{lem}[Expansion property of attributed \erdos--\renyi graph]
\label{lem:attri-er-expander}
    Suppose $G\sim\cg(n,p;m,q)$. Assume $np\ge \lambda_1(1+\epsilon)\log n$, $mq\ge \lambda_2(1+\epsilon)\log n$ for some constants $\epsilon,\lambda_1,\lambda_2$ with $\lambda_1+\lambda_2\ge 1$. Let $\kappa^\rmu_G(I,I^c)$ denote the number of user-user edges between $I\subseteq [n]$ and $I^c$, where $I^c\triangleq [n]\setminus I$. Let $\kappa^\rma_G(I,\va(G))$ denote the number of user-attribute edges between $I\subseteq [n]$ and $\va(G)$. Let $\eta_1$ denote the unique solution in $(0,1)$ to $f(\eta_1)=\frac{\lambda_1(1+\epsilon/8)\log n}{(1-\epsilon/16)np}$, and $\eta_2$ denote the unique solution in $(0,1)$ to $f(\eta_2)=\frac{\lambda_2(1+\epsilon/8)\log n}{mq}$. Then with probability at least $1-O(n^{-\epsilon/8})$, for all subsets $I\subseteq [n]$ with $|I|\le \frac{\epsilon}{16}n$, at least one of the following two events happens:
    \begin{enumerate}
        \item $\kappa^\rmu_G(I,I^c)\ge \eta_1|I||I^c|p$;
        \item $\kappa^\rma_G(I,\va(G))\ge \eta_2|I|mq$.
    \end{enumerate}
    Moreover, $\eta_1$ and $\eta_2$ satisfy lower bounds
    \begin{equation}
        \label{eq:eta1-lower}
        \eta_1\ge \max\left\{\epsilon/16,1-\sqrt{\frac{2\lambda_1(1+\epsilon/8)\log n}{(1-\epsilon/16)np}}\right\},
    \end{equation}
    \begin{equation}
        \label{eq:eta2-lower}
        \eta_2\ge \max\left\{\epsilon/16,1-\sqrt{\frac{2\lambda_2(1+\epsilon/8)\log n}{mq}}\right\}.
    \end{equation}
\end{lem}
\renewcommand*{\proofname}{Proof of Lemma~\ref{lem:attri-er-expander}.}
\begin{proof}
    When $|I|=\emptyset$, both events in Lemma~\ref{lem:attri-er-expander} happen. Therefore, we only consider the non-trivial cases of $1\le |I|\le \epsilon/16$. For some fixed $I$ with $1\le |I|\le \epsilon/16$, we observe that $\kappa^\rmu_G(I,I^c)\sim\mathrm{Binom}(|I||I^c|,p)$. Notice that
    $$f(\eta_1)=\frac{\lambda_1(1+\epsilon/8)\log n}{(1-\epsilon/16)np}\le \frac{1+\epsilon/8}{(1-\epsilon/16)(1+\epsilon)}<1.$$ Therefore, there indeed exists a unique solution $\eta_1\in (0,1)$.
    By the multiplicative Chernoff bound, we have
    \begin{align*}
        \P(\kappa^\rmu_G(I,I^c)\le \eta_1|I||I^c|p)
        &\le \exp(-f(\eta_1)|I||I^c|p)\\
        &=\exp\left(-\frac{\lambda_1(1+\epsilon/8)\log n}{(1-\epsilon/16)np}|I||I^c|p\right)\\
        &\stackrel{(a)}{\le} \left(-\frac{\lambda_1(1+\epsilon/8)\log n}{(1-\epsilon/16)np}|I|(1-\epsilon/16)np\right)\\
        &=\exp\left(-\lambda_1(1+\epsilon/8)|I|\log n\right),
    \end{align*}
    where (a) follows because $|I^c|\ge (1-\epsilon/16)n$.
    Similarly, we observe that $\kappa^\rma_G(I,\va(G))\sim\mathrm{Binom}(|I|m,q)$. Note that 
    $$f(\eta_2)=\frac{\lambda_2(1+\epsilon/8)\log n}{mq}\le \frac{1+\epsilon/8}{1+\epsilon}<1.$$
    Thus, there indeed exists a unique solution $\eta_2\in (0,1)$.
    By the multiplicative Chernoff bound, we have
    \begin{align*}
        \P(\kappa^\rma_G(I,\va(G))\le \eta_2|I|\va(G)q)
        &\le \exp(-f(\eta_2)|I|mq)\\
        &=\exp\left(-\frac{\lambda_2(1+\epsilon/8)\log n}{mq}|I|mq\right)\\
        &=\exp(-\lambda_2(1+\epsilon/8)|I|\log n).
    \end{align*}
    Notice that $\kappa^\rmu_G(I,I^c)$ and $\kappa^\rma_G(I,\va(G))$ are two independent random variables because $\kappa^\rmu_G(I,I^c)$ only counts user-user edges, while $\kappa^\rma_G(I,\va(G))$ only counts user-attribute edges. Hence, we have
    \begin{align*}
        &\P(\kappa^\rmu_G(I,I^c)\le \eta_1|I||I^c|p\text{ and }\kappa^\rma_G(I,\va(G))\le \eta_2|I|\va(G)q)\\&\le \exp(-(\lambda_1+\lambda_2)(1+\epsilon/8)|I|\log n)\\
        &\le n^{-(1+\epsilon/8)|I|},
    \end{align*}
    where the last inequality follows because $\lambda_1+\lambda_2\ge 1$. Applying a union bound over all $\binom{n}{k}$ choices of $|I|=l$ yields
    \begin{align*}
        &\P(\exists I\subseteq [n]:|I|=l ,\kappa^\rmu_G(I,I^c)\le \eta_1|I||I^c|p\text{ and }\kappa^\rma_G(I,\va(G))\le \eta_2|I|\va(G)q)\\&\le \binom{n}{l}n^{-(1+\epsilon/8)l}\le n^{-\epsilon l/8 }.
    \end{align*}
    Finally, applying a union bounder over $1\le l\le \lfloor\epsilon n/16\rceil$ yields
    \begin{align*}
        &\P(\exists I\subseteq [n]:1\le|I|\le \lfloor\epsilon n/16\rceil ,\kappa^\rmu_G(I,I^c)\le \eta_1|I||I^c|p\text{ and }\kappa^\rma_G(I,\va(G))\le \eta_2|I|\va(G)q)\\
        &\le \sum_{l=1}^{\lfloor\epsilon n/16\rceil} n^{-\epsilon l/8 }=O(n^{-\epsilon/8 }).
    \end{align*}
    To complete the proof, we still need show the two lower bounds for $\eta_1$ and $\eta_2$. For $\eta_1$, we notice that
    \begin{equation}
        f(\eta_1)=\frac{\lambda_1(1+\epsilon/8)\log n}{(1-\epsilon/16)np}\le \frac{1+\epsilon/8}{(1-\epsilon/16)(1+\epsilon)}\le f(\epsilon/16).
    \end{equation}
    Because $f(x)$ is monotone decreasing on $x\in (0,1)$, we know that $\eta_1\ge \epsilon/16$. Moreover, because $f(x)\ge (1-x)^2/2$ for any $x\in (0,1)$, it follows that
    $$\frac{\lambda_1(1+\epsilon/8)\log n}{(1-\epsilon/16)np}\ge \frac{(1-\eta_1)^2}{2}.$$
    This proves that $$\eta_1\ge 1-\sqrt{\frac{2\lambda_1(1+\epsilon/8)\log n}{(1-\epsilon/16)np}}.$$
    For $\eta_2$, similarly, we have 
    \begin{equation}
        f(\eta_2)=\frac{\lambda_2(1+\epsilon/8)\log n}{mq}\le \frac{1+\epsilon/8}{1+\epsilon}\le f(\epsilon/16),
    \end{equation}
    which implies that $\eta_2\ge \epsilon/16$.
    Finally, we have 
    $$\eta_2\ge 1-\sqrt{\frac{2\lambda_2(1+\epsilon/8)\log n}{mq}},$$
    because
    $$\frac{\lambda_2(1+\epsilon/8)\log n}{mq}\ge \frac{(1-\eta_2)^2}{2}.$$
\end{proof}
Now, we can proceed to prove Proposition~\ref{prop:refine-rich}. We assume without generality that the true permutation $\pi^*$ is the identity permutation. For a pair of user vertices $i\in \vuone$ and $j\in \vutwo$, let $N^\rmu(i,j)$ denote the number of user vertices $u\in [n]$ such that $u$ is connected to $i$ in $G_1$ and $u$ is connected to $j$ in $G_2'$, and let $N^\rma(i,j)$ denote the number of attribute vertices $a\in \{a_1,\ldots,a_m\}$ such that $a$ is connected to $i$ in $G_1$ and $a$ is connected to $j$ in $G_2'$. By the definition of \erdos--\renyi graph pair model, we have $N^\rmu(i,j)\sim\mathrm{Binom}(n-2,q_\rmu^2)$ and $N^\rma(i,j)\sim\mathrm{Binom}(m,q_\rma^2)$ for any $i\neq j$. By the multiplicative Chernoff bound, we have
\begin{equation*}
    \P(N^\rmu(i,j)\ge \gamma_2 (n-2)q_\rmu^2)\le \exp(-(n-2)q_\rmu^2f(\gamma_2))=n^{-3},
\end{equation*}
and 
\begin{equation*}
    \P(N^\rma(i,j)\ge \gamma_3 mq_\rma^2)\le \exp(-mq_\rma^2f(\gamma_3))=n^{-3}.
\end{equation*}
By the union bound, for a fixed pair $(i,j)$,
$$\P(N^\rmu(i,j)\ge \gamma_2 (n-2)q_\rmu^2\text{ or }N^\rma(i,j)\ge \gamma_3 mq_\rma^2)\le 2n^{-3}.$$
Moreover, by a union bound over all $\binom{n}{2}$ pairs, we have
\begin{equation}
\label{eq:rich-no-wrong}
    \P(\exists i,j\in [n], i\neq j:N^\rmu(i,j)\ge \gamma_2 (n-2)q_\rmu^2\text{ or }N^\rma(i,j)\ge \gamma_3 mq_\rma^2)\le \binom{n}{2}2n^{-3}\le 2n^{-1}.
\end{equation}
Now, we assume that $\{\nexists i,j\in [n], i\neq j:N^\rmu(i,j)\ge \gamma_2 (n-2)q_\rmu^2\text{ or }N^\rma(i,j)\ge \gamma_3 mq_\rma^2\}$. By a similar induction as in the proof of Proposition~\ref{prop:refine-sparse}, we have that $\tilde{\pi}=\pi^*|_J$ throughout out the process of the algorithm. Therefore, it suffices to show that when the algorithm terminates, we have $J=[n]$. Towards contradiction, we suppose that the algorithm ends with $|J|<n$. This suggests that for each $i\in J^c$, $N^\rmu_{\tilde{\pi}}(i,i)<\gamma_2(n-2)q_\rmu^2$ and $N^\rma(i,i)<\gamma_3mq_\rma^2$. These further imply that
\begin{equation}
    \kappa^\rmu_{G_1\cap G_2}(J^c,J)=\sum_{i\in J^c}N^\rmu_{\tilde{\pi}}(i,i)<\gamma_2(n-2)q_\rmu^2|J^c|,
\end{equation}
and
\begin{equation}
    \kappa^\rma_{G_1\cap G_2}(J^c,\va(G_1))=\sum_{i\in J^c}N^\rma(i,i)<\gamma_3mq_\rma^2|J^c|.
\end{equation}
\begin{sloppypar}
From the attributed \erdos--\renyi graph pair model, it is not hard to see that $G_1\cap G_2'\sim\cg(n,\qu s_\rmu;m,\qa s_\rma)$, where $s_\rmu\triangleq \qu+\ru(1-\qu)$ and $s_\rma\triangleq \qa+\ra(1-\qa)$. 
By conditions~\eqref{eq:rich-cond-3},~\eqref{eq:rich-cond-4} and~\eqref{eq:rich-cond-5}, we know that there exists constants $\lambda_1,\lambda_2,\epsilon$ with $\lambda_1+\lambda_2\ge 1$ such that $n\qu s_\rmu\ge \lambda_1(1+\epsilon)\log n$ and $m\qa s_\rma\ge \lambda_2(1+\epsilon)\log n$. Let $\eta_1$ denote the unique solution in $(0,1)$ to $f(\eta_1)=\frac{\lambda_1(1+\epsilon/8)\log n}{(1-\epsilon/16)n\qu s_\rmu}$, and $\eta_2$ denote the unique solution in $(0,1)$ to $f(\eta_2)=\frac{\lambda_2(1+\epsilon/8)\log n}{m\qa s_\rma}$. By Lemma~\ref{lem:attri-er-expander}, with probability at least $1-O(n^{-\epsilon/8})$, we have at least one of
$$\kappa^\rmu_{G_1\cap G_2}(J^c,J)\ge\eta_1 |J||J^c|\qu s_\rmu\ge \eta_1(1-\epsilon/16)n|J^c|\qu s_\rmu, $$
or 
$$\kappa^\rma_{G_1\cap G_2}(J^c,\va(G_1))\ge \eta_2|J^c|m\qa s_\rma.$$ 
\end{sloppypar}
To complete the proof, it suffices to show that 
\begin{equation}
\label{eq:rich-wts1}
\bar{\gamma}_2\triangleq \eta_1(1-\epsilon/16)\frac{s_\rmu}{\qu}\ge \gamma_2,
\end{equation}
and
\begin{equation}
\label{eq:rich-wts2}
\bar{\gamma}_3\triangleq \eta_2\frac{s_\rma}{\qa}\ge \gamma_3.
\end{equation}
We firstly show~\eqref{eq:rich-wts1} by considering two different cases.

\underline{\textbf{Case 1: $\qu=o(\ru):$}} Because $\eta_1\ge \epsilon/16$, we have
\begin{align*}
    \bar{\gamma}_2&\ge (1-\epsilon/16)\frac{\epsilon s_\rmu}{16\qu}\\
    &=\frac{\epsilon}{16}(1-\epsilon/16)\left(1+\frac{\ru(1-\qu)}{\qu}\right)\\
    &=\omega(1),
\end{align*}
where the last equality follows because $\qu=o(\ru)$. Therefore,
\begin{align}
    f(\bar{\gamma}_2)&>\bar{\gamma}_2(\log \bar{\gamma}_2 -1)\nonumber\\
    &=\omega(\bar{\gamma}_2)\nonumber\\
    &=\omega(s_\rmu/\qu),\label{eq:f-gamma-2}
\end{align}
where the penultimate equality follows because $\bar{\gamma}_2=\omega(1)$ and the last equality follows because $\epsilon=\Theta(1)$. Notice that~\eqref{eq:f-gamma-2} further implies that $f(\bar{\gamma}_2)>\frac{3\log n}{(n-2)q_\rmu^2}=f(\gamma_2)$ because $\frac{s_\rmu(n-2)\qu}{3\log n}=\Omega(1)$ by condition~\eqref{eq:rich-cond-4}. Therefore, we have $\bar{\gamma}_2>\gamma_2$.

\underline{\textbf{Case 2: $\qu=\Omega(\ru):$}} By Lemma~\ref{lem:attri-er-expander}, we have $\eta_1\ge 1-\sqrt{\frac{2(1+\epsilon/8)\log n\lambda_1}{(1-\epsilon/16)n\qu s_\rmu}}$. Because $n\qu s_\rmu\ge nq_\rmu^2=\Omega(n\rho_\rmu^2)=\omega(\log n)$ by~\eqref{eq:rich-cond-1}, we have $\eta_1\ge 1-\epsilon/16$. Hence, we have
\begin{align*}
    \bar{\gamma}_2&\ge (1-\epsilon/16)^2\frac{s_\rmu}{\qu}\\
    &=(1-\epsilon/16)^2\left(1+\frac{\ru(1-\qu)}{\qu}\right)\\
    &\ge (1-\epsilon/16)^2(1+\epsilon)\\
    &\ge 1+\frac{\epsilon}{2},
\end{align*}
where the penultimate inequality follows by assumption~\eqref{eq:rich-cond-6}. Finally, we have $$f(\bar{\gamma}_2)\ge f(1+\epsilon)\ge \frac{3\log n}{(n-2)q_\rmu^2},$$
where the last inequality follows because $f(1+\epsilon)=\Theta(1)$ and $(n-2)q_\rmu^2=\Omega(n\rho_\rmu^2)=\omega(\log n)$ by assumption~\eqref{eq:rich-cond-1}.

To show~\eqref{eq:rich-wts2}, we again consider two cases.

\underline{\textbf{Case 1: $\qa=o(\ra):$}} By Lemma~\ref{lem:attri-er-expander}, we have $\eta_2\ge \epsilon/16$. Therefore, we have 
$$\bar{\gamma}_3\ge\frac{\epsilon }{16}\left(1+\frac{\ra(1-\qa)}{\qa}\right)=\omega(1),$$
where the last equality follows because $\qa=o(\ra)$. It follows that
$$f(\bar{\gamma}_3)>\bar{\gamma}_3(\log \bar{\gamma}_3-1)=\omega(\bar{\gamma}_3)=\omega\left(\frac{\epsilon s_\rma}{16\qa}\right).$$
From here, we conclude that $f(\bar{\gamma}_3)>\frac{3\log n}{mq_\rma^2}=f(\gamma_3)$ because $m\qa s_\rma=\Omega(\log n)$. This completes the proof for $\bar{\gamma}_3>\gamma_3$.

\underline{\textbf{Case 2: $\qa=\Omega(\ra):$}} By Lemma~\ref{lem:attri-er-expander}, we have $\eta_2\ge 1-\sqrt{\frac{2\lambda_2(1+\epsilon/8)\log n}{m\qa s_\rma}}$. It follows that $\eta_2\ge 1-\epsilon/16$ because $m\qa s_\rma\ge mq_\rma^2=\Omega(m\rho_\rma^2)=\omega(\log n)$. Then by assumption~\eqref{eq:rich-cond-7}, we have $$\bar{\gamma}_3\ge (1-\epsilon/16)\frac{s_\rma}{\qa}= (1-\epsilon/16)\left(1+\frac{\ra(1-\qa)}{\qa}\right)\ge (1-\epsilon/16)(1+\epsilon)\ge 1+\epsilon/2.$$ Finally, we have
$$f(\bar{\gamma}_3)\ge f(1+\epsilon/2)>\frac{3\log n}{mq_\rma^2}=f(\gamma_3),$$
where the penultimate inequality follows because $f(1+\epsilon/2)=\Omega(1)$ and $mq_\rma^2=\Omega(m\rho_\rma^2)=\omega(\log n)$. This shows that $\bar{\gamma}_3>\gamma_3$ and completes the proof.

\section{Comparison to Existing works for Seeded Graph Alignment}\label{appd:seeded-comparison}
In this section, we compare the feasible region of the proposed algorithms to the best-known feasible region under the context of seeded graph alignment. 
In the seeded \erdos--\renyi graph pair model $\mathcal{G}(N,\alpha,p,s)$ by~\citet{Mossel-seeded}, the graph generation process is described as follows: First, a base graph $G$ is sampled from \erdos--\renyi model $\mathcal{G}(N,p)$. Next, two correlated graphs $G_1$ and $G_2$ are produced by independently retaining each edge from $G$ with a probability $s$. The graph $G_2'$ is then generated by applying a random permutation $\Pi^*$ to $G_2$. Let $\mathcal{V}_1$ and $\mathcal{V}_2$ denote the sets of vertices in $G_1$ and $G_2'$ respectively. A subset $\mathcal{V}_s\subset \mathcal{V}_1$ of size $\lfloor N\alpha\rfloor$ is then selected uniformly at random, and the vertex pairs $\mathcal{I}_0=\{(v,\Pi^*(v)):v\in\mathcal{V}_s\}$ are identified as the seed set. The graph pair $(G_1,G_2')$ and the seed set $\mathcal{I}_0$ are provided, and the goal is to recover the vertex correspondence for all the remaining vertices. 

In the attributed \erdos--\renyi graph pair model $\mathcal{G}(n,\qu,\ru;m,\qa,\ra)$, when we set $\qu=\qa$ and $\ru=\ra$, the $m$ attributes can be viewed as seeds, and graph pair can be viewed as generated from the seeded \erdos--\renyi graph pair model $\mathcal{G}(n+m,\frac{m}{n+m},\frac{\qu}{\qu+\ru(1-\qu)},\qu+\ru(1-\qu))$ except that the edges between the seeds are removed. Therefore, the feasible region of the proposed algorithms for attributed graph alignment directly implies feasible region for seeded graph alignment. 
% For simplicity, we denote $N\triangleq n+m$, $\alpha\triangleq\frac{m}{m+n}$, $p=\frac{\qu}{\qu+\ru(1-\qu)}$ and $s=\qu+\ru(1-\qu)$. 
In the following, we compare the feasible region of the proposed algorithms to that of the algorithms by~\cite{Mossel-seeded}, and demonstrate that the implied feasible region covers certain parameter regimes that are unknown from the literature. For simplicity, we denote $N\triangleq n+m$, $\alpha\triangleq\frac{m}{m+n}$, $p\triangleq\frac{\qu}{\qu+\ru(1-\qu)}$ and $s\triangleq\qu+\ru(1-\qu)$. Firstly, the feasible region by~\cite{Mossel-seeded} requires a constant correlation $s=\Theta(1)$, while our feasible region includes both cases of $s=o(1)$ and $s=\Theta(1)$. In the rest of the comparison, we consider constant correlation regime $s=\Theta(1)$. In this regime, we demonstrate that the seed fraction $\alpha$ required by the proposed algorithms is strictly smaller in the following two cases:

\noindent\textbf{Case 1: $np=n^c$ for some constant $1/2<c<1$.} In this case, one can check that conditions~\eqref{eq:almost-cond-1}, \eqref{eq:almost-cond-5} and \eqref{eq:almost-cond-2} in Theorem~\ref{thm:almost-exact}, and conditions~\eqref{eq:exact-cond-1} and \eqref{eq:exact-cond-2} in Theorem~\ref{thm:exact} are satisfied without any further assumptions. To satisfy condition~\eqref{eq:almost-cond-4}, we need to set $m=\omega(n^\epsilon)$ for an arbitrarily small $\epsilon$. This is equivalent as $\alpha=\omega(n^{\epsilon-1})$. In contrast, Theorem 4 in \cite{Mossel-seeded} requires at least $\alpha=\Omega(\frac{\log n}{n^c})$.

\noindent\textbf{Case 2: $np=n^c$ for some constant $1/3<c<1/2$.} In this case, one can check that conditions~\eqref{eq:almost-cond-1} and \eqref{eq:almost-cond-5} in Theorem~\ref{thm:almost-exact}, and conditions~\eqref{eq:exact-cond-1} and \eqref{eq:exact-cond-2} in Theorem~\ref{thm:exact} are satisfied without any further assumptions. For conditions~\eqref{eq:almost-cond-4} and~\eqref{eq:almost-cond-2}, we need to set $m=\omega(n^{1-2c})$. This is equivalent as $\alpha=\omega(n^{-2c})$. In contrast, Theorem 4 in \cite{Mossel-seeded} requires at least $\alpha=\Omega(\frac{\log n}{n^{2c}})$.

To provide a fair comparison, we point to that when $np=n^c$ for some $c=1/2$ or $c\le 1/3$ or $c=o(1)$, the algorithms by~\cite{Mossel-seeded} requires strictly less seeds to achieve exact recovery.

\section{Tightness of Propositions~\ref{prop:var-correct-pair} and~\ref{prop:var-wrong-pair}}\label{appd:tightness}
    The conditions presented in Propositions~\ref{prop:var-correct-pair} and~\ref{prop:var-wrong-pair} are not only sufficient, but also necessary for the variance lower bounds under the assumption that $k=\Theta(1)$. By the definition of $\Phi_{ij}$, we can expand its variance expression as
    \begin{align*}
        \var(\Phi_{ij})&=\var\left(\sum_{\cA\subseteq\va:|\cA|=k} W_{i,\cA}(G_1)W_{j,\cA}(G_2')\right)\nonumber\\
        &=\sum_{\cA,\cB\subseteq\va:|\cA|=|\cB|=k} \sum_{S_1\in\cg_{i,\cA}} \sum_{S_2\in\cg_{i,\cA}}\sum_{T_1\in\cg_{i,\cB}} \sum_{T_2\in\cg_{i,\cB}}\cov(\omega_1(S_1)\omega_2(S_2),\omega_1(T_1)\omega_2(T_2)).
    \end{align*}
    By considering specific overlapping patterns of the four subgraphs $S_1,S_2,T_1,T_2$ satisfying the conditions in the summation, we can obtain lower bounds on the variance.
    For a true pair of matched users, we have
    \begin{equation}
    \label{eq:true_lower}
        \frac{\var(\Phi_{ii})}
        {\E[\Phi_{ii}]^2}=\Omega\left(\max\left(\frac{1}{n\qu\ru},\frac{1}{n\rho_\rmu^2}\right)+
        \frac{1}{m\rho_\rmu^2\rho_\rma^2}+\max\left(\frac{1}{n\qu\ru m\qa\ra},
        \frac{1}{n\rho_\rmu^2m\qa\ra}\right)\right).
    \end{equation}
\begin{figure}[htbp]
% \hspace{-30pt}
\centering
\begin{subfigure}{0.2\textwidth}
    \def\svgwidth{1.1\columnwidth}
    %% Creator: Inkscape 1.2.2 (b0a84865, 2022-12-01), www.inkscape.org
%% Accompanies image file 'true_overlap1.eps' (pdf, eps, ps)
\begingroup%
  \makeatletter%
  \providecommand\color[2][]{%
    \errmessage{(Inkscape) Color is used for the text in Inkscape, but the package 'color.sty' is not loaded}%
    \renewcommand\color[2][]{}%
  }%
  \providecommand\transparent[1]{%
    \errmessage{(Inkscape) Transparency is used (non-zero) for the text in Inkscape, but the package 'transparent.sty' is not loaded}%
    \renewcommand\transparent[1]{}%
  }%
  \providecommand\rotatebox[2]{#2}%
  \newcommand*\fsize{\dimexpr\f@size pt\relax}%
  \newcommand*\lineheight[1]{\fontsize{\fsize}{#1\fsize}\selectfont}%
  \ifx\svgwidth\undefined%
    \setlength{\unitlength}{78.22458513bp}%
    \ifx\svgscale\undefined%
      \relax%
    \else%
      \setlength{\unitlength}{\unitlength * \real{\svgscale}}%
    \fi%
  \else%
    \setlength{\unitlength}{\svgwidth}%
  \fi%
  \global\let\svgwidth\undefined%
  \global\let\svgscale\undefined%
  \makeatother%
  \begin{picture}(1,1.41905778)%
    \lineheight{1}%
    \setlength\tabcolsep{0pt}%
    \put(0,0){\includegraphics[width=\unitlength]{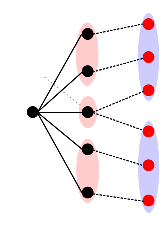}}%
    \put(0.56564955,1.30177585){\color[rgb]{0,0,0}\makebox(0,0)[rt]{\lineheight{1.25}\smash{\begin{tabular}[t]{r}$S_1,S_2$\end{tabular}}}}%
    \put(0.96846561,1.37341767){\color[rgb]{0,0,0}\makebox(0,0)[rt]{\lineheight{1.25}\smash{\begin{tabular}[t]{r}$S_1,S_2$\end{tabular}}}}%
    \put(0.95753487,0.02859839){\color[rgb]{0,0,0}\makebox(0,0)[rt]{\lineheight{1.25}\smash{\begin{tabular}[t]{r}$T_1,T_2$\end{tabular}}}}%
    \put(0.5639125,0.08358686){\color[rgb]{0,0,0}\makebox(0,0)[rt]{\lineheight{1.25}\smash{\begin{tabular}[t]{r}$T_1,T_2$\end{tabular}}}}%
    \put(0.2718584,0.93799796){\color[rgb]{0,0,0}\makebox(0,0)[rt]{\lineheight{1.25}\smash{\begin{tabular}[t]{r}$S_1,S_2,T_1,T_2$\end{tabular}}}}%
    \put(0.2071453,0.57828182){\color[rgb]{0,0,0}\makebox(0,0)[rt]{\lineheight{1.25}\smash{\begin{tabular}[t]{r}$i$\end{tabular}}}}%
  \end{picture}%
\endgroup%

    \caption{}
\label{fig:true_overlap1}
\end{subfigure}
\hspace{5em}
\begin{subfigure}{0.18\textwidth}
    \def\svgwidth{1.1\columnwidth}
    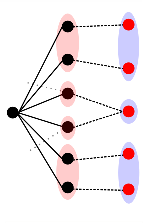
    \caption{}
    \label{fig:true_overlap2}
\end{subfigure}
\hspace{5em}
\begin{subfigure}{0.22\textwidth}
    \def\svgwidth{1.1\columnwidth}
    %% Creator: Inkscape 1.2.2 (b0a84865, 2022-12-01), www.inkscape.org
%% Accompanies image file 'true_overlap3.eps' (pdf, eps, ps)
\begingroup%
  \makeatletter%
  \providecommand\color[2][]{%
    \errmessage{(Inkscape) Color is used for the text in Inkscape, but the package 'color.sty' is not loaded}%
    \renewcommand\color[2][]{}%
  }%
  \providecommand\transparent[1]{%
    \errmessage{(Inkscape) Transparency is used (non-zero) for the text in Inkscape, but the package 'transparent.sty' is not loaded}%
    \renewcommand\transparent[1]{}%
  }%
  \providecommand\rotatebox[2]{#2}%
  \newcommand*\fsize{\dimexpr\f@size pt\relax}%
  \newcommand*\lineheight[1]{\fontsize{\fsize}{#1\fsize}\selectfont}%
  \ifx\svgwidth\undefined%
    \setlength{\unitlength}{78.22458513bp}%
    \ifx\svgscale\undefined%
      \relax%
    \else%
      \setlength{\unitlength}{\unitlength * \real{\svgscale}}%
    \fi%
  \else%
    \setlength{\unitlength}{\svgwidth}%
  \fi%
  \global\let\svgwidth\undefined%
  \global\let\svgscale\undefined%
  \makeatother%
  \begin{picture}(1,1.33239459)%
    \lineheight{1}%
    \setlength\tabcolsep{0pt}%
    \put(0,0){\includegraphics[width=\unitlength]{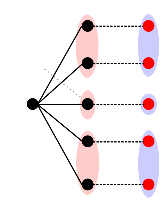}}%
    \put(0.56564962,1.26751353){\color[rgb]{0,0,0}\makebox(0,0)[rt]{\lineheight{1.25}\smash{\begin{tabular}[t]{r}$S_1,S_2$\end{tabular}}}}%
    \put(0.96846568,1.27667735){\color[rgb]{0,0,0}\makebox(0,0)[rt]{\lineheight{1.25}\smash{\begin{tabular}[t]{r}$S_1,S_2$\end{tabular}}}}%
    \put(0.95753494,0.04673638){\color[rgb]{0,0,0}\makebox(0,0)[rt]{\lineheight{1.25}\smash{\begin{tabular}[t]{r}$T_1,T_2$\end{tabular}}}}%
    \put(0.56391257,0.04932468){\color[rgb]{0,0,0}\makebox(0,0)[rt]{\lineheight{1.25}\smash{\begin{tabular}[t]{r}$T_1,T_2$\end{tabular}}}}%
    \put(0.27185847,0.90373551){\color[rgb]{0,0,0}\makebox(0,0)[rt]{\lineheight{1.25}\smash{\begin{tabular}[t]{r}$S_1,S_2,T_1,T_2$\end{tabular}}}}%
    \put(0.20714537,0.57997353){\color[rgb]{0,0,0}\makebox(0,0)[rt]{\lineheight{1.25}\smash{\begin{tabular}[t]{r}$i$\end{tabular}}}}%
    \put(0.65921788,0.56038767){\color[rgb]{0,0,0}\makebox(0,0)[lt]{\lineheight{1.25}\smash{\begin{tabular}[t]{l}$S_1,S_2,T_1,T_2$\end{tabular}}}}%
  \end{picture}%
\endgroup%

    \caption{}
    \label{fig:true_overlap3}
\end{subfigure}
\caption{Three overlapping patterns of $(S_1,S_2,T_1,T_2)$ for a correct pair of users $(i,i)$: In these figures we assume $k=3$. The black nodes represent users and the red nodes represent attributes. The solid lines represent user-user edges and the dotted lines represent user-attribute edges. The subgraph names in the figures indicate which subgraphs do the vertices belong to.}
\label{fig:true-overlap}
\end{figure}
    To see the first term, consider the overlapping pattern shown in Figure~\ref{fig:true_overlap1}. In this case, we have $S_1=S_2$, $T_1=T_2$ and $S_1,T_1$ overlap at only $1$ user-user edge. We can show that $$\cov(\omega_1(S_1)\omega_2(S_2),\omega_1(T_1)\omega_2(T_2))=\Omega(\sigma_\rmu^{4k}\sigma_\rma^{4k}\rho_\rmu^{2k-2}\rho_\rma^{2k}\max(\ru/\qu,1)),$$ and there are in total $\Omega(n^{2k-1}m^{2k})$ copies of $(S_1,S_2,T_1,T_2)$ satisfying this pattern. Together with the first moment expression~\eqref{eq:prop-1-order}, we have $ \frac{\var(\Phi_{ii})}
        {\E[\Phi_{ii}]^2}=\Omega\left(\max\left(\frac{1}{n\qu\ru},\frac{1}{n\rho_\rmu^2}\right)\right).$

\begin{sloppypar}
    For the second term in the lower bound, we consider the overlapping pattern shown in Figure~\ref{fig:true_overlap2}. In this case, we can show that 
    $$\cov(\omega_1(S_1)\omega_2(S_2),\omega_1(T_1)\omega_2(T_2))=\Omega(\sigma_\rmu^{4k}\sigma_\rma^{4k}\rho_\rmu^{2k-2}\rho_\rma^{2k-2}),$$ and $\Omega(n^{2k}m^{2k-1})$ copies of $(S_1,S_2,T_1,T_2)$ follow this pattern. This would lead to the lower bound $\frac{\var(\Phi_{ii})}
        {\E[\Phi_{ii}]^2}=\Omega\left(\frac{1}{m\rho_\rmu^2\rho_\rma^2}\right).$
\end{sloppypar}
    The third term in~\eqref{eq:true_lower} comes from the overlapping pattern shown in Figure~\ref{fig:true_overlap3}. In this case, we can show that $$\cov(\omega_1(S_1)\omega_2(S_2),\omega_1(T_1)\omega_2(T_2))=\Omega\left(\sigma_\rmu^{4k}\sigma_\rma^{4k}\rho_\rmu^{2k-2}\rho_\rma^{2k-2}\frac{\ra}{\qa}\max(\ru/\qu,1)\right),$$ and there are $\Omega(n^{2k-1}m^{2k-1})$ copies of $(S_1,S_2,T_1,T_2)$ following this pattern. With the first moment expression~\eqref{eq:prop-1-order}, we can get the lower bound $\frac{\var(\Phi_{ii})}
        {\E[\Phi_{ii}]^2}=\Omega\left(\max\left(\frac{1}{n\qu\ru m\qa\ra},
        \frac{1}{n\rho_\rmu^2m\qa\ra}\right)\right).$

        \begin{figure}[htbp]
% \hspace{-30pt}
\centering
\begin{subfigure}{0.2\textwidth}
    \def\svgwidth{1.1\columnwidth}
    %% Creator: Inkscape 1.2.2 (b0a84865, 2022-12-01), www.inkscape.org
%% Accompanies image file 'wrong_overlap1.eps' (pdf, eps, ps)
\begingroup%
  \makeatletter%
  \providecommand\color[2][]{%
    \errmessage{(Inkscape) Color is used for the text in Inkscape, but the package 'color.sty' is not loaded}%
    \renewcommand\color[2][]{}%
  }%
  \providecommand\transparent[1]{%
    \errmessage{(Inkscape) Transparency is used (non-zero) for the text in Inkscape, but the package 'transparent.sty' is not loaded}%
    \renewcommand\transparent[1]{}%
  }%
  \providecommand\rotatebox[2]{#2}%
  \newcommand*\fsize{\dimexpr\f@size pt\relax}%
  \newcommand*\lineheight[1]{\fontsize{\fsize}{#1\fsize}\selectfont}%
  \ifx\svgwidth\undefined%
    \setlength{\unitlength}{65.87772489bp}%
    \ifx\svgscale\undefined%
      \relax%
    \else%
      \setlength{\unitlength}{\unitlength * \real{\svgscale}}%
    \fi%
  \else%
    \setlength{\unitlength}{\svgwidth}%
  \fi%
  \global\let\svgwidth\undefined%
  \global\let\svgscale\undefined%
  \makeatother%
  \begin{picture}(1,1.65572345)%
    \lineheight{1}%
    \setlength\tabcolsep{0pt}%
    \put(0,0){\includegraphics[width=\unitlength]{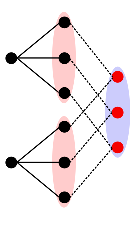}}%
    \put(0.49958636,1.60002292){\color[rgb]{0,0,0}\makebox(0,0)[rt]{\lineheight{1.25}\smash{\begin{tabular}[t]{r}$S_1,T_1$\end{tabular}}}}%
    \put(0.47928,0.04318859){\color[rgb]{0,0,0}\makebox(0,0)[rt]{\lineheight{1.25}\smash{\begin{tabular}[t]{r}$S_2,T_2$\end{tabular}}}}%
    \put(0.09027971,1.08213916){\color[rgb]{0,0,0}\makebox(0,0)[rt]{\lineheight{1.25}\smash{\begin{tabular}[t]{r}$i$\end{tabular}}}}%
    \put(0.08689477,0.3138755){\color[rgb]{0,0,0}\makebox(0,0)[rt]{\lineheight{1.25}\smash{\begin{tabular}[t]{r}$j$\end{tabular}}}}%
    \put(0.72316118,0.37495872){\color[rgb]{0,0,0}\makebox(0,0)[lt]{\lineheight{1.25}\smash{\begin{tabular}[t]{l}$$\end{tabular}}}}%
    \put(0.69683605,0.3845315){\color[rgb]{0,0,0}\makebox(0,0)[lt]{\lineheight{1.25}\smash{\begin{tabular}[t]{l}$S_1,S_2,T_1,T_2$\end{tabular}}}}%
  \end{picture}%
\endgroup%

    \caption{}
\label{fig:wrong_overlap1}
\end{subfigure}
\hspace{5em}
\begin{subfigure}{0.2\textwidth}
    \def\svgwidth{1.1\columnwidth}
    %% Creator: Inkscape 1.2.2 (b0a84865, 2022-12-01), www.inkscape.org
%% Accompanies image file 'wrong_overlap2.eps' (pdf, eps, ps)
\begingroup%
  \makeatletter%
  \providecommand\color[2][]{%
    \errmessage{(Inkscape) Color is used for the text in Inkscape, but the package 'color.sty' is not loaded}%
    \renewcommand\color[2][]{}%
  }%
  \providecommand\transparent[1]{%
    \errmessage{(Inkscape) Transparency is used (non-zero) for the text in Inkscape, but the package 'transparent.sty' is not loaded}%
    \renewcommand\transparent[1]{}%
  }%
  \providecommand\rotatebox[2]{#2}%
  \newcommand*\fsize{\dimexpr\f@size pt\relax}%
  \newcommand*\lineheight[1]{\fontsize{\fsize}{#1\fsize}\selectfont}%
  \ifx\svgwidth\undefined%
    \setlength{\unitlength}{65.87772489bp}%
    \ifx\svgscale\undefined%
      \relax%
    \else%
      \setlength{\unitlength}{\unitlength * \real{\svgscale}}%
    \fi%
  \else%
    \setlength{\unitlength}{\svgwidth}%
  \fi%
  \global\let\svgwidth\undefined%
  \global\let\svgscale\undefined%
  \makeatother%
  \begin{picture}(1,1.65572312)%
    \lineheight{1}%
    \setlength\tabcolsep{0pt}%
    \put(0,0){\includegraphics[width=\unitlength]{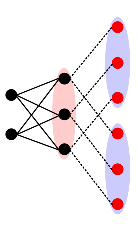}}%
    \put(0.90163562,1.55215938){\color[rgb]{0,0,0}\makebox(0,0)[rt]{\lineheight{1.25}\smash{\begin{tabular}[t]{r}$S_1,S_2$\end{tabular}}}}%
    \put(0.89760199,0.02517431){\color[rgb]{0,0,0}\makebox(0,0)[rt]{\lineheight{1.25}\smash{\begin{tabular}[t]{r}$S_1,S_2$\end{tabular}}}}%
    \put(0.09027971,0.81410675){\color[rgb]{0,0,0}\makebox(0,0)[rt]{\lineheight{1.25}\smash{\begin{tabular}[t]{r}$i$\end{tabular}}}}%
    \put(0.08689477,0.51968619){\color[rgb]{0,0,0}\makebox(0,0)[rt]{\lineheight{1.25}\smash{\begin{tabular}[t]{r}$j$\end{tabular}}}}%
    \put(0.72316118,0.37495872){\color[rgb]{0,0,0}\makebox(0,0)[lt]{\lineheight{1.25}\smash{\begin{tabular}[t]{l}$$\end{tabular}}}}%
    \put(0.19427529,1.19580829){\color[rgb]{0,0,0}\makebox(0,0)[lt]{\lineheight{1.25}\smash{\begin{tabular}[t]{l}$S_1,S_2,T_1,T_2$\end{tabular}}}}%
  \end{picture}%
\endgroup%

    \caption{}
    \label{fig:wrong_overlap2}
\end{subfigure}
\hspace{5em}
\begin{subfigure}{0.22\textwidth}
    \def\svgwidth{1.1\columnwidth}
    %% Creator: Inkscape 1.2.2 (b0a84865, 2022-12-01), www.inkscape.org
%% Accompanies image file 'wrong_overlap3.eps' (pdf, eps, ps)
\begingroup%
  \makeatletter%
  \providecommand\color[2][]{%
    \errmessage{(Inkscape) Color is used for the text in Inkscape, but the package 'color.sty' is not loaded}%
    \renewcommand\color[2][]{}%
  }%
  \providecommand\transparent[1]{%
    \errmessage{(Inkscape) Transparency is used (non-zero) for the text in Inkscape, but the package 'transparent.sty' is not loaded}%
    \renewcommand\transparent[1]{}%
  }%
  \providecommand\rotatebox[2]{#2}%
  \newcommand*\fsize{\dimexpr\f@size pt\relax}%
  \newcommand*\lineheight[1]{\fontsize{\fsize}{#1\fsize}\selectfont}%
  \ifx\svgwidth\undefined%
    \setlength{\unitlength}{65.87772489bp}%
    \ifx\svgscale\undefined%
      \relax%
    \else%
      \setlength{\unitlength}{\unitlength * \real{\svgscale}}%
    \fi%
  \else%
    \setlength{\unitlength}{\svgwidth}%
  \fi%
  \global\let\svgwidth\undefined%
  \global\let\svgscale\undefined%
  \makeatother%
  \begin{picture}(1,1.65572312)%
    \lineheight{1}%
    \setlength\tabcolsep{0pt}%
    \put(0,0){\includegraphics[width=\unitlength]{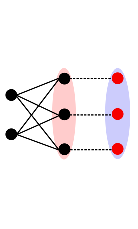}}%
    \put(0.09027971,0.81410675){\color[rgb]{0,0,0}\makebox(0,0)[rt]{\lineheight{1.25}\smash{\begin{tabular}[t]{r}$i$\end{tabular}}}}%
    \put(0.08689477,0.51968619){\color[rgb]{0,0,0}\makebox(0,0)[rt]{\lineheight{1.25}\smash{\begin{tabular}[t]{r}$j$\end{tabular}}}}%
    \put(0.12008761,1.19580829){\color[rgb]{0,0,0}\makebox(0,0)[lt]{\lineheight{1.25}\smash{\begin{tabular}[t]{l}$S_1,S_2,T_1,T_2$\end{tabular}}}}%
    \put(0.68961215,1.19542354){\color[rgb]{0,0,0}\makebox(0,0)[lt]{\lineheight{1.25}\smash{\begin{tabular}[t]{l}$S_1,S_2,T_1,T_2$\end{tabular}}}}%
  \end{picture}%
\endgroup%

    \caption{}
    \label{fig:wrong_overlap3}
\end{subfigure}
\caption{Three overlapping patterns of $(S_1,S_2,T_1,T_2)$ for a wrong pair of users $(i,j)$: In these figures we assume $k=3$. The black nodes represent users and the red nodes represent attributes. The solid lines represent user-user edges and the dotted lines represent user-attribute edges. The subgraph names in the figures indicate which subgraphs do the vertices belong to.}
\label{fig:wrong-overlap}
\end{figure}
        
%         \begin{figure}[htbp]
% % \hspace{-30pt}
% \centering
% \begin{subfigure}{0.2\textwidth}
%     \def\svgwidth{1.1\columnwidth}
%     \input{./images/wrong_overlap1.eps_tex}
%     \caption{}
% \label{fig:wrong_overlap1}
% \end{subfigure}
% \hspace{5em}
% \begin{subfigure}{0.2\textwidth}
%     \def\svgwidth{1.1\columnwidth}
%     \input{./images/wrong_overlap2.eps_tex}
%     \caption{}
%     \label{fig:wrong_overlap2}
% \end{subfigure}
% \hspace{5em}
% \begin{subfigure}{0.22\textwidth}
%     \def\svgwidth{1.1\columnwidth}
%     \input{./images/wrong_overlap3.eps_tex}
%     \caption{}
%     \label{fig:wrong_overlap3}
% \end{subfigure}
% \caption{Three overlapping patterns of $(S_1,S_2,T_1,T_2)$ for a wrong pair of users $(i,j)$: In these figures we assume $k=3$. The black nodes represent users and the red nodes represent attributes. The solid lines represent user-user edges and the dotted lines represent user-attribute edges. The subgraph names in the figures indicate which subgraphs do the vertices belong to.}
% \label{fig:wrong-overlap}
% \end{figure}

For a wrong pair of matching $(i,j)$, we can similarly show that
    \begin{equation}
    \label{eq:wrong_lower}
        \frac{\var(\Phi_{ij})}
        {\E[\Phi_{ii}]^2}=\Omega\left(\frac{1}{(m\rho_\rmu^2\rho_\rma^2)^k}+
        \frac{1}{(n\rho_\rmu^2)^k}+\frac{1}{(n\rho_\rmu^2 m\qa\ra)^k}\right),
    \end{equation}
    where the three terms in the lower bound come from the three overlapping patterns shown in Figure~\ref{fig:wrong-overlap} respectively. Note the condition $n^2\qu=\Omega(1)$ in Propositions~\ref{prop:var-wrong-pair} is not shown to be necessary by~\eqref{eq:wrong_lower}. However, $n^2\qu=\Omega(1)$ is a mild condition ensuring the existence of at least one user-user edge in graphs $G_1$ and $G_2'$. Moreover, it is implied by condition $n\qu\ru=\omega(1)$ in Proposition~\ref{prop:var-correct-pair}, so it does not affect the overall feasible region of the algorithm.
\end{document}